\documentclass{tCPH2e}

\usepackage{epstopdf}

\theoremstyle{plain}
\newtheorem{theorem}{Theorem}[section]

\newtheorem{lemma}[theorem]{Lemma}

\usepackage{hyperref}

\usepackage{bm}
\def\bm#1{\textbf{\em #1}}

\usepackage{mathptmx}

\usepackage{amssymb}
\usepackage{amsfonts}
\usepackage{amsmath}

\usepackage{todonotes}
\usepackage{enumitem}
\usepackage{booktabs}
\usepackage{graphicx}
\usepackage{caption}
\usepackage{subcaption}
\usepackage{rotating}

\theoremstyle{definition}

\theoremstyle{remark}

\newtheorem{Teorema}{Theorem}

\newtheorem{Definicao}{Definition}

\usepackage{changes}

\definechangesauthor[name={Deniz Eroglu}, color=blue]{DE}

\begin{document}


\articletype{}

\title{\textit{Synchronization of Chaos}}

\author{
\name{Deniz Eroglu\textsuperscript{a,b}, Jeroen Lamb\textsuperscript{b} and 
Tiago Pereira\textsuperscript{a}$^{\ast}$\thanks{$^\ast$Corresponding author. Email: tiago@icmc.usp.br}}
\affil{\textsuperscript{a} Instituto de Ci\^encias Matem\'aticas e Computa\c{c}\~ao, Universidade de S\~ao Paulo, Brazil; \\
\textsuperscript{b}Department of Mathematics, Imperial College London, UK}
\received{March 22, 2017}
}

\maketitle

\begin{abstract}
Dynamical networks are important models for the behaviour of complex systems, modelling physical, biological and societal systems, including the brain, food webs, epidemic disease in populations, power grids and many other. Such dynamical networks can exhibit behaviour in which deterministic chaos, exhibiting unpredictability and disorder, coexists with synchronization, a classical paradigm of order. We survey  the main theory behind complete, generalized and phase synchronization phenomena in simple as well as complex networks and discuss applications to secure communications, parameter estimation and the anticipation of chaos.

%
\end{abstract}

\begin{keywords}
synchronization; interaction; networks; stability; coupled systems
\end{keywords}

\tableofcontents

\vspace{-0.5cm}

\newpage
\section{Introduction} \label{sec:introduction}
This survey provides an introduction to the phenomenon of synchronization in coupled chaotic dynamical systems. Both chaos and synchronization are important concepts in science, from a philosophical as well as a practical point of view.

Synchronization expresses a notion of strong correlations between coupled systems. In its most elementary and intuitive form, synchronization refers to the tendency to have the same dynamical behaviour. Scientists also recognize weaker forms of synchronization, where some key aspects of dynamical behaviour are the same - like frequencies - or where coupled dynamical behaviours satisfy a specific spatiotemporal relationship - like a constant phase lag. 

Synchronization is fundamental to our understanding of a wide range of natural phenomena, from cosmology and natural rhythms like heart beating \cite{strogatz2003} and hand clapping \cite{neda2000} to superconductors \cite{wiesenfeld1996}. While synchronization is often beneficial, some pathologies of the brain such as Parkinson disease \cite{hammond2007, tass1998} and epilepsy \cite{dominguez2005} are also related to this phenomenon. In ecology, synchronization of predators can lead to extinction \cite{earn1998, earn2000} while improving the quality of synchronized behaviour of prey can increase the odds to survive~\cite{bode2010}. In epidemiology, synchronization in measles outbreaks can cause social catastrophes \cite{grenfell2001}.  

Synchronization
is also relevant to technology. Lasers form an important example. The stability of a laser generally decreases when its power increases. A successful way to create a high-power laser system is by combining many low-power stable lasers. A key challenge is to make sure that the lasers synchronize \cite{winful1990, oliva2001, hirosawa2013}, as without synchronization destructive interference diminishes power.  

synchronization can also cause engineering problems.  A recent well-publicized example concerns the {\it London Millennium Bridge}, traversing the River Thames. On the opening day in 2000, the bridge attracted 90,000 visitors, holding up to 2000 visitors on the bridge at the same time. Lateral motion caused by the pedestrians made the bridge lurches to one side, as a result of which the pedestrians would adjust their rhythm to keep from falling over. In turn, this led to increased oscillations of the bridge due to the synchronization between the bridge's oscillations and pedestrians' gait  \cite{eckhardt2007,strogatz2005, belykh2016}. Eventually, the oscillations became so extensive that the bridge was closed for safety reasons. The bridge was only opened to the public again after a redesign where dampers were installed to increase energy dissipation and thereby impede synchronization between bridge and pedestrians. 

The above examples of synchronization in coupled systems describe a spontaneous transition to order because of the interaction.  Coupled systems are modelled as networks of interacting elements. We often have a detailed understanding about the dynamics of the individual uncoupled elements. For example, we have reasonably good models for  individual superconducting Josephson junctions, heart cells, neurons, lasers and even pedestrians. 
In the systems we consider here, the coupling between elements is assumed to be built up from bilateral interactions between pairs of elements, and a network structure indicating which pairs of elements interact with each other. First, the way the individuals talk to each other.  For example, in the neurons the interaction is mediated by synapses and in heart cells by electrical diffusion. Second, the linking structure describing who is influencing whom. 
So, it is the network structure that provides the interaction among individual elements. The collective behavior emerges from the collaboration and competition of many elements mediated by the network structure.

Synchronization can be effectively used to create secure communication schemes \cite{kocarev1995,stankovski2014,ren2013}. And it can help developing new technologies. Synchronization is also used for model calibration, that is, the synchronized regime between data and equations can reveal the parameter of the equations \cite{parlitz1996b, yu2008}. 

Chaos in dynamics is one of the scientific revolutions of the twentieth century that has deepened our understanding of the nature of unpredictability. Initiated by Henri Poincar\'e in the late 19th century, the chaos revolution took off in the 1980s when computers with which chaotic dynamics can be studied and illustrated, became more widely available. Chaos normally arises when recurrent dynamical behaviour has locally dispersing characteristics, as measured by a positive Lyapunov exponent. In this review, we will not discuss any details of chaotic dynamics in detail. For a comprehensive monograph on this topic, see for instance \cite{katok1997}.

At first sight it may appear that the concept of synchronization, as an expression of order, and the concept of chaos, associated with disorder, could not be more distant from one another. Hence, it was quite a surprise when physicists realised that coupled chaotic systems also could  spontaneously synchronize \cite{fujisaka1983,afraimovich1986}. Despite many years of studies into this phenomenon and its applications, many fundamental problems remain open. This survey is meant to provide a concise overview of some of the most important theoretical insights underlying our current understanding of synchronization of chaos as well as highlighting some of the many remaining challenges.   

In this review, we will discuss the basic results for synchronization of chaotic systems. 
The interaction can make these systems adapt and display a complicated unpredictable dynamics while behaving in a synchronous manner.    Synchronization in these systems can appear in hierarchy depending on the details of the individual elements and the network structure. We will first discuss this hierarchy in two coupled chaotic oscillators and latter generalize to complex networks. 
The review is organised as follows. In Section~\ref{sec:two_systems} we discuss the synchronization scenarious between two coupled oscillators. In Section~\ref{sec:applications} we discuss such applications where the synchronization phenomenon can be used for prediction and parameter estimation. In Section \ref{sec:networks} we discuss synchronization in complex networks. 


\section{Synchronization between two coupled systems}
\label{sec:two_systems}
\subsection{Synchronization of linear systems}

Before touching upon more topical and interesting settings in which synchronization is observed in nonlinear systems, as an introduction we consider the elementary example of synchronization between two linearly coupled linear systems. Although simple, this example bears the main ideas of the general case of synchronization between two or more nonlinearly coupled systems. 

We consider two identical linear systems
$$
\dot{x}_i := \frac{dx_i}{dt} = ax_i,~~i=1,2
$$
with $a$ a non-zero constant. The solution of these linear differential equations with initial condition $x_i(0)$ is $x_i(t)  = e^{a t}x_i(0)$ so that the ensuing dynamics is simple and all solutions converge exponentially fast to zero if $a<0$, or diverge to infinity if $a>0$ unless $x(0)=0$.

We now consider these linear systems coupled in the following way
\begin{eqnarray}
\dot{x_1} = a x_1 + \alpha ( x_2 - x_1) \nonumber \\
\dot{x_2} = a x_2 + \alpha ( x_1 - x_2)
\label{eq:coupled_linear}
\end{eqnarray}
and $\alpha$  is called the {\it coupling parameter}.
  
In the context of this model, we speak of {\textit{Complete Synchronization} (CS) if $x_1(t)$ and $x_2(t)$ converge to each other as $t\to\infty$. 
In order to study this phenomenon, it is natural to consider the new variable
$$
z:= x_1 - x_2.
$$
In terms of this variable, synchronization corresponds to the fact that $\lim_{t\to\infty} z(t)=0$. As $\dot z = \dot x_1 - \dot x_2$, we find directly by substitution from Eq.~(\ref{eq:coupled_linear}) that 
$$
\dot z = (a - 2 \alpha) z,
$$
which has the explicit solution $ z(t) = z(0) e^{( a - 2 \alpha) t} $. Hence, we find that $\lim_{t\to\infty} z(t)=0$ if and only if $a-2\alpha<0$ (unless the initial condition is already synchronized, i.e.~$z(0)=0$). 
Defining the {\it critical} coupling value $\alpha_c$ as
\begin{equation}
\label{cdif}
\alpha_c := \frac{a}{2} 
\end{equation}
we thus obtain synchronization if the coupling parameter exceeds the critical value: $\alpha>\alpha_c$. 

We finally note that as the system synchronizes, the coupling term converges to zero and the solution of each of the components behaves in accordance with the underlying uncoupled linear system: to be precise, $\lim_{t\to\infty}\left(x_i(t)-\frac{x_1(0)+x_2(0)}{2}e^{at}\right)=0$ for $i=1,2$. It is important to note that the sign of the parameter $a$ here determines a difference between synchronization to the trivial equilibrium (if $a<0$) or to an exponentially growing solution (if $a>0$).

In view of later generalizations, we will go through the above analysis again, exploiting more the linear structure of the problem so that we can appreciate synchronization in terms of spectral properties of the coupling term. 

With
$
\bm{x} :=
\left( 
\begin{array}{c}
x_1 \\
x_2
\end{array}
\right),
$
(\ref{eq:coupled_linear}) can be written as
\begin{equation}
\dot{\bm{x}} = \left[  a \bm{I} - \alpha \bm{L} \right] \bm{x}
\label{eq:odelinear}
\end{equation}
where $$
\bm{I} = 
\left( 
\begin{array}{cc}
1 & 0 \\
0 & 1
\end{array}
\right)~~\mbox{and}~~
\bm{L} = 
\left( 
\begin{array}{rr}
1 & -1 \\
-1 & 1
\end{array}
\right).
$$
$L$ is known as the {\it Laplacian matrix}. In Section~\ref{sec:networks}, we will generalize it to any network. 
The solution of (\ref{eq:odelinear}) with initial condition $\bm{x}(0)$ is
\begin{equation}
\bm{x}(t) = e^{\left[  a \bm{I} - \alpha \bm{L} \right]  t  } \bm{x}(0), ~~\mbox{where}~~e^{At}:=\sum_{n=0}^\infty \frac{t^n}{n!} A^n .
\label{aL}
\end{equation}


To solve (\ref{aL}) we note that since $\bm{I}$ and $L$ commute, 
\[
e^{\left[  a \bm{I} - \alpha \bm{L} \right]  t  }=e^{a \bm{I}t}e^{ - \alpha \bm{L}t }
\]
and $e^{a \bm{I}t}=e^{at}\bm{I}$. In order to evaluate $e^{ - \alpha \bm{L}t }$, it is useful to observe that $\bm{v}_1 = (1,1)^*$ and $\bm{v}_2 = (1,-1)^*$ are the eigenvectors of $L$ for its corresponding eigenvalues $\lambda_1=0$ and $\lambda_2  = 2$. As $\{ \bm{v}_1, \bm{v}_2 \}$ is a basis of $\mathbb{R}^2$, we
may write any initial condition as $\bm{x}(0) = c_1 \bm{v}_1 + c_2 \bm{v}_2$ with $c_1,c_2\in\mathbb{R}$, so that 
\[
e^{ - \alpha \bm{L}t }\bm{x}(0)=c_1 \bm{v}_1 + c_2 e^{-\alpha \lambda_2 t } \bm{v}_2
\]
and
\begin{equation}
\bm{x}(t) =e^{\left[  a \bm{I} - \alpha \bm{L} \right]  t  } \bm{x}(0)=c_1 e^{a t } \bm{v}_1 + c_2 e^{( a - \alpha \lambda_2 ) t } \bm{v}_2.
\end{equation}
%
%
%
%
Synchronization corresponds to the phenomenon that $\bm{x}(t)$ converges to the {\it synchronization subspace} generated by $\bm{v}_1$. This only happens if $\lim_{t \rightarrow \infty} c_2 e^{( a - \alpha \lambda_2 ) t } \bm{v}_2  = 0$, i.e.~if $\alpha>\frac{a}{\lambda_2}$. Thus in view of (\ref{cdif}), we define the {\em critical coupling} value 
\[
\alpha_c=\frac{a}{\lambda_2}=\frac{a}{2}.
\]
 We note that the critical coupling value is expressed in terms of the gap between the lowest eigenvalue $0$ and smallest nonzero (and positive) eigenvalue of the Laplacian $L$. 


\subsection{Complete synchronization of nonlinear systems}
\label{sec:complete_sync}
\label{sec:complete_sync}
%
%
%


We now consider two fully diffusively coupled identical nonlinear $n$-dimensional systems
\begin{equation}
\begin{array}{rcl}
\dot{\bm{x}}_1 &=& \bm{f}(\bm{x}_1) + \alpha \bm{H}(\bm{x}_2 - \bm{x}_1)  \\
\dot{\bm{x}}_2 &=& \bm{f}(\bm{x}_2) + \alpha \bm{H}(\bm{x}_1- \bm{x}_2) 
\end{array}
\label{eq:nonlinear}\end{equation}
where $\bm{f}:\mathbb{R}^n \rightarrow \mathbb{R}^n$ is in general nonlinear and $\bm{H}:\mathbb{R}^n \rightarrow \mathbb{R}^n$ is a smooth coupling function. We assume that $\bm{H}(\bm{0})=\bm{0}$ so that the synchronization subspace 
$
\bm{x}_1=\bm{x}_2
$
is invariant for all coupling strengths $\alpha$. Meaning that for any synchronized initial condition the entire solution remains synchronized: as in the synchronized state the diffusive coupling term vanishes, the dynamics is identical to that of the uncoupled system (with $\alpha=0$). 
Consequently, the coupling has no influence on the synchronized motion. In particular, it could be the case that the synchronized motion is chaotic, if the uncoupled systems exhibit such behaviour.

We aim to show that if the coupling is sufficiently strong, the system Eq.~(\ref{eq:nonlinear}) will synchronize $\bm{x}_1(t)-\bm{x}_2(t) \to 0$ as $t \to \infty$. We consider $\bm{H}=\bm{I}$ (the identity matrix) then the term reads as
\[
\alpha\bm{H}({\bm x}_2-{\bm x}_1) = \alpha({\bm x}_2-{\bm x}_1).
\]

To analyze stability, we consider -- as before -- the evolution of the difference variable
$\bm{z} := \bm{x}_1 - \bm{x}_2$ in terms of which the synchronization subspace is characterized as $\bm{z}=0$:  
\begin{eqnarray}
\dot{\bm{z}} &=& \dot{\bm{x}}_1 - \dot{\bm{x}}_2 \\
& = &  \bm{f}(\bm{x}_1) - \bm{f}(\bm{x}_2) -  2 \alpha \bm{z}
\label{eq:nonlinearz}
\end{eqnarray}
The aim is to identify sufficient conditions for the coupling parameter $\alpha$ to
guarantee that locally near $\bm{z}=0$ we have $\lim_{t \rightarrow \infty } \bm{z}(t) =\bm{0}$. To this end, we linearize the equations of motion Eq.~(\ref{eq:nonlinearz}) near $\bm{z}=0$. We note to this extent that near $\bm{x}_1=\bm{x}_2$ we obtain by Taylor expansion that
\begin{eqnarray*}
\bm{f}(\bm{x}_2(t)) &=& \bm{f}(\bm{x}_1(t))-D\bm{f}(\bm{x}_1(t))(\bm{x}_2(t)-\bm{x}_1(t))+O(\|\bm{x}_1(t)-\bm{x}_2(t)\|^2)\\
&=& \bm{f}(\bm{x}_1(t))-D\bm{f}(\bm{x}_1(t))\bm{z}(t)+O(\|\bm{z}(t)\|^2).
\end{eqnarray*} 
Here $D\bm{f}(\bm{x}_1(t))$ is the derivative (Jacobian matrix of $\bm{f}(\bm{x})$) at $\bm{x}=\bm{x}_1(t)$. We use this to write Eq.~(\ref{eq:nonlinearz}) near $\bm{z}=0$ as
%
\begin{equation}
\label{var2}
\frac{d \bm{z}}{dt } = [  D \bm{f}(\bm{x}_1(t))  - 2 \alpha \bm{I} ] \bm{z}+O(\|\bm{z}\|^2).
\end{equation}
The linear part of this equation, obtained by ignoring the $O(|\bm{z}|^2)$ term in Eq.~(\ref{var2}), is commonly known as the {\it first variational equation} .
It should be noted that this equation is nonautonomous as it depends explicitly on the reference solution $\bm{x}_1(t)$. In general it is not easy to analyze nonautonomous differential equations, not even linear ones. Fortunately, we are able to achieve insights without solving this equation because the coupling is rather convenient adding an extra damping term $-\alpha \bm z$.

To simplify the analysis, we introduce a new variable
\begin{equation}\label{ansatz}
\bm{w}(t)=e^{2\alpha t} \bm{z}(t)
\end{equation}
in terms of which the linear part of Eq.~(\ref{var2}) becomes precisely the variational equation for the solution $\bm{x}_1(t)$ of the uncoupled equation of motion $\dot{\bm{x}}=\bm f(\bm{x})$:
\begin{eqnarray}
 \dot{\bm{w}}(t) &=& 2 \alpha e^{2\alpha t}\bm{z}(t) + e^{2\alpha t}\dot{\bm{z}}(t) \nonumber \\
 &=& 2 \alpha \bm{w} + [  D \bm{f}(\bm{x}_1(t))  - 2 \alpha \bm{I} ] e^{2\alpha t} \bm{z} \nonumber \\ 
 &=& [  D \bm{f}(\bm{x}_1(t)) ] \bm{w}.
\end{eqnarray}

Let 
$\Phi(\bm{x}_1(t))$ be the fundamental matrix for the variational equation, so that any solution of this nonautonomous equation can be written as $\bm{z}(t)=\Phi(\bm{x}_1(t))\bm{z}(0)$. Let  $\{\lambda_j((\bm{x}_1(t))\}_{j=1}^n$ be the set of positive square roots of the eigenvalues of the symmetric matrix $\Phi(\bm{x}_1(t))^*\Phi(\bm{x}_1(t))$ (where $^*$ denotes transpose).
Then we define
\begin{equation}
\Lambda:=\max_j\lim_{t\to\infty} \frac{1}{t}  \lambda_j(\bm{x}_1(t)).
\end{equation} 
$\Lambda$ is known as the {\it Lyapunov exponent} of the orbit $\bm{x}_1(t)$ and it measures the infinitesimal asymptotic divergence rate near this trajectory. We refer to 
Appendix~\ref{sec:lyapunov_exp} for more details about the Lyapunov exponent.

The assertion now is that if the orbit $\bm{x}_1(t)$ has Lyapunov exponent $\Lambda$, then there exists a constant $C>0$ such that
\begin{equation}\label{expbound}
 ||\bm{w}(t)||\leq C e^{\Lambda t}.
\end{equation}


From Eq. (\ref{expbound}) and using Eq.(\ref{ansatz}) we obtain that 
$$
\| \bm{z}(t) \| \le C e^{(\Lambda - 2\alpha ) t}.
$$
Hence,
$$
\alpha_c := \frac{\Lambda}{2} 
$$
is a critical coupling strength for synchronization, above which observe synchronization.

A complication with the above analysis, is that the Lyapunov exponent $\Lambda$ and constant $C$ may depend on the chosen trajectory $\bm{x}_1(t)$. The probabilistic (ergodic) theory of dynamical systems, which we will not dwell on here, asserts that often the Lyapunov exponent is constant for almost all trajectories on a given attractor. However, the constant $C$ may still vary per trajectory, which leads to {\em non-uniform} synchronization, implying the potential of large variation of transit times until synchronization occurs. For similar phenomena, see also \cite{ashwin1994, ashwin1998}.  
We now proceed to apply the above to the examples of coupled Lorenz and R\"ossler systems. 

%

\noindent \textbf{Lorenz system.} The Lorenz system was introduced by Edward Lorenz in 1963 as a simplified model for atmospheric convection: 
\begin{equation}
\begin{array}{rcl}
\dot{x}&=&\sigma ( y -x ),  \\ \dot{y}&=&x ( \rho  - z) - y, \\ \dot{z}&=&-\beta z + xy,
\end{array}
\label{eq:lorenz_system}
\end{equation}
where the three coordinates $x$, $y$ and $z$ represent the state of the system and $\sigma$, $\rho$, $\beta$ are parameters. When parameter values are chosen as $\sigma=10$,  $\rho= 28$ and $\beta = 8/3$, the equations display unpredictable (chaotic) dynamics. 
Lorenz used this choice of parameters in his original paper \cite{lorenz1963}. We use these parameter values as well.

\begin{figure}[h]
\centering
\begin{subfigure}[t]{0.4\textwidth}
\includegraphics[width=\linewidth]{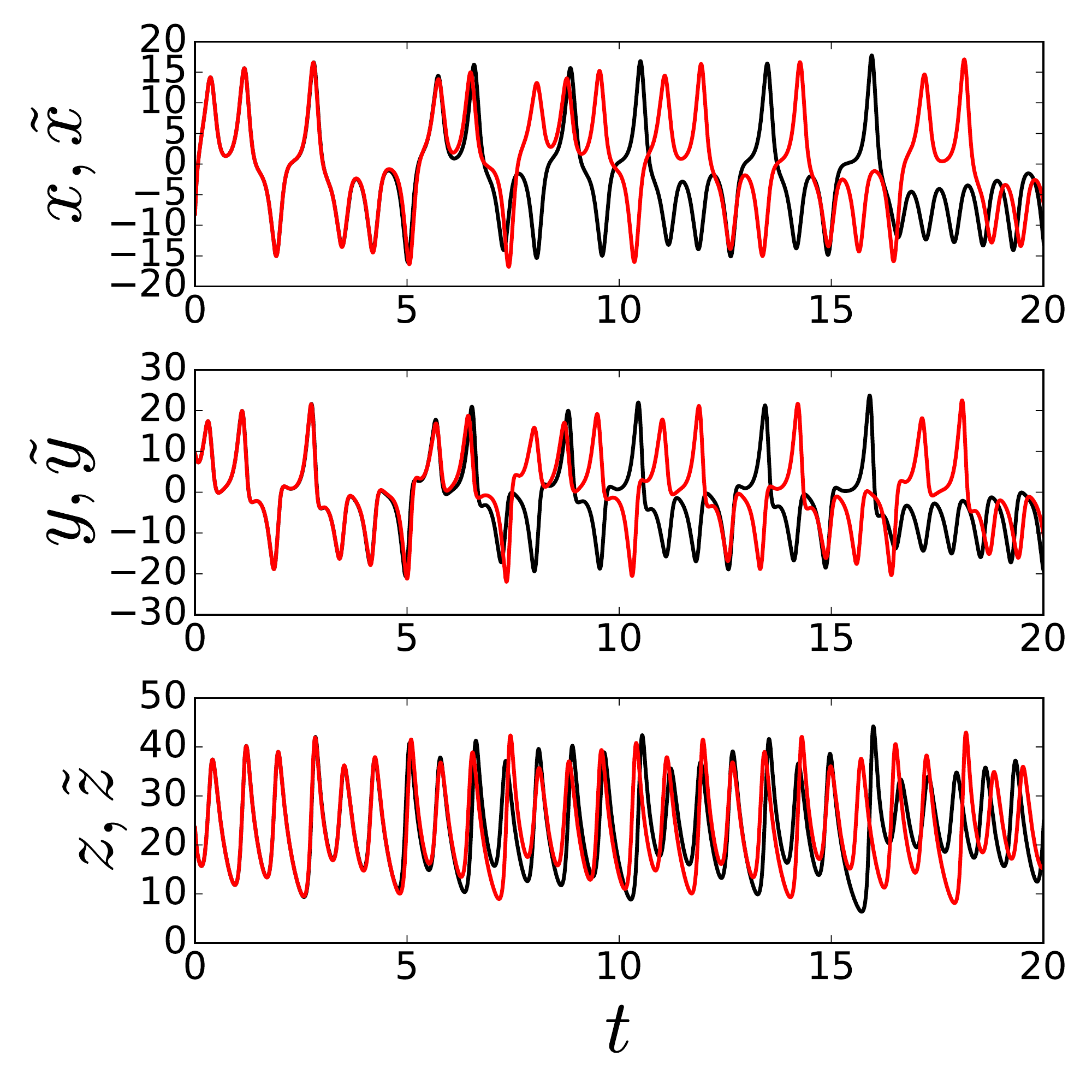}
\caption{Time series of components}
\label{fig:lorenz_ts}
\end{subfigure}
\hspace{0.1\textwidth}
\captionsetup[subfigure]{oneside,margin={-2cm,0cm}}
\begin{subfigure}[t]{0.4\textwidth}
\includegraphics[width=0.75\linewidth]{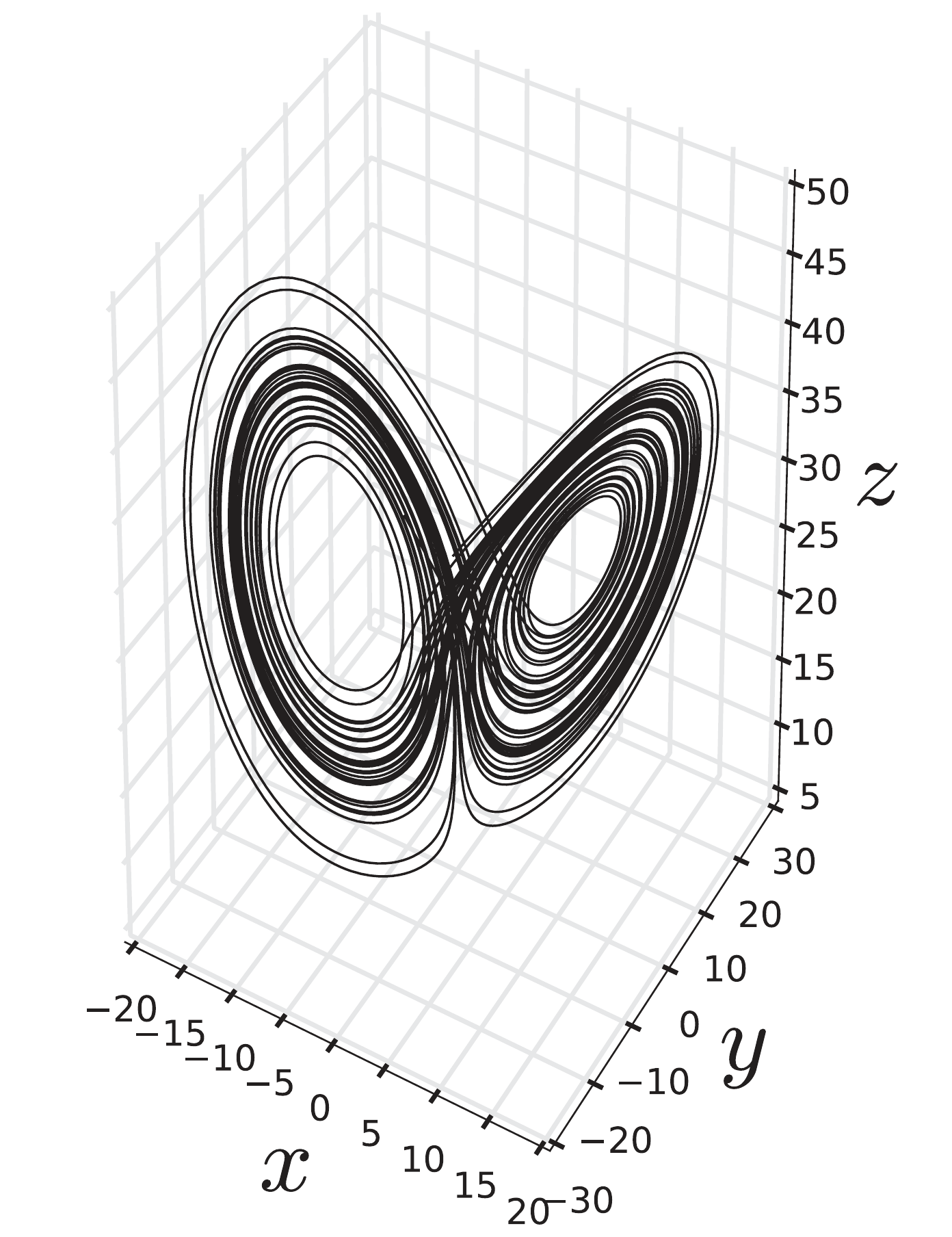}
\caption{Trajectory presentation in phase space}
\label{fig:lorenz_att}
\end{subfigure}
\caption{Illustration of the chaotic dynamics of the Lorenz system (\ref{eq:lorenz_system}) with parameter values $\sigma=10$, $\rho= 28$ and $\beta = 8/3$. (a) Two simulations for the Lorenz system starting from two slightly different initial conditions $(x,y,z) = (-10,10,25)$ and  $(\tilde x,\tilde y,\tilde z) = (-10.01,10,25)$. The Lorenz attractor has a positive Lyapunov exponent and the trajectories diverge from each other. (b) Representation of the trajectory of $(x,y,z) = (-10,10,25)$ in the phase space. The shape of the attractor resembles a butterfly.}
\end{figure}

The Lorenz equations are dissipative and all trajectories eventually enter the absorbing domain 
$$
\Omega = \left\{ \bm{x} \in \mathbb{R}^3 \, \, : \, \, \rho x^2 + \sigma y^2 + \sigma (z - 2\rho )^2 < \frac{\beta^2  \rho^2}{\beta-1}  \right\},
$$
see Appendix C or Ref. \cite{sparrow1982}.  For the classical parameters, $\sigma=10, \rho = 28$ and $\beta = 8/3$, inside $\Omega$, trajectory accumulates on the chaotic Lorenz attractor  \cite{viana2000}, as depicted in Fig. \ref{fig:lorenz_att}. Close to the attractor nearby trajectories diverge. To see this, we simulate two trajectories with nearly the same initial condition. The initial  $10^{-2}$ difference  grows to roughly $10^{2}$ in a matter of only six cycles, see Fig. \ref{fig:lorenz_ts}. Using numerical simulations, we estimate the maximal divergence rate of nearby trajectories $\Lambda \approx 0.906$. 



We consider two coupled chaotic Lorenz oscillators, as in Eq.~(\ref{eq:nonlinear}). We derived above that the critical coupling $\alpha_c$ for synchronization depends on the Lyapunov exponent 
$\Lambda$. Using the  numerical results for $\Lambda$ we obtain  
\[
\alpha_c=\frac{\Lambda}{2}\approx 0.453.
\]
%
%

Simulation confirms that this critical coupling is sharp. Indeed, 
for $\alpha=0.4$ there is no synchronization, and trajectories do not move together, see Fig.~\ref{fig:lorenz_no_sync}. On the other hand, when $\alpha=0.5$, the trajectories synchronize, see Fig.~\ref{fig:lorenz_sync}).

\begin{figure}[h]
\centering
\begin{subfigure}[t]{0.4\textwidth}
{\includegraphics[width=\linewidth]{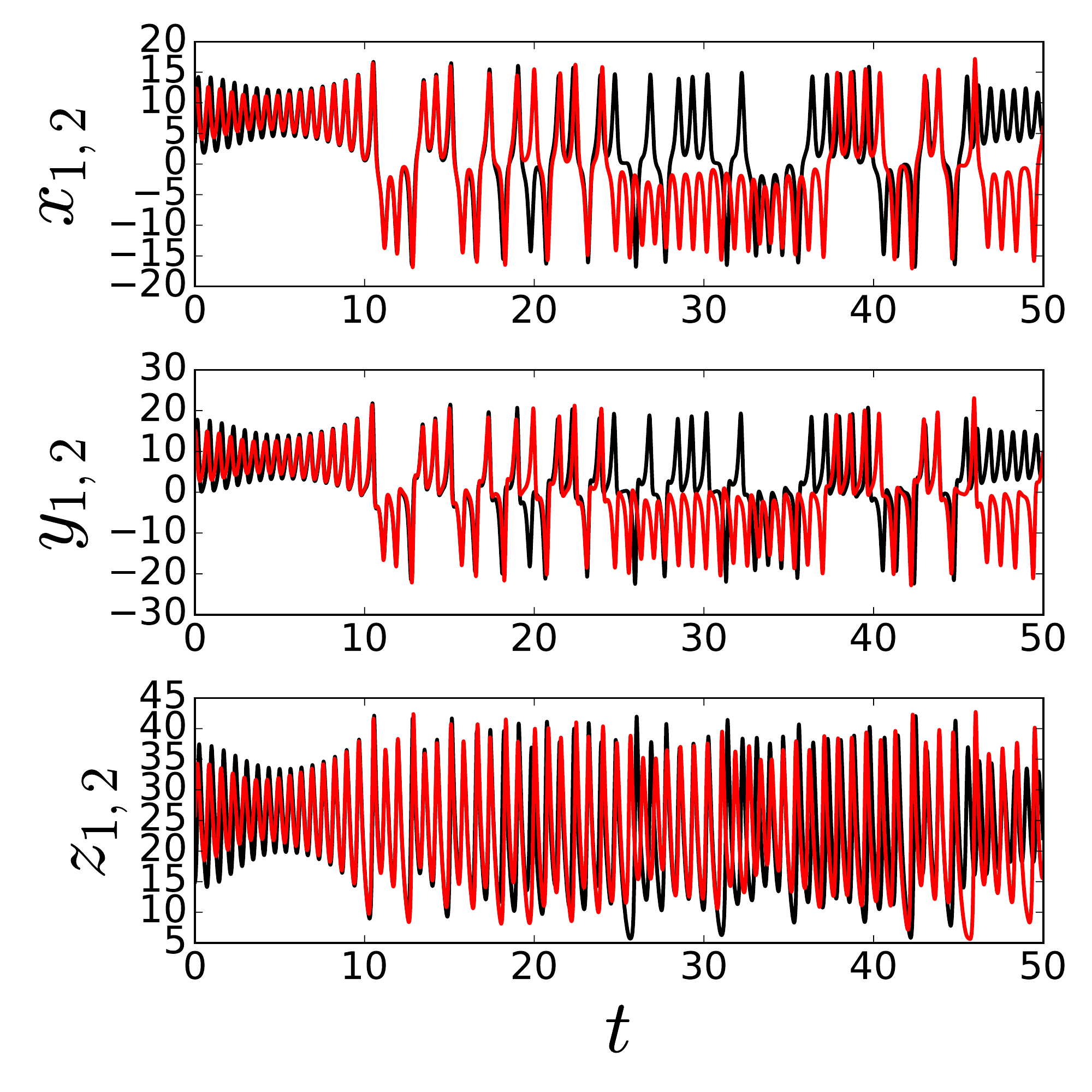}}
\caption{No synchronization}
\label{fig:lorenz_no_sync}
\end{subfigure}
~
\begin{subfigure}[t]{0.4\textwidth}
\includegraphics[width=\linewidth]{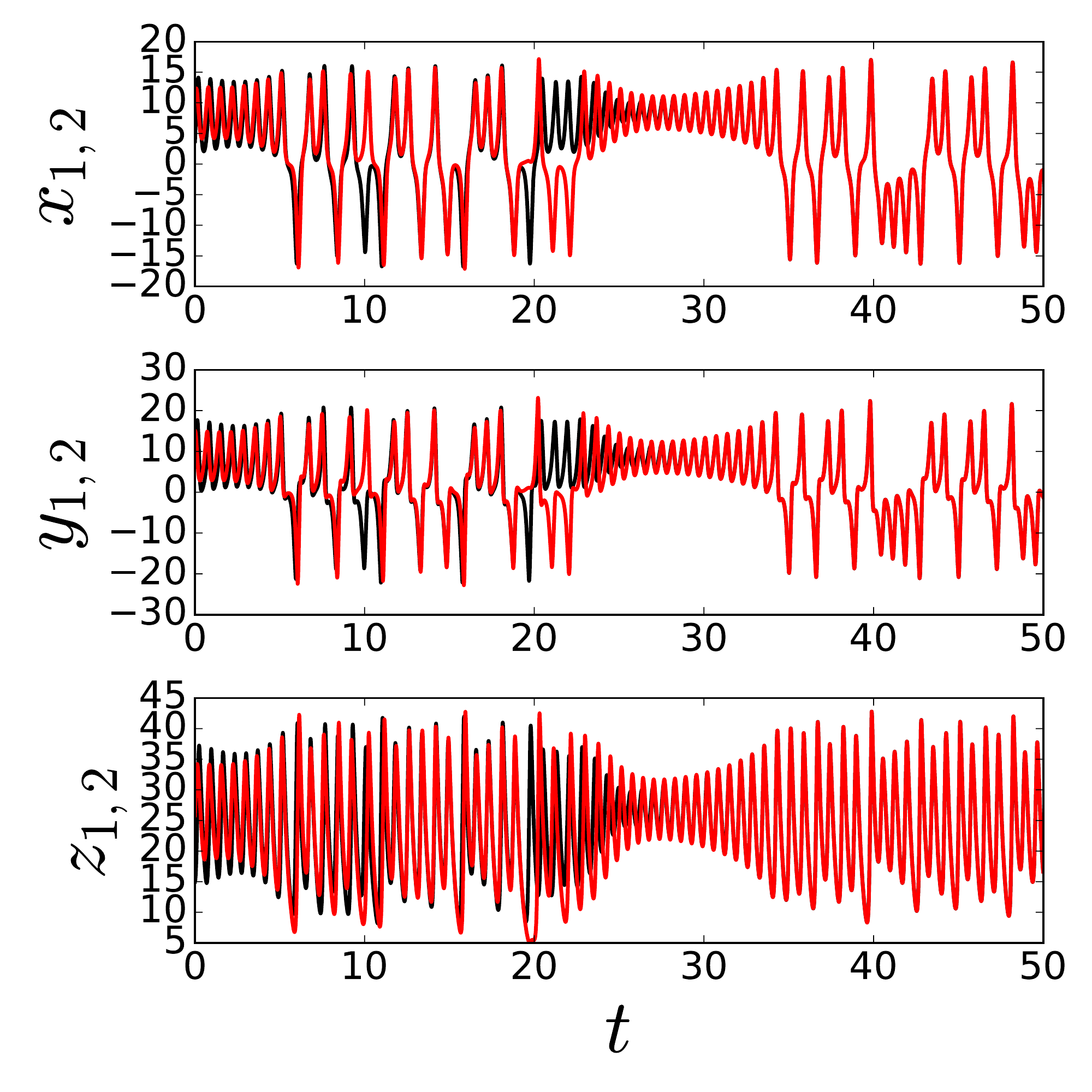}
\caption{Synchronization}
\label{fig:lorenz_sync}
\end{subfigure}
\caption{Comparison of trajectories of two initial conditions for the system of two coupled Lorenz systems. The critical transition coupling is $\alpha_c\approx0.453$ for the classical parameters. The initial conditions are selected as $(x_1,y_1,z_1)=(3,10,15)$ and $(x_2,y_2,z_2)=(10,15,25)$. (a) When $\alpha = 0.4<\alpha_c$, there is no synchronization. (b) If $\alpha = 0.5>\alpha_c$ one observes synchronization of trajectories.}
\end{figure}

To compare the amount of synchronization at different parameter values, we may consider the average deviation from synchronization during a time-interval of length $T$ as
\begin{equation}
E = \frac{1}{T}\int_{t=0}^{T}\|\bm{x}_1(t)-\bm{x}_2(t)\| dt
\label{eq:sync_err}
\end{equation}
In Fig.~\ref{lorenz_sync_diagram} we present a synchronization diagram where we plot $E$ against the coupling strength $\alpha$. We observe a good correspondence with the derived value of $\alpha_c$. The synchronization error depends on initial conditions so that we compute the synchronization diagram via averaging over some realizations.



\begin{figure}[h]
\centering
\begin{subfigure}[t]{0.4\textwidth}
{\includegraphics[width=\linewidth]{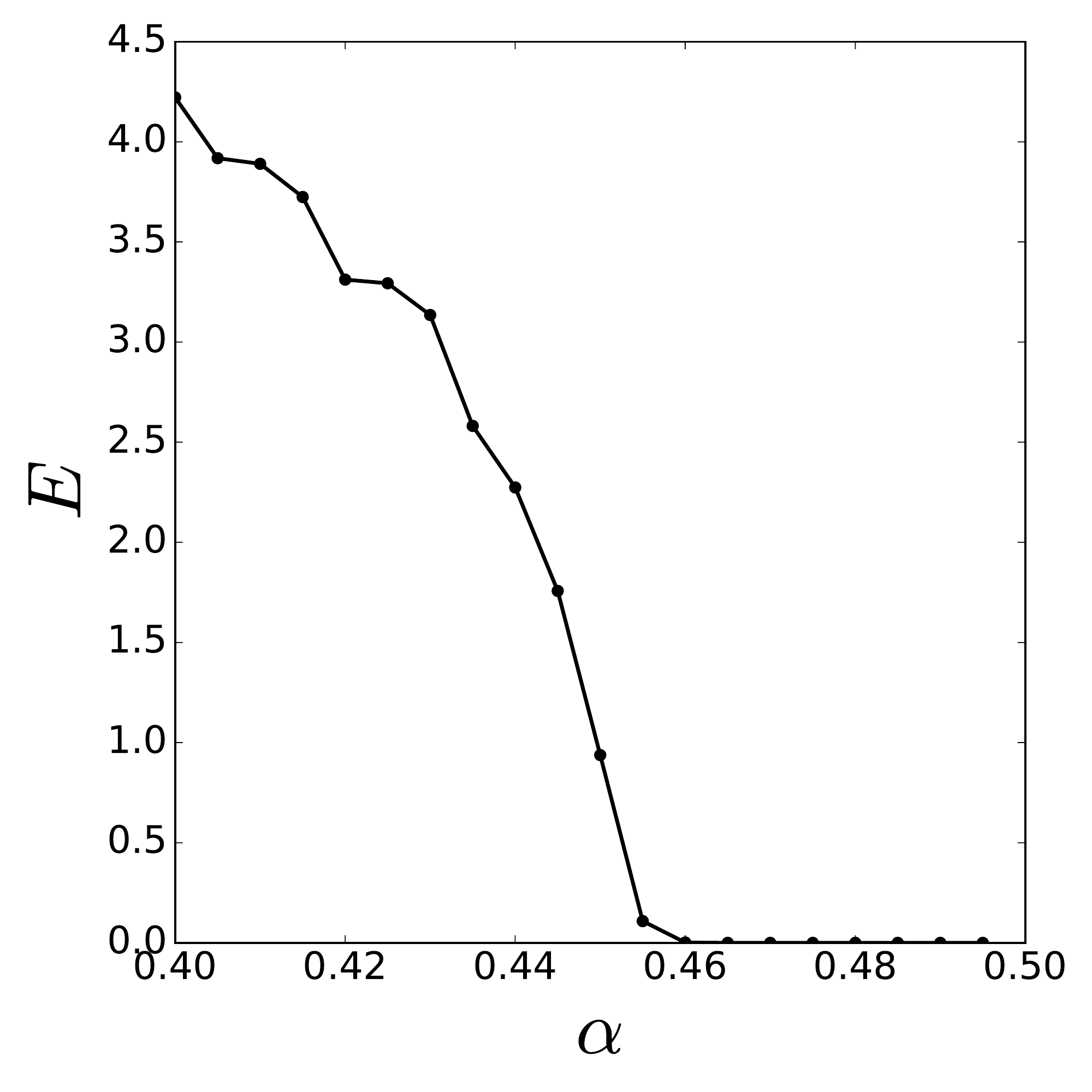}}
\caption{Identity coupling}
\label{lorenz_sync_diagram}
\end{subfigure}
~
\begin{subfigure}[t]{0.4\textwidth}
\includegraphics[width=\linewidth]{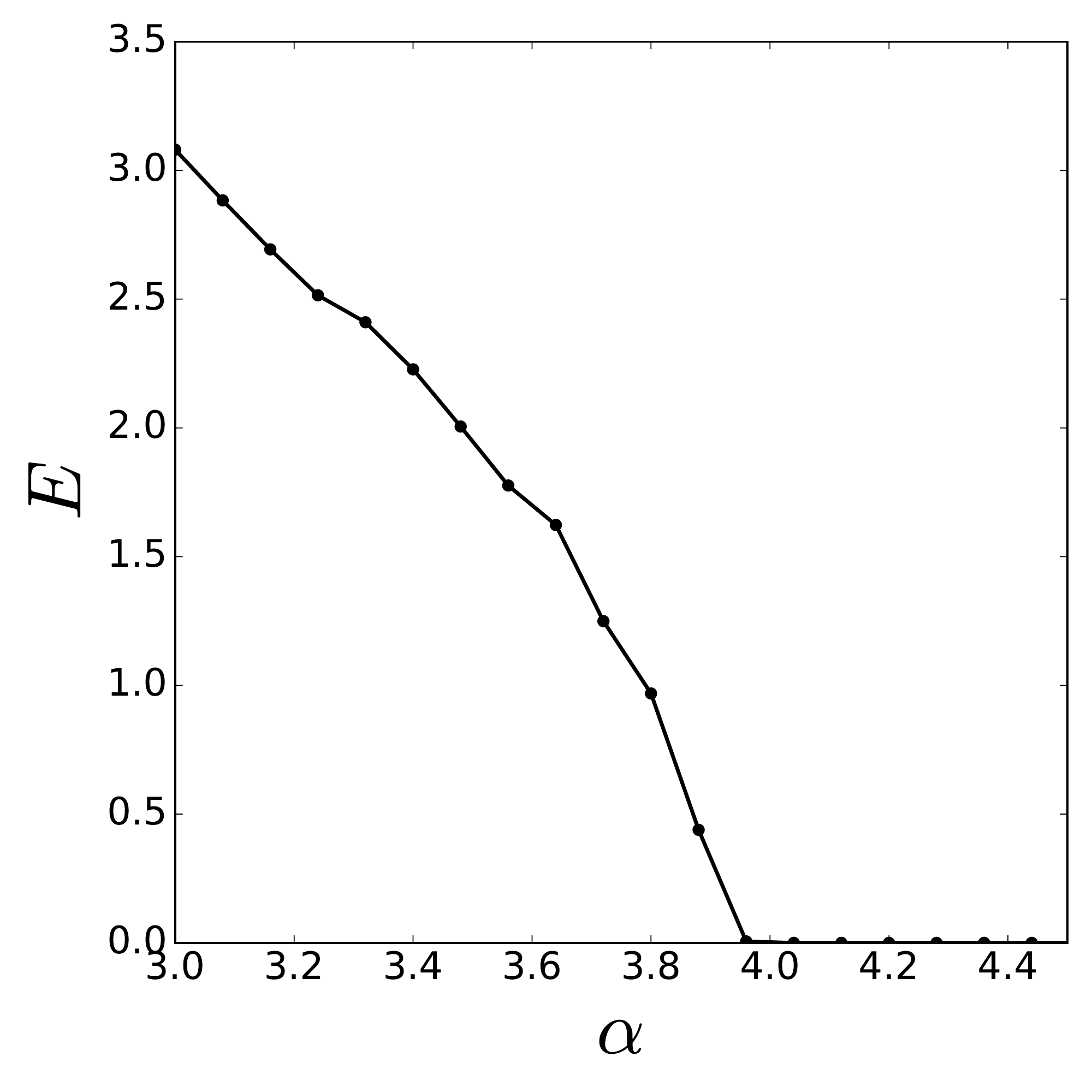}
\caption{$x$-coupling}
\label{lorenz_sync_diagram_x}
\end{subfigure}
\caption{Synchronization diagram of two coupled Lorenz systems, (a) with coupling matrix $\bm{H}=\bm{I}$ and (b) with coupling matrix $\bm{H}$ as in Eq.~(\ref{Hxcoupling}). When $\bm{H}=\bm{I}$, the observed critical coupling constant  corresponds to the theoretically derived value $\alpha_c\sim0.453$. With coupling matrix  (\ref{Hxcoupling}), sychronisation is observed to set in for coupling strengths larger than $\sim3.75$. The synchronization error $E$ was averaged over 300 realisations. Each realisation is simulated by a fourth order Runge-Kutta scheme for 2000 seconds with 0.01 time step.
}
\end{figure} 
~

\noindent
\textbf{Examples on different coupling functions.} 
It is worth mentioning that this above analysis works when the coupling adds a damping term $\alpha z$ in other words when $\bm{H} = \bm{I}$. Indeed, the damping term in general form is $\alpha \bm{H}z$ and in this case the above results can no longer be applied. Therefore the synchronization depends on the coupling function, we here just illustrate the effect on synchronization if $\bm{H}$ is chosen to be
\begin{equation}\label{Hxcoupling}
\bm{H} = 
\left( 
\begin{array}{ccc}
1 & 0 & 0 \\
0 & 0 & 0 \\
0 & 0 & 0
\end{array}
\right),
\end{equation}
implying that the coupling arises only via the first coordinate $x$. The corresponding synchronization diagram shows that the critical coupling $\alpha_c$ for $x$-coupling increases as a result, see Fig.~\ref{lorenz_sync_diagram_x}. Importanly, when $\bm{H}$ does not commute with the Jacobian matrix along the trajectory, we cannot use the ansatz of
Eq.~(\ref{ansatz}). In that case we need a different approach to derive the critical coupling, which will be discussed in Section~\ref{sec:networks} that deals with synchronization in complex networks.

\textbf{R\"ossler System.} As a final example, we consider a system introduced by Otto R\"ossler in 1976:
%
\begin{eqnarray}
\begin{array}{rcl}
\dot{x}&=&-y -z ,  \\ \dot{y}&=&x + ay, \\ \dot{z}&=&b  + z( x-c),
\end{array}
\label{eq:roessler_system}
\end{eqnarray}
where $a, b$ and $c$ denote parameters. 

We consider two coupled R\"ossler systems with identity coupling function $\bm{H} = \bm{I}$ and coupling parameter $\alpha$, as in Eq.~(\ref{eq:nonlinear}). We consider parameter values $a=0.2$, $b=0.2$ and $c=5.7$. We numerically find that the corresponding attractor has a Lyapunov exponent $\Lambda \approx 0.071$. Hence, the expected critical coupling for synchronization is
\begin{equation}
\alpha_c = \frac{\Lambda}{\lambda_2}\approx0.0355
\label{eq:roessler_critical}
\end{equation}
This is in excellent agreement with the numerical results is shown in the synchronization diagram Fig.\ref{roessler_sync_diagram}.

\begin{figure}[h]
\centerline{\includegraphics[width=0.4\linewidth]{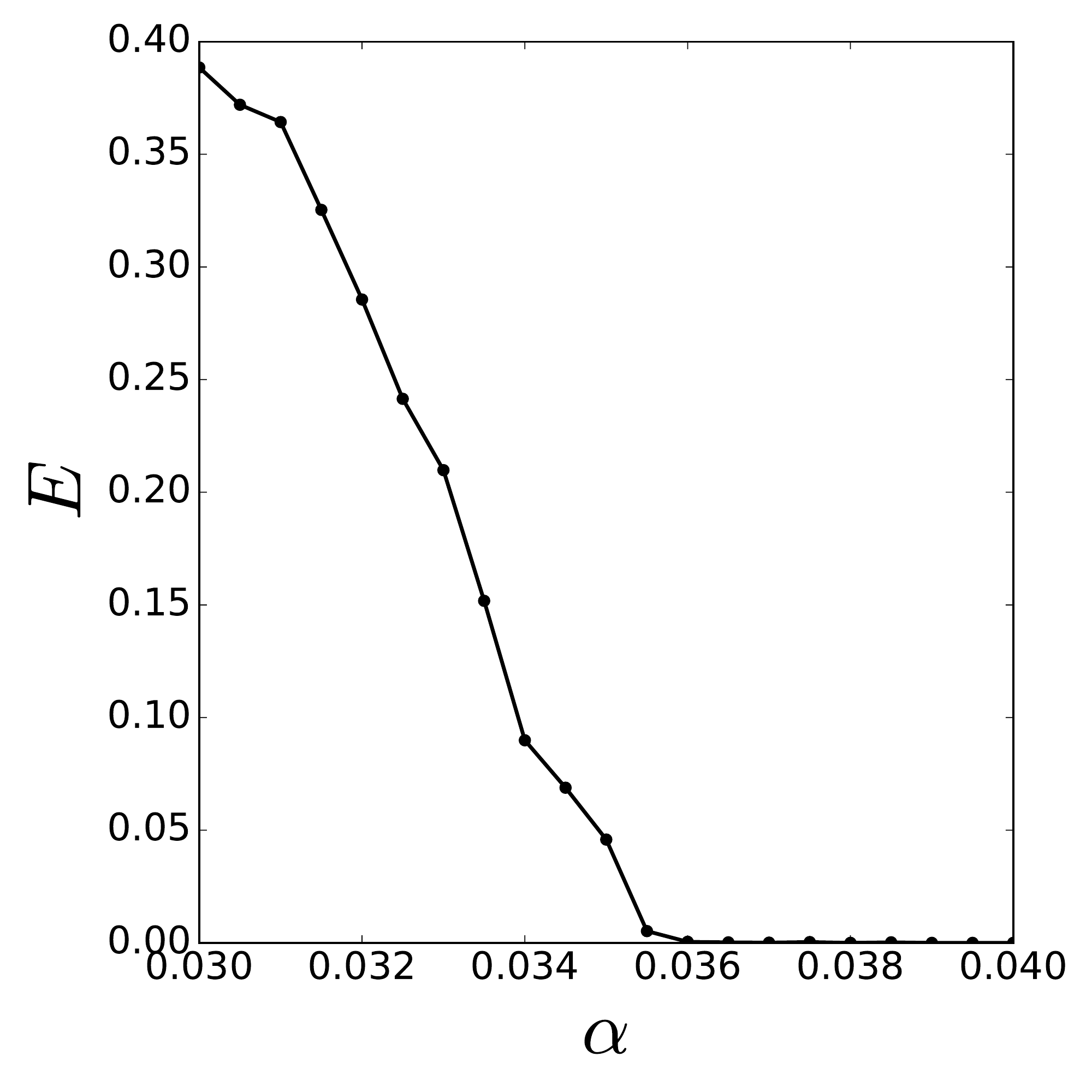}}
\caption{The synchronization diagram of two coupled R\"ossler systems with the coupling function $\bm{H}=\bm{I}$. The theoretical critical coupling constant 
$\alpha_c\approx0.0355$ indeed corresponds to the numerically observed one. The synchronization error $E$ was averaged over 300 realisations. Each realisation is simulated by a fourth order Runge-Kutta scheme for 2000 seconds with 0.01 time step.}
\label{roessler_sync_diagram}
\end{figure}

\subsubsection{CS  in driven systems}
\label{sec:master_slave}

Another possibility is that we use certain sets of variables to drive a subsystem. For appropriate choices we can  observe synchronization
\cite{pecora1990}. We illustrate this scheme in the  Lorenz system where $x$-component can be driving signal of another identical system Fig.~\ref{fig:ms_scheme}.

\begin{figure}[h]
\centerline{\includegraphics[width=0.4\linewidth]{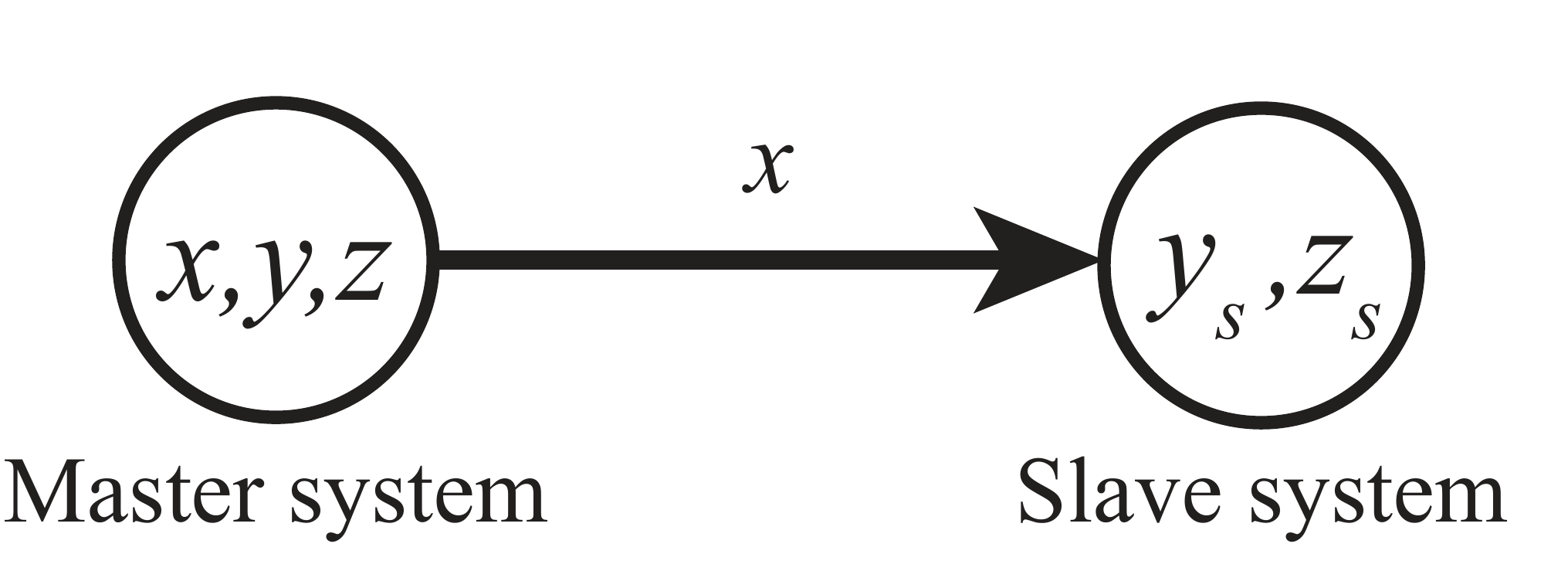}}
\caption{Master-Slave configuration where, $x$-variable is made identical to the response and thereby it drives the response subsystem.}
\label{fig:ms_scheme}
\end{figure}

In this scheme we consider the variable $x$ for the master the same as in the slave. That is, the 
$x$- variable of the master is fully replaced to the $x$ variable in the slave
\begin{eqnarray}
\begin{array}{lcc}
\dot{x}= \sigma(y-x) & \\
\dot{y}= x(\rho-z)-y & \quad \dot{y_s} = x(\rho-z_s)-y_s \\
\dot{z}= -\beta z +xy & \dot{z_s}=-\beta z_s +xy_s
\end{array}
\label{eq:dr_lorenz_x}
\end{eqnarray}
where $(x,y,z)$ are the states of the master system and $(y_s,z_s)$ are the states of the slave system.
In order to check the behaviour of the trajectories, we track the simultaneous variation of the trajectories by $\Delta_y(t) = y(t)-y_s(t)$ and  $\Delta_z(t)= z(t)-z_s(t)$. For given initial conditions $(x,y,z,y_s,z_s) = (-10.1,10.1,10.1,0.1,0.1)$, $\Delta y$ and $\Delta z$ goes to zero (Fig.~\ref{fig:dr_sims}). 

\begin{figure}[h]
\centerline{\includegraphics[width=0.4\linewidth]{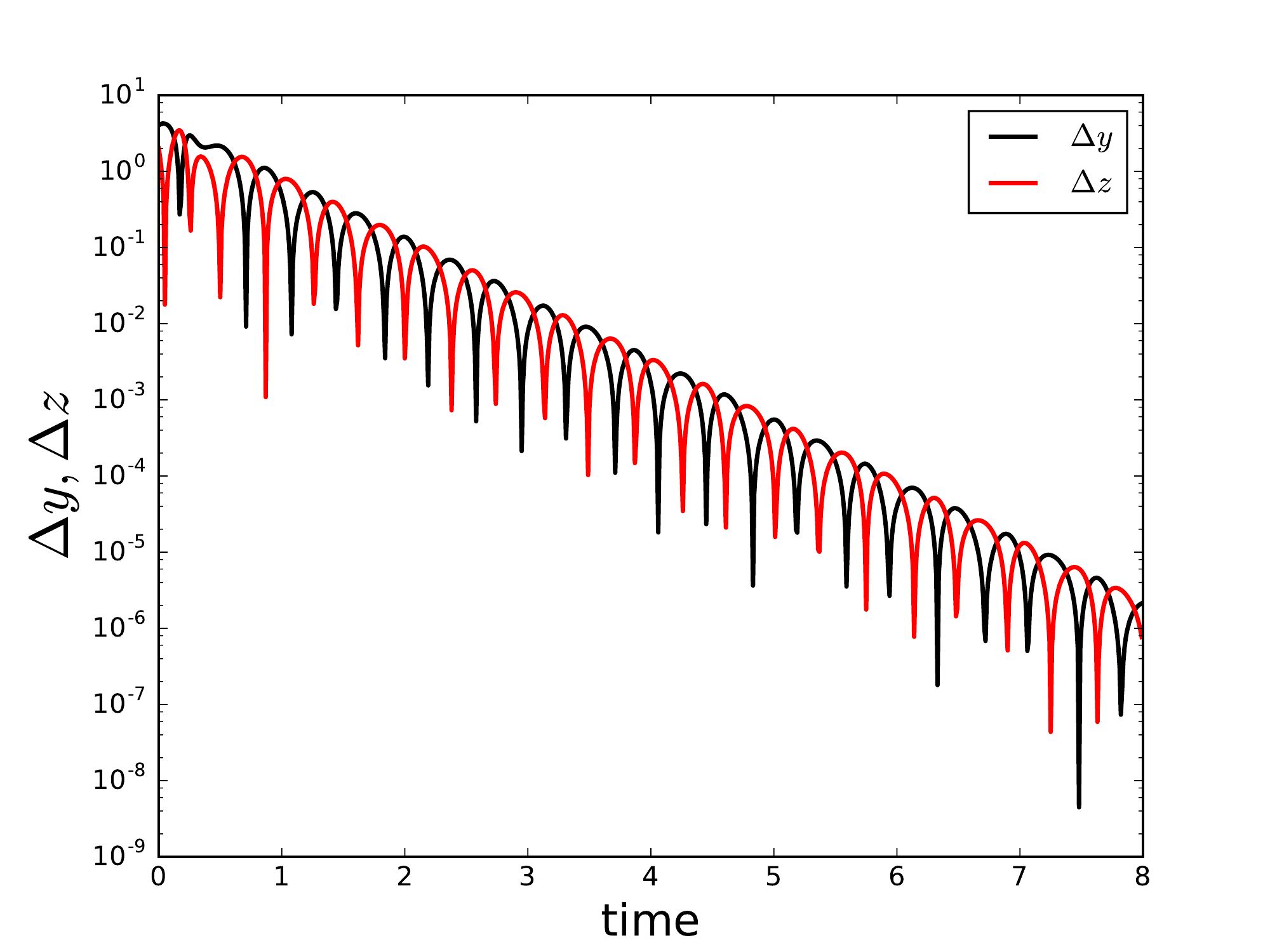}}
\caption{Simulation of master-slave type of coupling}
\label{fig:dr_sims}
\end{figure}

For this particular choice of subsystem it is possible to construct a Lyapunov function for the 
displacements
 $\Delta_y = y -  y_s$ and $\Delta_z= z - z_s$ for $x$-driven system (Eq.~(\ref{eq:dr_lorenz_x})).We obtain
\begin{equation}
\begin{array}{rcl}
\dot{\Delta}_y &=& -\Delta_y -x \Delta_z \\
\dot{\Delta}_z &=& x \Delta_y - \beta \Delta_z.
\end{array}
\label{eq:dr_variational}
\end{equation}
Next consider the Lyapunov function
$$
V = \frac 1 2 \left( \Delta_y^2 +  \Delta_y^2 \right),
$$
and along solutions of the subsystem we obtain $\dot V = \Delta_y \dot \Delta_y + \Delta_z \dot \Delta_z$, after some manupulation we obtain 
$$
\dot{V} = - \Delta^2_y - \beta\Delta^2_z.
$$
Since $V$ is positive and $\dot V$ negative $\Delta_y$ and $\Delta_z$ will converge to zero. So, the slave subsystem will have the same dynamics as the master.

%

%
%
%

%
%
%
%
%
%
%
%

\subsection{Phase synchronization}
\label{sec:phase_sync}
If there are small mismatches between the systems another type of synchronization can appear for very small coupling strengths: \textit{Phase Synchronization} (PS) -- which corresponds to a locking of phases of chaotic oscillators 
$$| m\phi_1 (t)- n\phi_2 (t) |< C $$ 
where $\phi$ is the phase of the chaotic oscillators, $m, n$ and $C$ are constants. When this holds we have phase synchronization between the two systems \cite{rosenblum1996,rosenblum1997}. We are considering the phases on the lift, that is, diverging steadily as opposed to consider the phase mod $2\pi$. The phase difference won't be precisely zero because of the chaotic nature of the system. We could consider higher relations of phase locking, however, the higher the relation $m:n$ more difficult is to observe the phase synchronization. Therefore, our examples will be for $1:1$ phase synchronization. 

Phase synchronization is also vast research periodic oscillators \cite{pikovsky2001,rodrigues2016,tonjes2009,tonjes2010}. In this case, the phases may be perfectly locked. If we are considering periodic oscillators the phase reduction approach will lead to a description of the interaction in terms of the phases alone \cite{kuramoto1975}. The simplest equation in this setting 
is 
$$
\dot{\phi}_{1,2} = \omega_{1,2} + \alpha \sin (\phi_{2,1} - \phi_{1,2}) 
$$
where $\phi$ is the phase along the periodic orbit. Introducing the phase difference $\Phi = \phi_1 - \phi_2$  and $\Delta = \omega_1 - \omega_2$ we obtain
$$
\dot{\Phi} = \Delta - 2\alpha \sin \Phi
$$
this equation has a stable fixed point $\Phi = \phi_1 - \phi_2 =$ constant if $\alpha> \alpha_c =  |\Delta| /2 $. 

For a chaotic oscillator if coupling strength is small, the amplitudes will remain chaotic but the phase difference will be bounded. Though, it will oscillate as a result of the coupling to the amplitude.
In general, it is not straightforward to introduce a phase for a chaotic attractor \cite{josic2001,pikovsky1997,pereira2007,baptista2005, baptista2006}.  For a suitable class of attractors it is possible to define a phase in a useful way. 

We focus on coupled two nonidentical R\"ossler oscillators, the equation is given by
\begin{equation}
\begin{array}{rcl}
\dot{x}_{1,2} &=& -\omega_{1,2}y_{1,2} - z_{1,2} + \alpha(x_{2,1}-x_{1,2}) \\
\dot{y}_{1,2} &=& \omega_{1,2}x_{1,2} + ay_{1,2}  \\
\dot{z}_{1,2} &=& b + z_{1,2}(x_{1,2}-c) 
\end{array}
\label{eq:two_roessler}
\end{equation}
where $a=0.165$, $b=0.2$ and $c=10$ are the constants of the R\"ossler system. $\omega$ is the mismatch parameter to make the oscillators \textit{nonidentical} and given as $\omega_{1,2}=\omega_0 \pm \Delta$ where $\omega_0 =0.97$ and $\Delta = 0.02$. $\alpha$ is the coupling constant, the system is coupled over $x$ components ($x$-coupling).  For certain values of the parameter  $a$, the projection of the attractor on $x-y$ plane resembles a limit cycle and the trajectories rotates around the origin (see Fig.~\ref{fig:roessler_xy_project}), and  phase and  amplitudes are given by
\begin{equation}
\phi_{1,2} = \arctan \left(\frac{y_{1,2}}{x_{1,2}}\right)
\label{eq:phase_roessler}
\end{equation}
\begin{equation}
A_{1,2} = \sqrt{x_{1,2}^{2} + y_{1,2}^{2}}.
\label{eq:amplitude_roessler}
\end{equation}
We consider the phase on the lift (growing in time without taking the mod).
\begin{figure}[h]
\centerline{\includegraphics[width=0.4\linewidth]{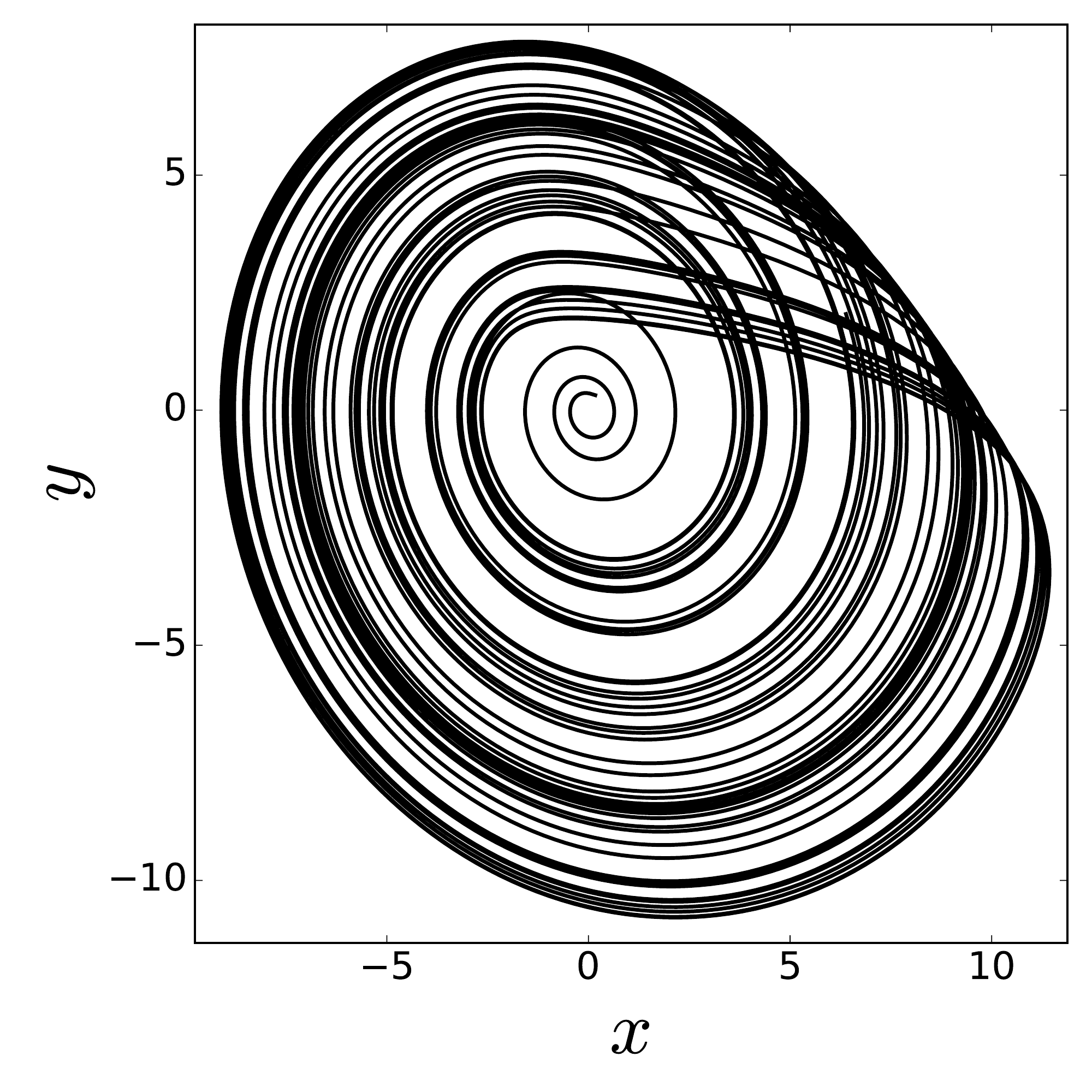}}
\caption{Projection of the R\"ossler attractor on $x-y$ plane for $a=0.165$.}
\label{fig:roessler_xy_project}
\end{figure}

To gain insight on the adjustment of rhythm  leading to  phase synchronization, we analyze the average frequencies 
defined as
 \begin{equation}
\Omega_{1,2} = \lim_{T \rightarrow \infty} \frac{\phi_{1,2}{(T)}-\phi_{1,2}(0)}{T}.
\label{eq:mean_freq}
\end{equation}
And the frequency mismatch is 
 \begin{equation}
\Delta\Omega = \Omega_2 - \Omega_1
\label{eq:phase_sync_err}
\end{equation}
When phase synchronization occurs 
$
| \phi_1(t) - \phi_2(t) | \le C, 
$
 the average frequency is the same $\Delta \Omega = 0$.
The phase difference will not be tend to a constant as the phase nature of the amplitudes acts as a noise in the phases causing mismatches. The comparison of the  amplitude difference (Eq.~(\ref{eq:sync_err})) and the phase (Eq.~(\ref{eq:phase_sync_err})) is given in Fig.~\ref{fig:phase_sync}. If we increase the coupling constant $\alpha$.

\begin{figure}[h]
\centerline{\includegraphics[width=0.4\linewidth]{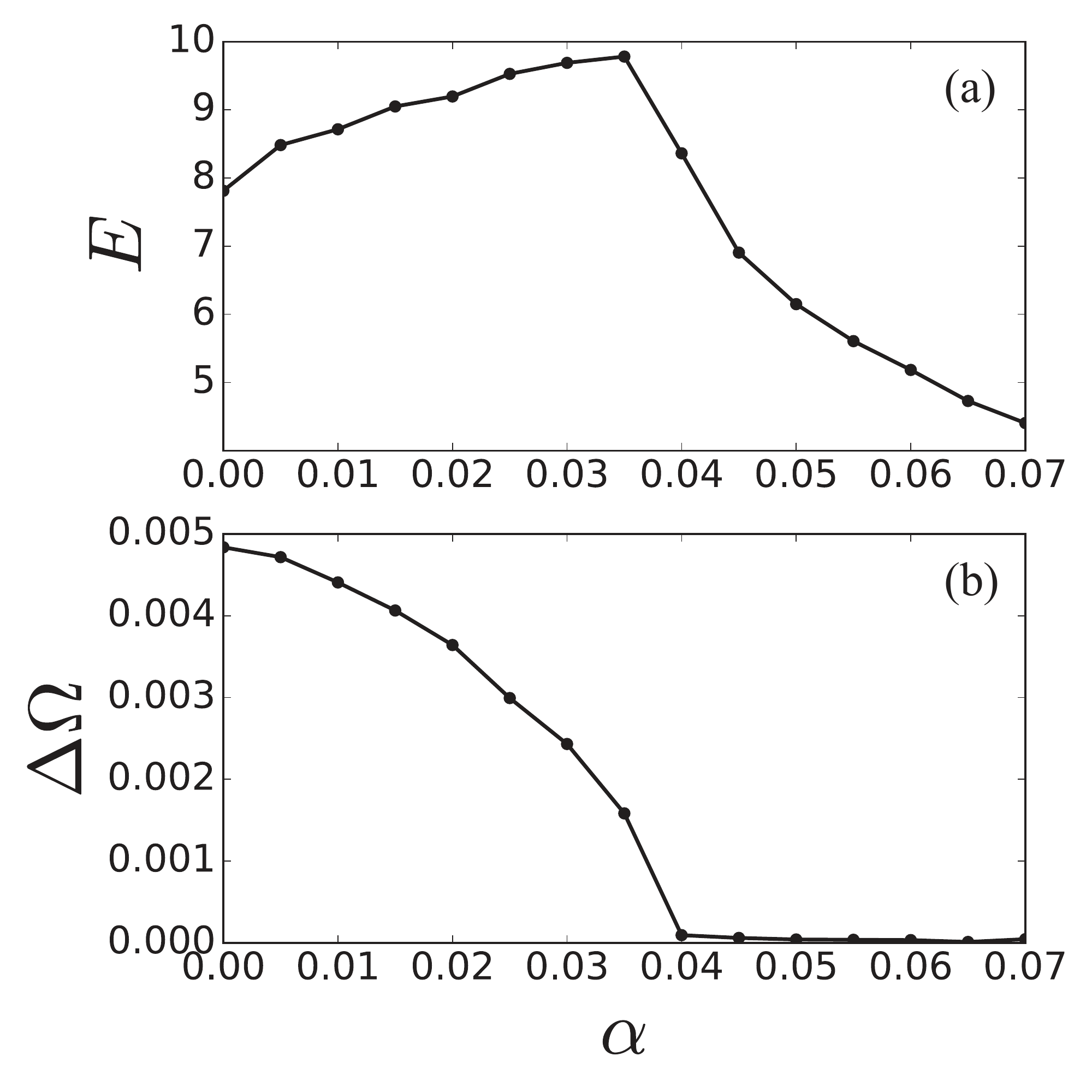}}
\caption{For a weak coupling constant: Although there is no synchrony for the difference of the amplitudes Eq.~(\ref{eq:sync_err})(top panel), there is a tendency towards to phase synchronization while the coupling constant $\alpha$ increases Eq.~(\ref{eq:mean_freq}) (bottom panel). The synchronization error $E$ and frequency mismatch $\Delta \Omega$ were averaged over 300 realisations. Each realisation is simulated by a fourth order Runge-Kutta scheme for 2000 seconds with 0.01 time step.}
\label{fig:phase_sync}
\end{figure}

An approximate theory of phase synchronization can be obtained by averaging \cite{sanders2007}. We write the model Eq.~(\ref{eq:two_roessler}) in terms of the phase Eqs.~(\ref{eq:phase_roessler}) as

\begin{equation}
\dot{\phi}_{1,2} =  \frac{x_{1,2}\dot{y}_{1,2}-y_{1,2}\dot{x}_{1,2}}{A^2}
\label{eq:ode_phase_roessler}
\end{equation}
In this form, using polar coordinates we have
$
x_{1,2} = A_{1,2}\cos\phi_{1,2} $ and $y_{1,2} = A_{1,2}\sin\phi_{1,2} 
$, and using this representation in Eq.~(\ref{eq:ode_phase_roessler}), we obtain

\begin{equation}
\dot{\phi}_{1,2} = \omega_{1,2}+a\sin\phi_{1,2}\cos\phi_{1,2}+\frac{z_{1,2}}{A_{1,2}} \sin\phi_{1,2}-\alpha \left(\frac{A_{2,1}}{A_{1,2}} \cos\phi_{2,1}\sin\phi_{1,2}-\cos\phi_{1,2}\sin\phi_{1,2}\right)  
\end{equation}

The idea now is that since the mismatch is small, both phases behave nearly the same. So we can split the dynamics of the phases as an overall increasing trend $\omega_0 t$ and a slow phase dynamics $\theta$. This split is very clear in Fig. \ref{fig:ps_demo}. So, we write
$$
\phi_{1,2} = \omega_0 t + \theta_{1,2},
$$
To obtain an equation for $\theta$ (simpler than the one for $\phi$) we use the fact that $\theta$ is a slow variable. That is, while $\omega_0 t$ grows a lot $\theta$ is nearly constant. Hence, we will average out the contribution of $\omega_0 t.$ 
So we average the phases over $\omega_0 t$ over a period $\frac{2\pi}{\omega_0}$ and keep $\theta_{1,2}$ fixed. After some laborious manipulation we obtain
\begin{equation}
\frac{d}{dt}(\theta_1-\theta_2)=2\Delta-\frac{\alpha}{2}\left(\frac{A_2}{A_1}+\frac{A_1}{A_2}\right)\sin(\theta_1-\theta_2) 
\label{eq:slow_phase}
\end{equation}

Both amplitudes $A_{1,2}$ depend on time and display a chaotic behaviour. Lets assume for a moment that they are constant. Then for the phase locking of the R\"ossler systems, $\frac{d}{dt}(\theta_1-\theta_2)=0$, the equation has a stable fixed point,
\begin{equation}
\theta_1-\theta_2=\arcsin \frac{4\Delta A_1 A_2}{\alpha(A_1^2+A_2^2)}.
\label{eq:phase_locking}
\end{equation}
This fixed point only exists when the argument of the arcsin has modulus less than 1. 
Therefore, we obtain the critical transition coupling 
$$\alpha_c\approx 2\Delta.$$
For the given parameters ($\Delta=0.02$) we find $\alpha_c \approx 0.04$, in agreement with the numerical analysis Fig.~\ref{fig:phase_sync}. The chaotic behaviour of the amplitudes leads to fluctuations of the phases around the stable fixed point, and so the phases different will not be identically zero.
Close to the critical coupling strength the frequencies exhibit a critical behaviour
$
\Delta \Omega \propto |\alpha - \alpha_c|^{1/2}
$
as observed  observed in Fig. \ref{fig:phase_sync}.

\subsection{Generalized synchronization}
\label{sec:generalized_sync}

When the interacting systems are  different, either because of a large parameter mismatch or the systems have distinct dynamics, these two can still exhibit synchronization in a generalized sense.  
\textit{Generalized Synchronization} (GS) can be observed in mutually coupled systems as well as unidirectionally coupled system \cite{pecora1990, he1992, kowalski1990, sugawara1994}. Surprisingly, GS can be a mapped to a complete synchronization (CS) problem!

Here we will focus on the dynamics of unidirectionally coupled systems. The master $\bm{x}$ and the slave $\bm{y}$ systems coupled as
\begin{eqnarray}
\dot{\bm{x}}&=&\bm{f}(\bm{x}) \nonumber \\
\dot{\bm{y}}&=&\bm{g}(\bm{y},\bm{h}(x)) 
\label{eq:drive_response}
\end{eqnarray}
where $\bm{x}\in \mathbb{R}^n$, $\bm{y}\in \mathbb{R}^m$ and $\bm{h}(x)$ is the coupling.
For certain coupling strengths, the dynamics  of  system $y$ is totally determined by the dynamics of  system $x$. That is, the solutions of, say $x$ can be mapped into solutions of $y$. 
$$
y = \psi (x)
$$
where $\psi$ is a function from the phase space of the system $x$ to the phase space of system $y$. When this happens we have generalized synchronization between these two systems. 
CS is a particular case of GS when $\psi$ is the identity. 

To detect a functional relation between two systems in generalized synchronization, Rulkov and co-workers proposed a technique called mutual false nearest neighbours \cite{rulkov1995}. 
The main idea is the see how nearby points are mapped under the dynamics. By studying the properties of nearby points one can infer the existence of the mapping $\psi$. Here, we focus on another approach that turns the GS problem into a CS problem.
This is the auxiliary system approach  \cite{abarbanel1996, kocarev1996}.  The master system drives the slave system and an auxiliary system (copy of the slave). If the two copies of the slave exhibit  CS then the master and slave are in GS.\cite{abarbanel1996, kocarev1996}.
An illustration of this scheme can be found in See Fig.~\ref{fig:auxiliary_scheme}. 
\begin{figure}[h]
\centerline{\includegraphics[width=0.4\linewidth]{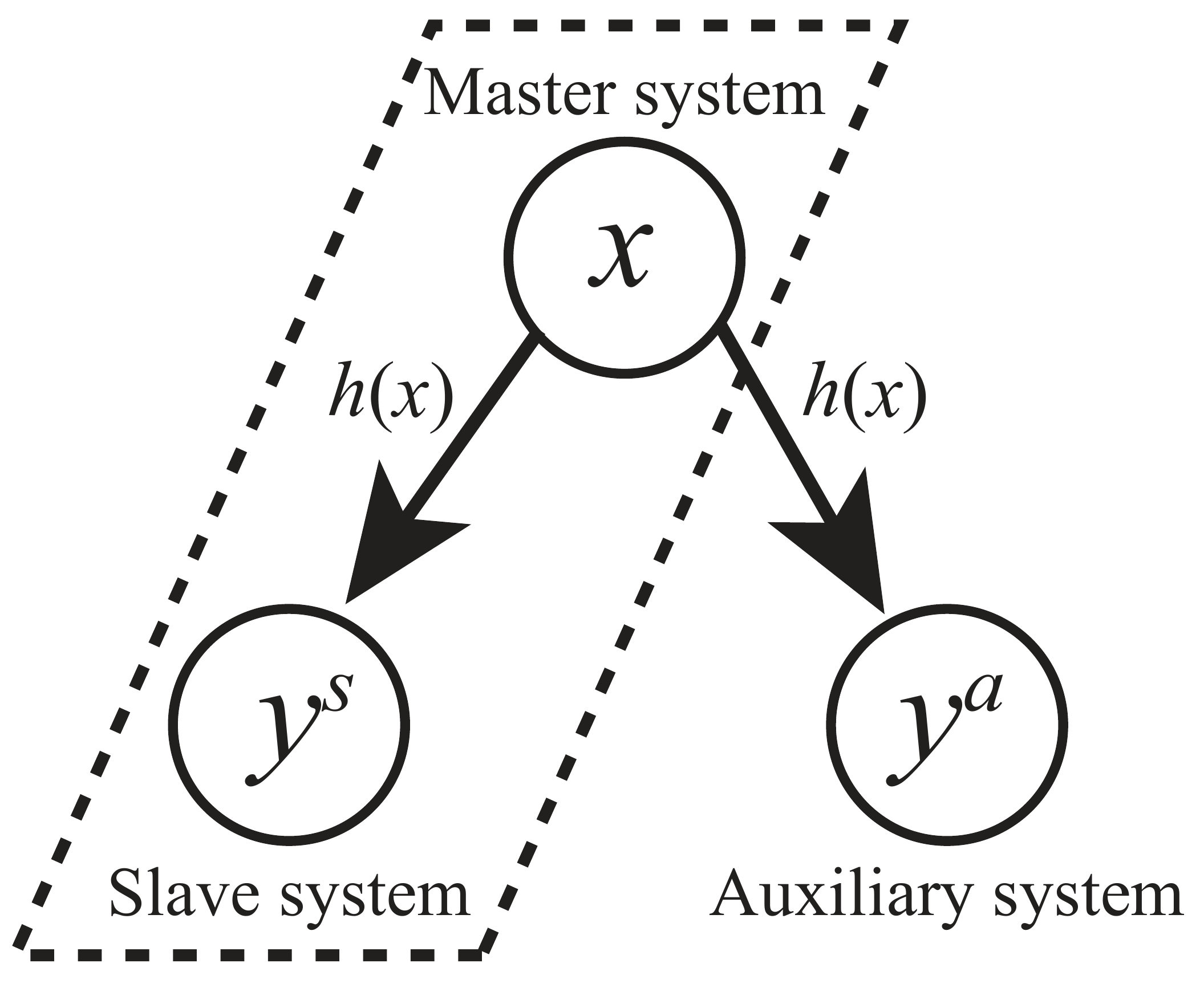}}
\caption{Scheme of the auxiliary system approach for the generalized synchronization. Originally we only have the system in the dashed line box which is master-slave system as in Sec.~\ref{sec:master_slave}. Then we add an auxiliary (helper) system $y^a$. If there is CS between $y^s$ and $y^a$, then the GS occurs between $\bm{x}$ and $\bm{y}^{a,s}$.}
\label{fig:auxiliary_scheme}
\end{figure}


Necessary conditions for the occurrence of GS for the system given by Eq.~(\ref{eq:drive_response}) is introduced by Kocarev and Parlitz as  following:  for all $(\bm{x}_0,\bm{y}_0) \in B$, where  $\bm{x}_0$ and $\bm{y}_0$ are states for the master-slave systems at time $t=0$ and $B$ is the basin where all the trajectories approach to a manifold 
$$
M_{\psi} = \{(\bm{x},\bm{y}): \bm{y} = {\psi}(\bm{x})\}.
$$
If $M_{\psi}$ is attractive, different trajectories of the slave system will converge to the trajectory lying in $M$ and it is determined only by $x$. 
In other words, if the master drives a slave $\bm{y}^s_0$ and an auxiliary (copy of slave) $\bm{y}^a_0$ systems simultaneously, the driven ones must be completely synchronized $\forall \bm{y}^s_0, \bm{y}^a_0 \in B_y$ we have  
$$\lim_{t \to \infty}  \| \bm{y}^s_t  - \bm{y}^a_t \| = 0.$$

{\bf Example:} Consider two identical R\"ossler systems (Eq.~(\ref{eq:roessler_system})) with the parameters ($a=0.2$, $b=0.2$ and $c=5.7$) are driven by a Lorenz system (Eq.~(\ref{eq:lorenz_system})) with the classical parameters ($\rho=28, \sigma=10$ and $\beta=8/3$) via $x$-components. We used the auxiliary system approach the detect the critical coupling for GS. Indeed, numerical results showed that  $\alpha_c \approx 0.12$ as seen in the synchronization diagram Fig.~\ref{fig:generalized_sync}}. For given $\alpha=0.06$ CS is not observed between the slave systems therefore there is no GS between master and slave systems as well Fig.~\ref{lorenz_no_sync}. For a coupling constant larger than the critical one $\alpha=0.2$ the slave and the auxiliary system display CS.  Hence GS can be observed between master-slave system Fig.~\ref{lorenz_sync}.

\begin{figure}[h]
\centerline{\includegraphics[width=0.4\linewidth]{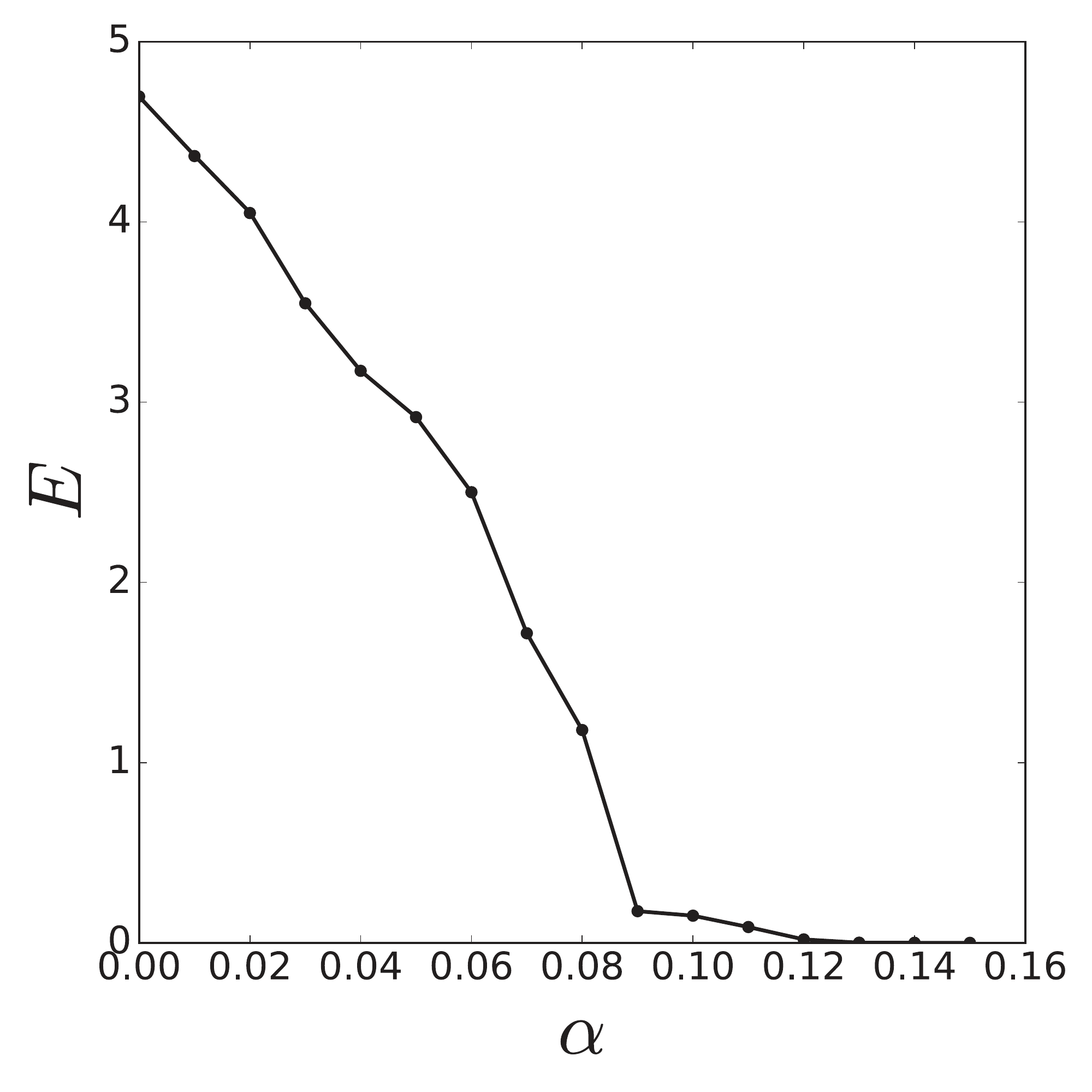}}
\caption{Generalized Synchronization: Averaged over 300 realisations, time=4000 and time step=0.01}
\label{fig:generalized_sync}
\end{figure}

\begin{figure}[h]
\centering
\begin{subfigure}[t]{0.4\textwidth}
{\includegraphics[width=\linewidth]{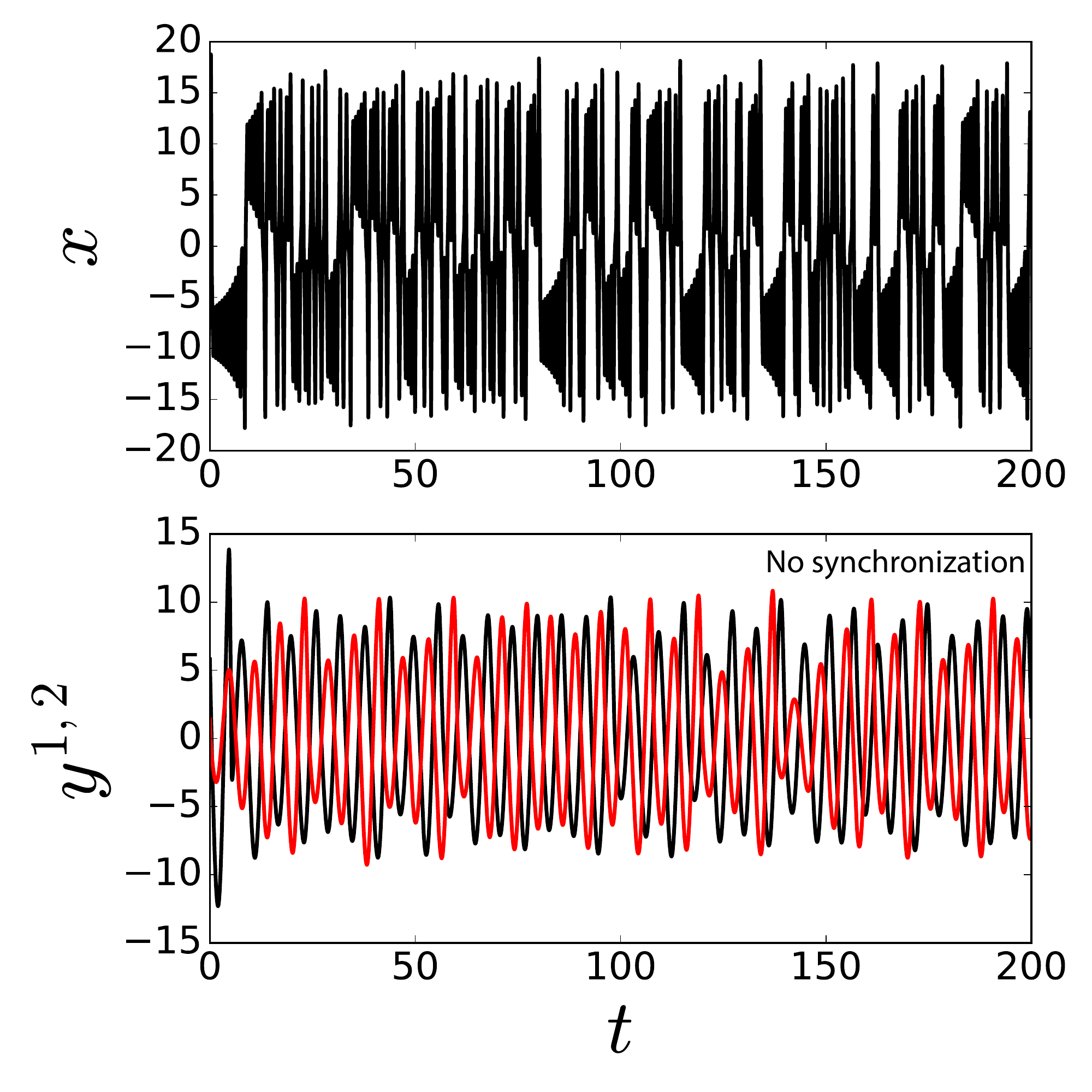}}
\caption{No Synchronization}
\label{lorenz_no_sync}
\end{subfigure}
~
\begin{subfigure}[t]{0.4\textwidth}
\includegraphics[width=\linewidth]{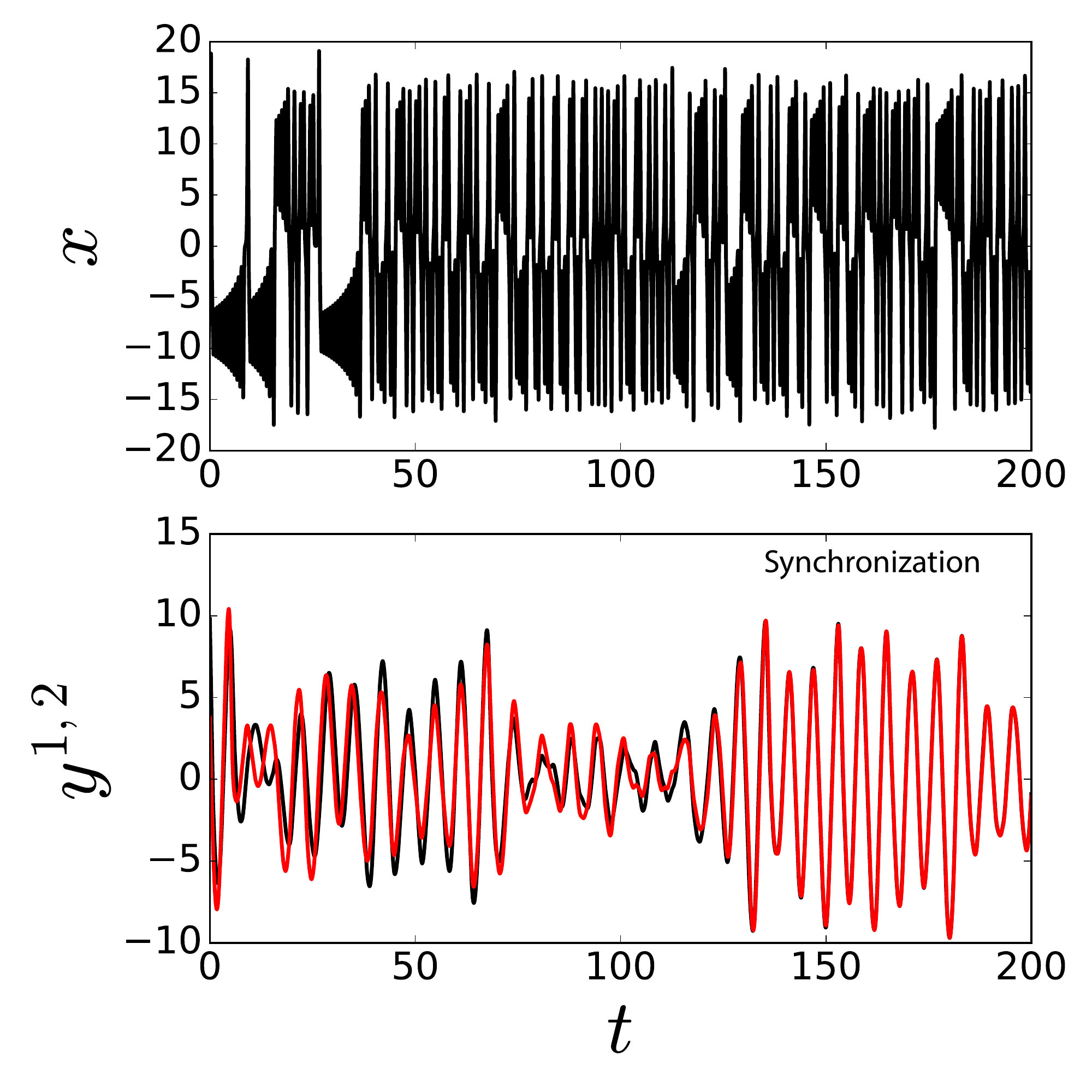}
\caption{Synchronization}
\label{lorenz_sync}
\end{subfigure}
\caption{Two simulations for the generalized coupling scheme: a Lorenz system $\bm x$ drives two R\"ossler systems $\bm y^{s,a}$. The critical transition coupling is $\alpha_c\approx0.12$. (a) For the coupling constant $\alpha = 0.06$, there is no synchronization since $\alpha<\alpha_c$ (b) for $\alpha = 0.2$ synchronization is obtained since $\alpha>\alpha_c$.}
\end{figure}

\subsubsection{Generalized Synchronization between diffusively coupled oscillators}

To gain some insight 
on this,  we will use some ideas put forward by \cite{hramov2004,hramov2005}. 
This approach allows us to obtain a analytical understanding of the critical coupling associated with the transition to GS. 
Consider a master-slave system diffusively coupled.

\begin{eqnarray}
\dot{\bm{x}}&=&\bm{f}(\bm{x}) \nonumber \\
\dot{\bm{y}}&=&\bm{g}(\bm{y}) + \alpha \bm{H}( x - y) 
\label{eq:drive_response}
\end{eqnarray}
where $\bm{H}$ is a positive definite matrix. We can write the slave equation as
$$
\dot{\bm{y}}= \bar{\bm{g}}(\bm{y}) + \alpha \bm{H} \bm{x} 
$$
where
$\bar{\bm{g}}(\bm{y}) = \bm{g}(\bm{y}) - \alpha \bm{H} \bm{y}.$ The equation then splits into 
contributions solely coming from the slave and the driver. 
Now consider two copies of the slaves $\bm{y}_1$ and $\bm{y}_2$. Because we know the system will exhibit GS when the copies of the slaves synchronize, we introduce a variable $\bm{z} = \bm{y}_1 - \bm{y}_2$.  The system will undergo GS when $\bm{z}\rightarrow \bm{0}$. Differentiating we  obtain
\begin{equation}\label{zGS}
\dot{\bm{z}} = \bm{U}(t)\bm{z} - \alpha \bm{H} \bm{z},  
\end{equation}
where we used the mean value theorem \cite{jeffreys1988} to express 
$$
\bm{U}(t)\bm{z} = \bm{g}(\bm{y}_1(t)) - \bm{g}(\bm{y}_1(t)+\bm{z}(t))= 
\left(\int_0^1 D\bm{G}(\bm{y}_1(t) +  s \bm{y}_2(t)) ds \right) \bm{z}(t) 
$$

Notice that for the difference $\bm{z}$ the driving term $\bm{H}\bm{x}$ disappears as it is common for both copies of the slave $\bm{y}_1$ and $\bm{y}_2$. 
The only part of the coupling remaining is the term $-\alpha \bm{H} \bm{z}$, which adds an extra damping and provides dissipation. The trivial solution of $\bm{z}$ is globally stable if the coupling is large enough. Indeed, we can construct a Lyapunov function for $\bm{z}$. Indeed, consider 
$$
V (\bm{z}) = \frac 1 2 \langle \bm{z}, \bm{z}\rangle,
$$
and differentiating the Lyapunov function along the solution $\bm{z}(t)$ of Eq. (\ref{zGS}) we obtain 
\begin{eqnarray}
\frac{dV(\bm{z}(t))}{dt} &=& \langle \dot{\bm{z}}(t), \bm{z}(t)\rangle \\
&\le& (\| D \bm{g} \|  - \alpha \lambda_{min}(\bm{H}))  \| \bm y \|^2
\end{eqnarray}
where $\lambda_{min}(\bm{H})$ is the minimum eigenvalue of $\bm{H}$. In this last passage, we used the Cauchy Schwartz inequality $|\langle \bm{U}(t) \bm{z} , \bm{z}\rangle | \le \| \bm{U}(t)\bm{z}\| \| \bm{z}\| \le \|\bm{U} \| \| \bm{z}\|^2$, and noticed that 
$\| \bm{U} \| \le \|D \bm{g}\|$\footnote{We are using the uniform operator norm $\| \bm{U} \| = \sup_{t\ge0}\| \bm{U}(t) \|$.}. We also used the fact that $\bm{H}$ is positive $\langle \bm{H} \bm{z} ,\bm{z} \rangle \ge \lambda_{min} (\bm{H}) \| \bm{z} \|^2$. Hence,  for 
$$
\alpha > \alpha_c = \frac{\| D \bm{g} \|}{\lambda_{min}(\bm{H})}
$$
the derivative of the Lyapunov function is negative and every solution of the system sinks to zero.

{\bf What did we learn?} When $\alpha > \alpha_c$ we have GS.  Any two trajectories of the slave $\bm{y}_1$ and $\bm{y}_2$ will converge towards the same asymptotic state. This happens because the coupling terms adds extra dissipation. The convergence rate is exponential   $\| \bm{y}_1(t) - \bm{y}_2(t)\| \le K e^{-\eta t}$ because  $\dot V \le \eta V $ and   $\eta$ is uniform on the trajectories $\bm{y}_1, \bm{y}_2$ and $\bm{x}$. As a conclusion, there a function $\psi$ such that the dynamics of the slave can be as $\bm{y} = \psi(\bm{x})$. 

In the literature, a typical way to estimate whether one has GS is to compute the Lyapunov exponents of the slaves. Since we are assuming that the uncoupled systems are chaotic, for $\alpha=0$ the slave will have a positive Lyapunov exponents. As we increase $\alpha$ the maximum Lyapunov exponent may become negative for a value $\alpha_c$. We use this $\alpha_c$ as an estimate for the critical coupling for GS.

\subsection{Summary of Synchronization types}

\begin{figure}[h]
\centerline{\includegraphics[width=0.3\linewidth]{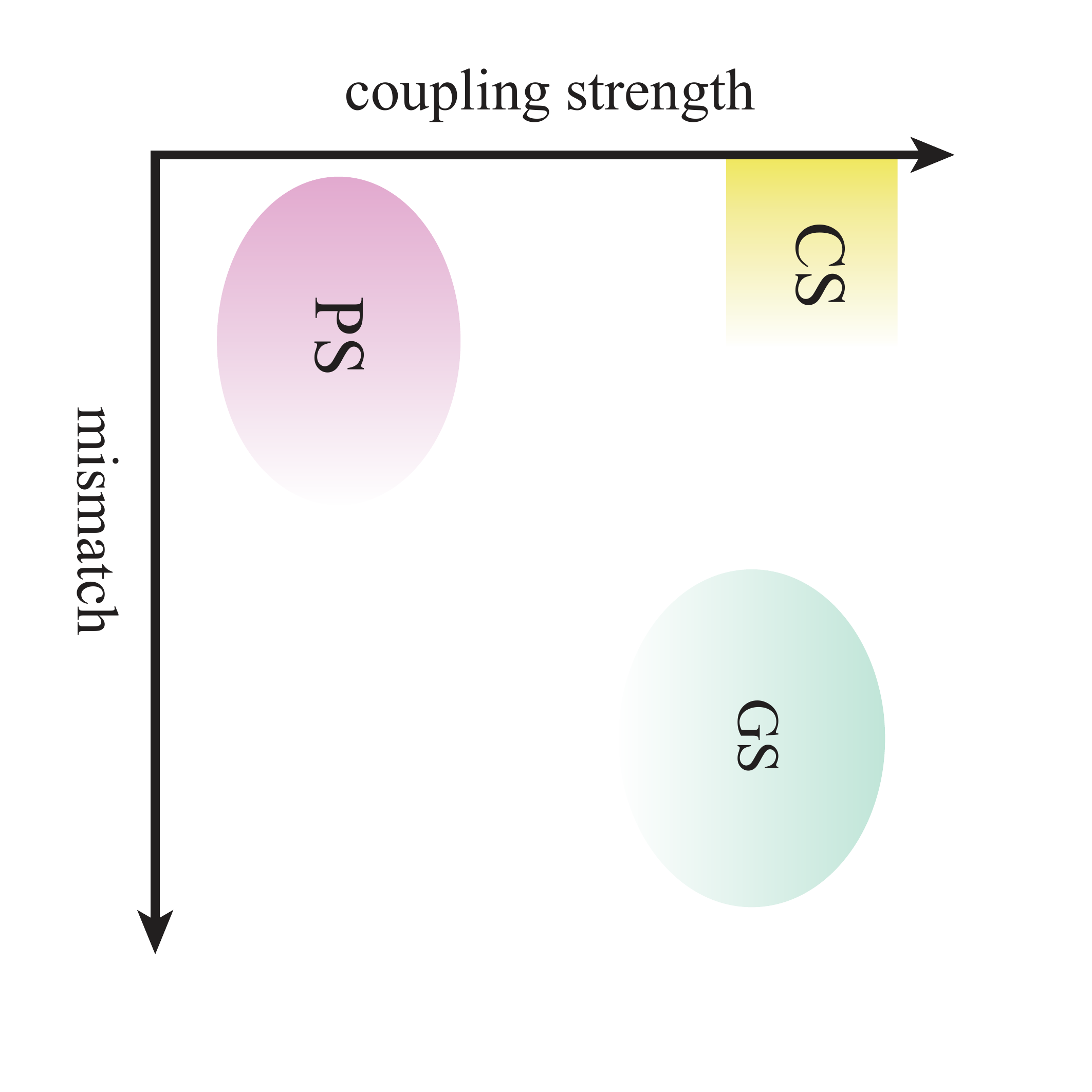}}
\caption{Diagram of synchronization types for diffusively coupled systems. The horizontal axis depicts the mismatch between the isolated dynamics ($\bm{f}_1$ and $\bm{f}_2$) and the vertical axis the coupling constant. The diagram shows the typical balance between mismatch and coupling strength and  to achieve a certain kind of  synchronization. Complete synchronization (CS) occurs for identical chaotic systems ($\bm{f}_1=\bm{f}_2$) and large enough coupling strengths. Phase synchronization (PS) is observed between slightly different systems for small coupling strengths. Generalized synchronization (GS) is a result of master-slave system and can occur for large mismatch parameters or even between distinct systems when the coupling strength large enough.}
\label{fig:sync_diagram}
\end{figure}

We discuss the three cases commonly found in applications. A schematic representation of the cases is found in  Fig.~\ref{fig:sync_diagram}. \\

\begin{figure}[h]
\centerline{\includegraphics[width=0.5\linewidth]{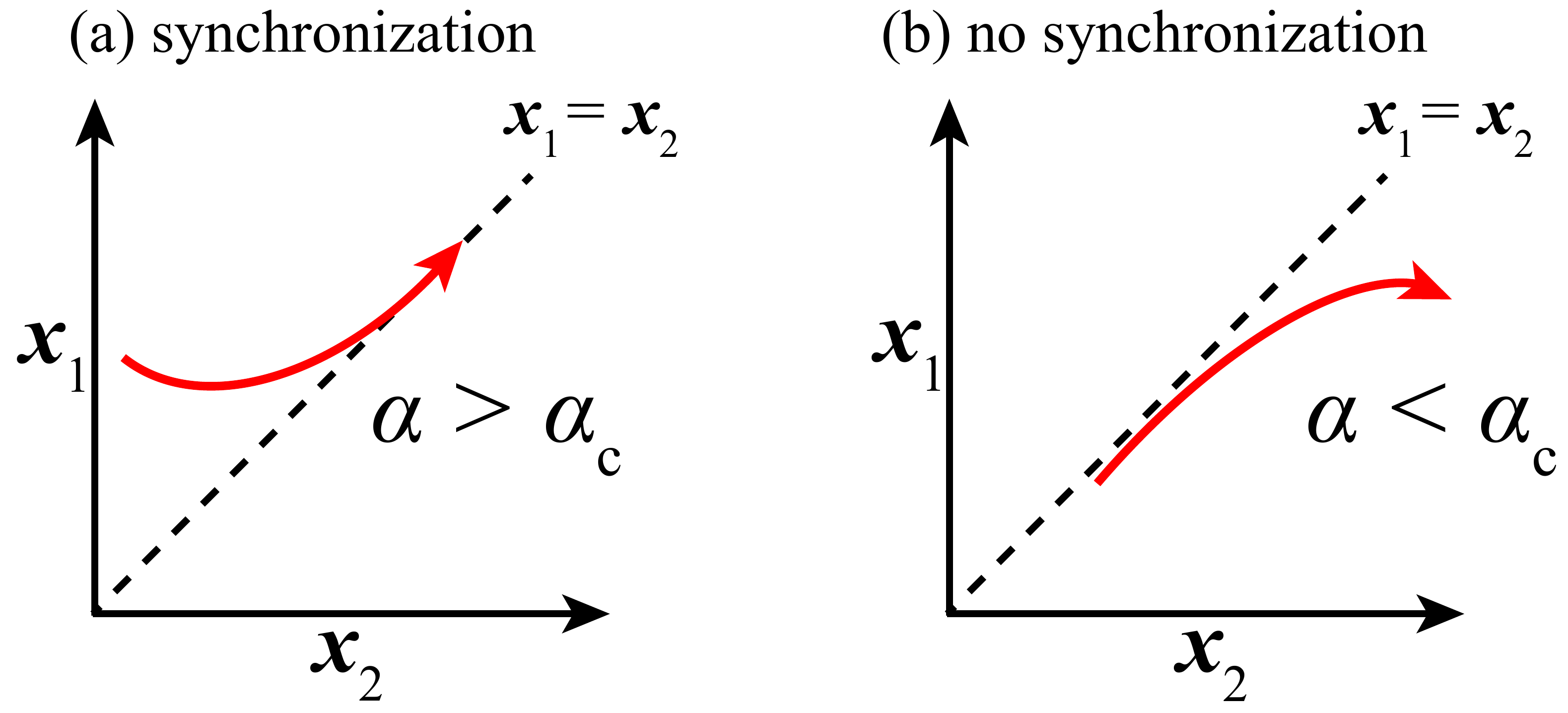}}
\caption{Illustration of complete synchronization. (a) If the coupling strength is large enough ($\alpha>\alpha_c$), the systems converge to invariant synchronization manifold ($\bm{x}_1 = \bm{x}_2$), (b) otherwise ($\alpha<\alpha_c$), they diverge hence no synchronization.}
\label{fig:cs_demo}
\end{figure}

{\bf Complete synchronization} in identical systems. If the isolated dynamics are identical, $\bm{f}_{1}=\bm{f}_{2}$ and diffusively coupled, hence,   
the subspace 
$
\bm{x}_{1}=\bm{x}_{2}
$
is invariant under Eq.~(\ref{eq:nonlinear}). Indeed, the coupling vanishes and both systems will oscillate in unison for all coupling strengths $\alpha$ and all times. Such collective motion is called complete synchronization (CS). The  question is whether CS is attractive, that is, if the oscillators state are nearly the same $\bm{x}_1(0) \approx \bm{x}_2(0)$, will they synchronize? Meaning that
$$
\lim_{t\rightarrow \infty} \| \bm{x}_1(t) - \bm{x}_2(t) \| = 0.
$$
See  Fig.~\ref{fig:cs_demo} for an illustration. In Sec~\ref{sec:complete_sync}, the CS was discussed in detail. 
\vspace{-0cm}

\begin{figure}[h]
\centerline{\includegraphics[width=0.35\linewidth]{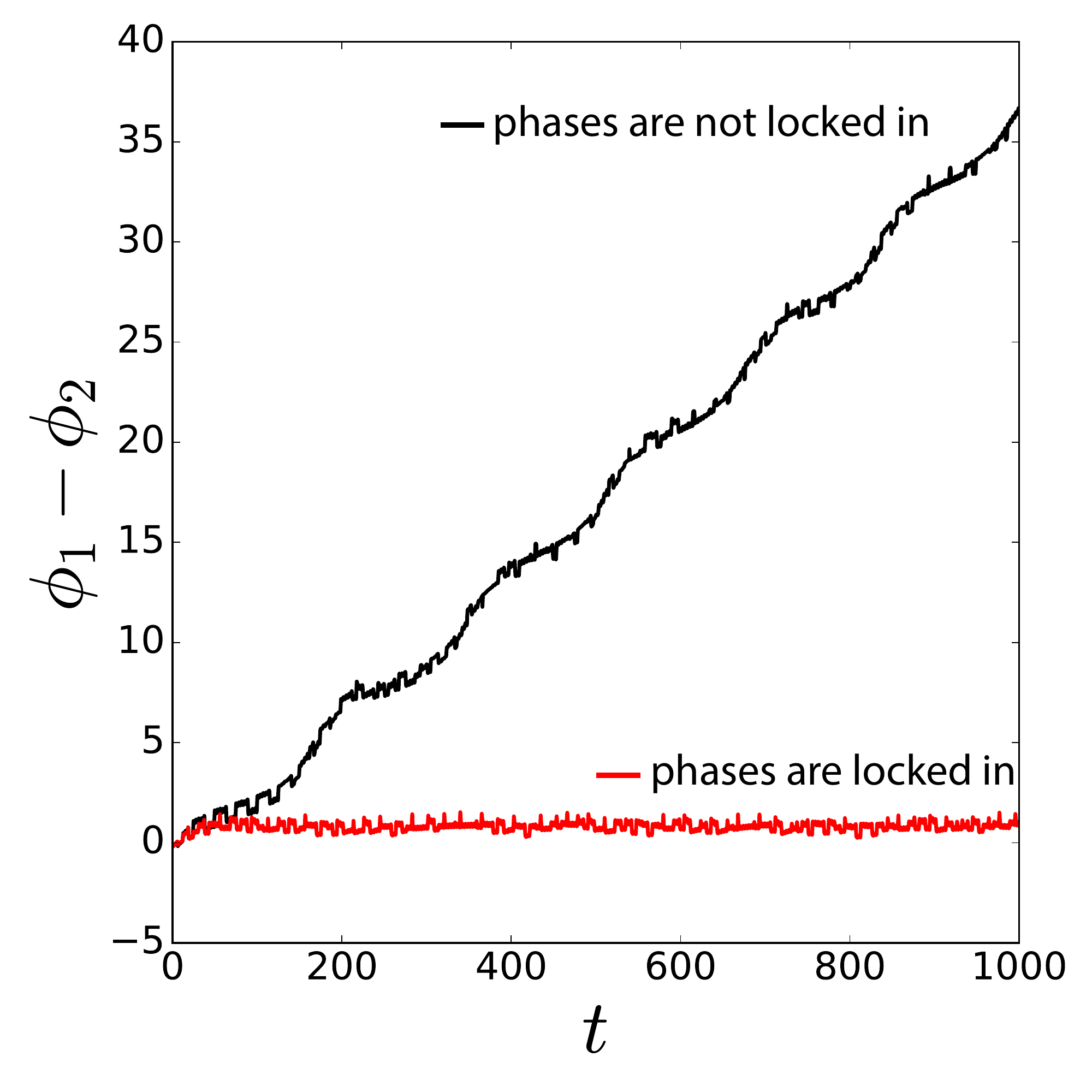}}
\caption{Illustration of phase synchronization for two coupled slightly different and chaotic systems ($\bm{f}_{1} \ne \bm{f}_{2}$).  The evolution of the phase differences between the systems for two different coupling constants $\alpha < \alpha_c$ and $\alpha > \alpha_c$.}
\label{fig:ps_demo}
\end{figure}

{\bf Phase synchronization} (PS) when $\bm f_1 \approx \bm f_2$. In this situation the subspace $\bm{x}_1 = \bm{x}_2$ is not invariant. And each system will have its own frequency given by their phase dynamics $\phi_{1,2}$. For small coupling strengths the phases can be locked $\phi_1 \approx \phi_2$ while the amplitudes remain uncorrelated Fig.~\ref{fig:ps_demo}. This phenomenon is called phase synchronization. Typically, the critical coupling for PS is proportional to the mismatch $\bm f_1 - \bm f_2$, as illustrated in Fig.~\ref{fig:sync_diagram}. Further details were given in Sec.~\ref{sec:phase_sync}. \\

\begin{figure}[h]
\centerline{\includegraphics[width=0.5\linewidth]{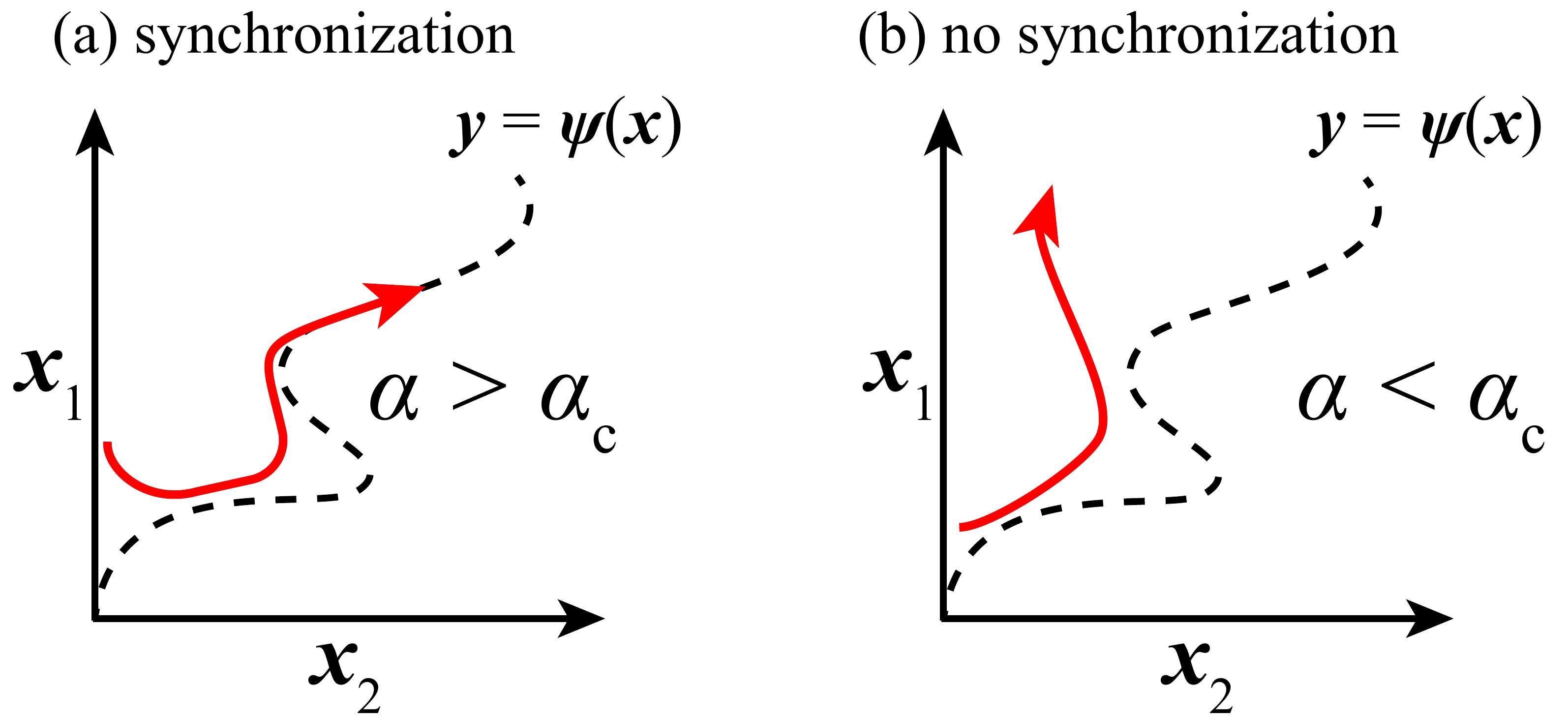}}
\caption{Illustration of generalized synchronization. If the coupling strength is large enough, a functional relationship ($\bm{y} =\psi(\bm{x})$) exhibits between the dynamical variables $\bm{x}_1$ and $\bm{x}_2$. (a) If $\alpha > \alpha_c$, the generalized synchronization is observed (b) otherwise $\alpha < \alpha_c$, there is no generalized synchronization.}
\label{fig:gs_demo}
\end{figure}

{\bf  Generalized synchronization} in master slave configurations.
If the vector fields are different $\bm{f}_{1} = \bm{f}$ and $\bm{f}_{2} = \bm{g}$, the systems can  synchronize, but in a generalized sense. We consider systems coupled in a master-slave configuration.  For certain coupling strengths, the dynamics of the master $\bm x$ can determine the dynamics of the slave
$\bm y$, that is $\bm y = \psi(\bm x)$, see Fig.~\ref{fig:gs_demo}. This is called \textit{Generalized Synchronization} (GS). Further details for GS was given in Sec.~\ref{sec:generalized_sync}.

\newpage

\subsection{Historical Notes}

Studies on synchronization dates back to Christiaan Huygens who studied coupled pendulums. In this case, the pendulums are periodic and have distinct frequencies, but due to interaction they adjust their rhythm. In the seventies, thanks to the works of Winfree \cite{winfree1967} and Kuramoto \cite{kuramoto1975} the area experienced a boom. In early 2000's many excellent books and reviews were devoted to this subject \cite{strogatz2003,pikovsky2001,balanov2009,kapitaniak2014,arenas2008,rodrigues2016}.

Chaotic synchronization on the other hand is younger. To begin with, the establishment and full acceptance of the chaotic nature of dynamics is fairly new \cite{gleick2011}.  The role of chaos in nature was object of intense debate in the seventies when Ruelle and Takens proposed that  turbulence was generated by chaos. 

The chaotic dynamics can be fairly complicated. Typically, the evolution never repeats itself, nearby points drift apart exponentially fast, but in the long run the dynamics return arbitrarily close to its initial state. Such dynamics is so unpredictable that modern approach tackles it from a probabilistic perspective. 

Given this complexity many researchers thought it is unlikely one could possibly synchronize two chaotic systems. How could a system with exponential divergence of nearby trajectories have a state were trajectories come together while keeping their chaotic nature? That seemed paradoxal. 
Chaos and synchronization should not come together. This view was proven wrong in the late eighties. In fact,  we have come to think it as rather natural. Funny enough, before this view was accepted synchronization of chaos had to be rediscovered a few times. 

Back in the eighties,  Fujisaka and Yamada had the first results on synchronization of chaos \cite{fujisaka1983, yamada1983, yamada1984}. They publish it in Japan, but their results went fairly unnotice in the west. Just two years later mathematicians and physicists from Nizhny Novgorod exposed many of the concepts necessary for analyzing synchronous chaos \cite{afraimovich1986}. This paper is now famous, but back then it also went largely unnoticed. 

Only some years later the study of synchronization of chaos had its boom, largely as a result of the works by Pecora and Carroll \cite{pecora1990}. Lou Pecora and co-workers went systematically tackling two coupled systems and then moved on to study chaotic systems coupled on periodic lattices \cite{heagy1995, carroll1996}. These early results were relying on ideas from Nuclear physics to diagonalize the lattice and stability theory (the Lyapunov methods) to analyze synchronization. 

The nineties proved prolific for synchronization! Two groups proposed an extension of synchronization, the so-called generalized synchronization \cite{rulkov1995, abarbanel1996, kocarev1996,kocarev1996b,kocarev1996c}.  Generalized synchronization {\it only} asked for a functional relationship between the states, that is, the dynamics of one system is fully determined by the dynamics of the other.  Also in the mid nineties, Rosenblum, Pikovsky and Kurths put forward the concept of chaotic phase synchronization. Here two nearly identical chaotic oscillators can have their phase difference bounded while the amplitudes remain uncorrelated \cite{rosenblum1996,rosenblum1997}.  

A few years down the road, Pecora and Carroll were able to generalize their approach to undirected networks of diffusively coupled systems \cite{pecora1998}. They also wrote a review about their approach \ref{ChaosPecora}. These results open the door to the understanding of the role of the linking structure on the stability of synchronization. Barahona and Pecora \cite{barahona2002} showed that small-world networks are easier to globally synchronize than regular networks. Motter and coworkers \cite{nishikawa2003, motter2005a,motter2005b, zhou2006} showed that heterogeneity in the network structure may hinder global synchronization. On the other hand, Pereira showed such heterogeneity may enhanced synchronization of highly connected nodes \cite{pereira2010}. 


\section{Applications}
\label{sec:applications}
In this section, we discuss the role of synchronization phenomena in various applications including secure communication approaches, parameter estimation of a model from data and prediction.

\subsection{Secure Communication based on Complete Synchronization}
\label{sec:app_cs}



\begin{figure}[h]
\centerline{\includegraphics[width=0.5\linewidth]{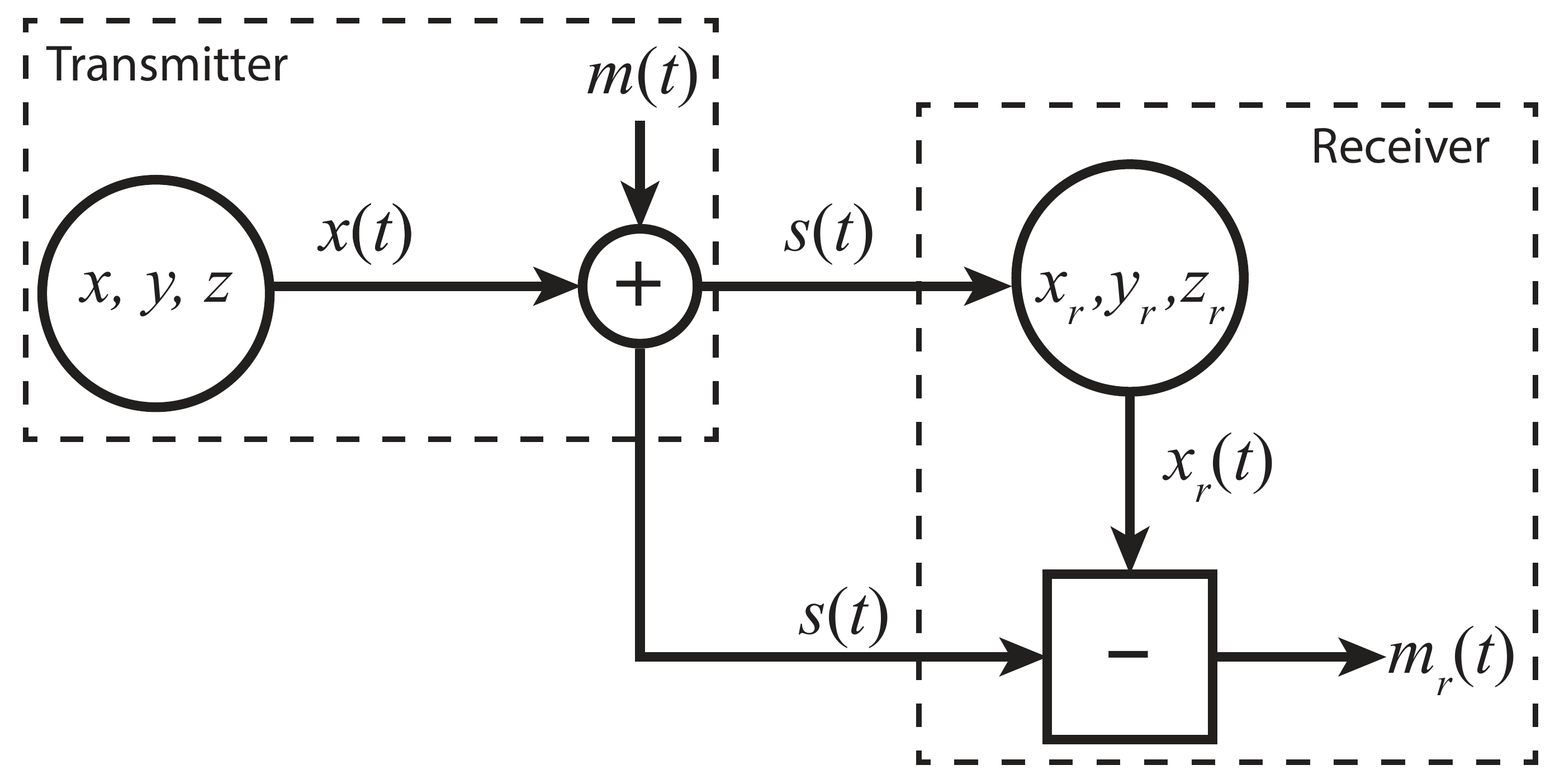}}
\caption{Illustration of the message masking on bearing signal scheme. Transmitter generates a chaotic signal $x(t)$ and add the message $m(t)$ on it. This combined two signals $s(t) = x(t) + m(t)$ is sent to the receiver and both systems synchronize. Discarding the synced signal $x_r$ from the $s$, the message $m_r\sim m$ is restored.}
\label{fig:app_scheme_masking}
\end{figure}

The first approach is to send an \textit{analog} message \cite{oppenheim1992}. The key idea is the following: the sender adds the message $m(t)$ on a chaotic signal $x(t)$ and generate a new signal $s(t)=m(t)+x(t)$ (Fig.~\ref{fig:app_scheme_masking}). The assumption is that the amplitude of $x(t)$ is much larger than the amplitude of $m(t)$. This method is called the masking information on bearing signals. Because chaotic signals are  noise-like and broadband (have many frequencies),  it is difficult to read the message. One could then retrieve the message using synchronization. 


Masking of messages on bearing signals  does not require encryption. Here is the keystone is selection of the transmitter and receiver systems such that they synchronize.
They are also assumed to be identical (this means that the receiver knows the parameters of the transmitter). One can retrieve the message if the parameters are known. 
So, the parameters play a role of encryption key.

\begin{figure}[h]
\centerline{\includegraphics[width=0.4\linewidth]{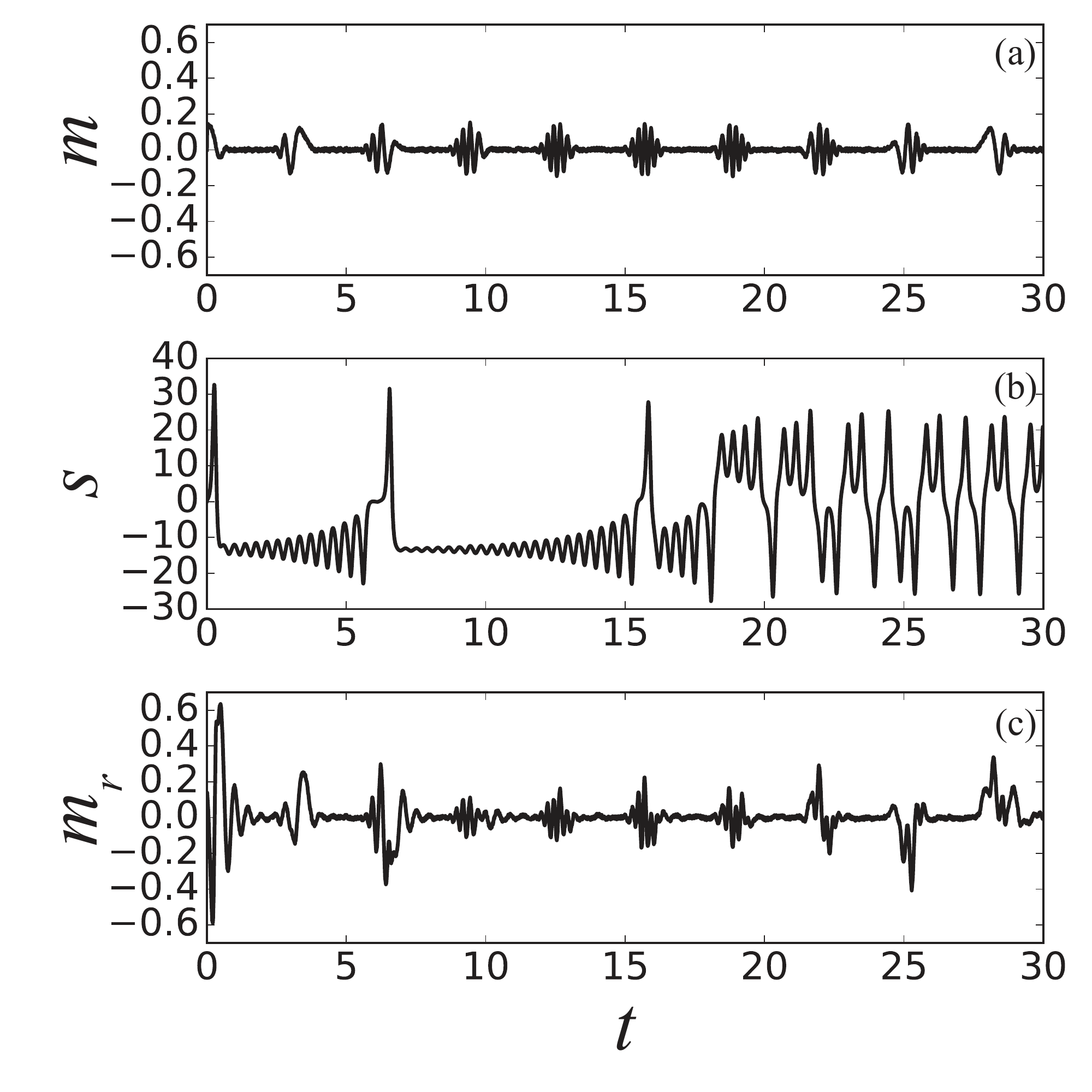}}
\caption{The masking an analog message on bearing chaotic signal. (a) the low-amplitude message, (b) the message embedded into high-amplitude chaotic signal and (c) restored message from the transmitted signal.}
\label{fig:app_res_masking}
\end{figure}

We illustrate this communication scheme using the Lorenz system Eq. (\ref{eq:lorenz_system}).
The Lorenz system has the particularity that it divided into subsystems $(x,z)$ and $(y,z)$. 
We can use the variables $x$ or $y$ of a subsystem to drive the other subsystem. In this driving setting the synchronization between driver and slave is exponentially stable provided that the parameters $\sigma$, $\rho$ and $\beta$ are identical. Here, we have chosen $x$ component to act as a driver. 
The message $s(t)$  drives the receiver system as
\begin{equation}
\begin{array}{rcl}
\dot{x}_r &=& \sigma ( y_r -x_r )   \\
\dot{y}_r &=& s ( \rho  - z_r) - y_r   \\
\dot{z}_r&=&-\beta z_r + sy_r. 
\end{array}
\label{eq:driven_lorenz_system_app}
\end{equation}
Since the synchronization of chaotic systems is exponentially stable for such system~\cite{pecora1990,pecora1991} under low amplitude of noise the synchronization (coherence) still occurs. Then the chaotic signal $x_r$ can be obtained from Eq.~\ref{eq:driven_lorenz_system_app}. Therefore the message can be regenerated by $m(t) = s(t) - x_r(t)$ (Fig.~\ref{fig:app_scheme_masking}).  

as a message we use the signal 
$$
m(t) = 0.1\frac{\sin(1.2\pi\sin^2(t))}{\pi\sin^2(t)}\cos(10\pi\cos{(0.9t)})+\xi
$$
where $\xi $ is a white noise Fig.~\ref{fig:app_res_masking}(a). We attach this message on $x$-component of the Lorenz system Eq.~(\ref{eq:lorenz_system_app}) with parameters $\sigma=16.0$, $\rho=45.2$ and $\beta=4.0$ then the information is masked on bearing signal $s$ Fig.~\ref{fig:app_res_masking}(b). By synchronization we restore the message $m_r\sim m$ (Fig.~\ref{fig:app_res_masking}(c)).  This secure communication application is also experimentally demonstrated  by using Chua's circuits~\cite{kocarev1992, parlitz1992} and Lorenz-like circuit~\cite{cuomo1993b}. 

\begin{figure}[h]
\centerline{\includegraphics[width=0.4\linewidth]{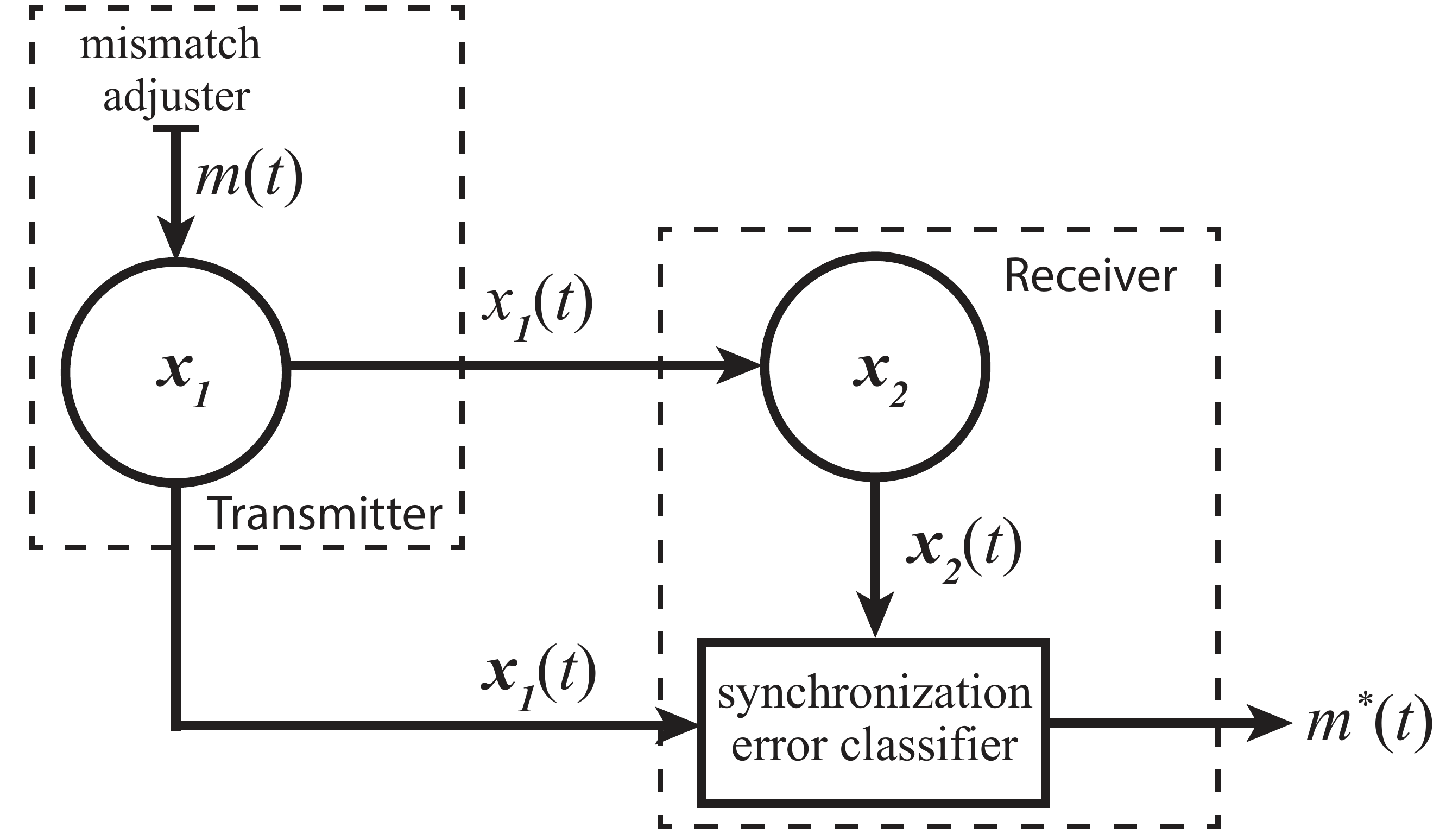}}
\caption{A secure communication scheme: hiding a digital message on a chaotic signal. Changing a parameter of transmitter causes different level of synchronization errors between the transmitter and the receiver. The amplitude of the error $E$ brings the message out.}
\label{fig:app_scheme_binary}
\end{figure}

The second approach is the modulation of the parameters for the \textit{digital} communication. In this case, the message $m(t)$ only carries binary-valued signals.  The setup is similar to the masking approach but the message is included in the transmitter parameters. The transmitter system has an adjustable  parameter $\sigma_a(t) = \sigma + \delta m(t)$ such that we can tune the system into synchronization when $m(t) = 0$ and out synchronization when $m(t) = 1$ Fig.~\ref{fig:app_scheme_binary}. We retrieve the message $m(t)$ by the  synchronization and desynchronization pattern.

The dynamics of transmitter and the receiver is given by
\begin{eqnarray}
\dot{x}_1 &=& \sigma_a ( y_1 -x_1 ) \nonumber \\
\dot{y}_1 &=& x_1 ( \rho  - z_1) - y_1 \nonumber \\
\dot{z}_1&=&-\beta z_1 + x_1y_1 \nonumber \\
\dot{x}_2 &=& \sigma ( y_2 -x_2 ) \nonumber \\
\dot{y}_2 &=& x_1 ( \rho  - z_2) - y_2 \nonumber \\
\dot{z}_2&=&-\beta z_2 + x_1y_2 \nonumber
\label{eq:lorenz_system_app_binary}
\end{eqnarray}

where $\sigma_a = \sigma+\delta m(t)$ is the adjustable parameter.  Again the key aspect is having mismatch between the transmitter and the receiver does not allow systems to synchronize.  Modulating the parameter $\sigma_a$ by the message $m(t)$ we can produce different levels of synchronization errors. Choosing the parameters of the transmitter and the receiver identical gives the synchronization error $E \sim$ 0 (CS), this can be assigned binary 0 by $m(t)=0$. The large mismatch $\delta$ causes a certain amount of synchronization error $E>0$, this can be assigned binary signal 1 by $m(t)=1$. Then the digital communication can be set between sender and receiver~\cite{cuomo1993a,cuomo1993b}.   

\begin{figure}[h]
\centerline{\includegraphics[width=0.4\linewidth]{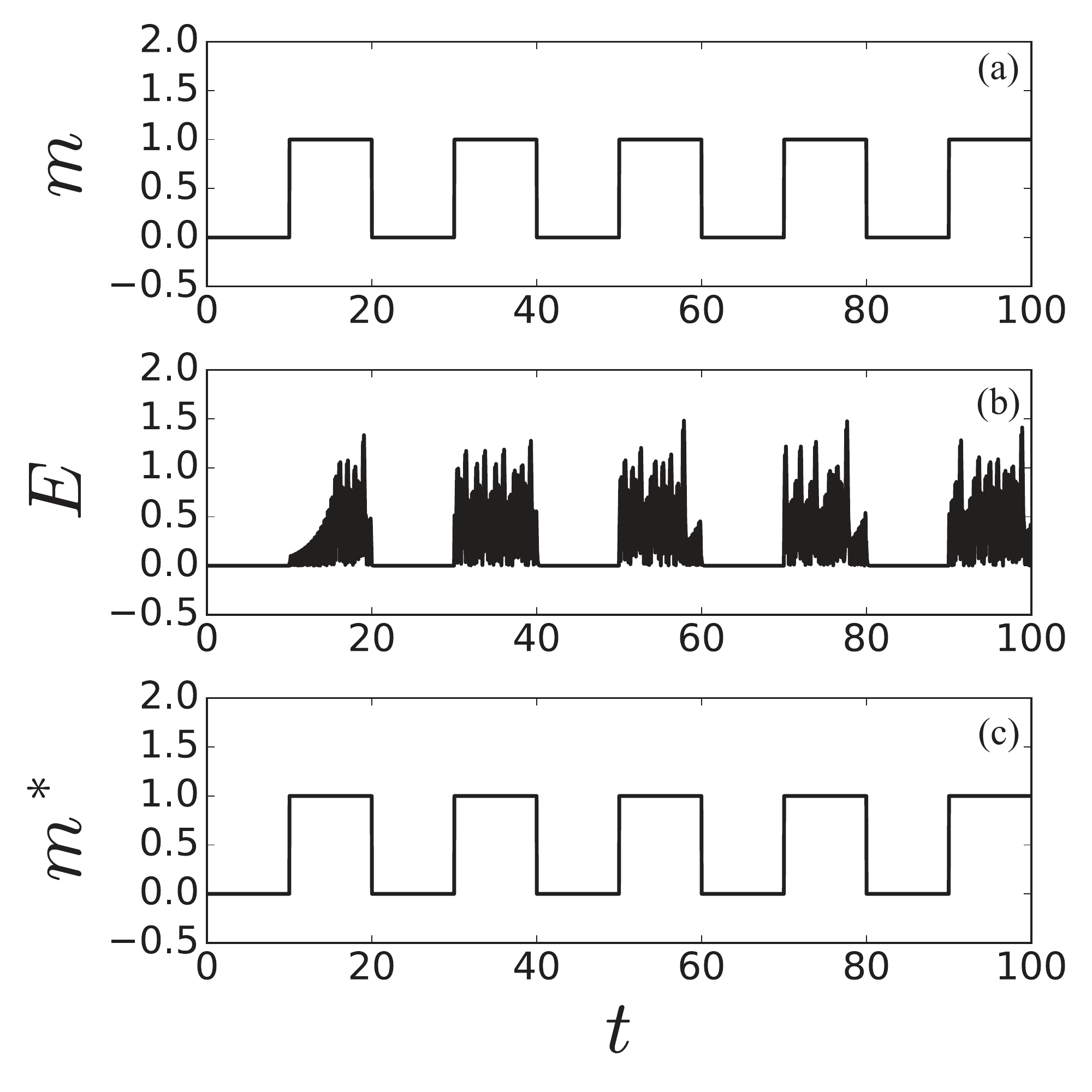}}
\caption{Manipulating the parameter of a transmitter allows digital secure communication. In this application 10 time units are used for a single digit. (a) Digital message (0101010101), (b) the synchronization error and  (c) restored message.}
\label{fig:app_results_binary}
\end{figure}

Using same parameters as in the previous application ($\sigma=16.0$, $\rho=45.2$ and $\beta=4.0$), we illustrate this digital communication.  For this example, a digit of the message is set for 10 time units and the message is 0101010101 (Fig.~\ref{fig:app_results_binary}(a)). For each message time, we change $\sigma_a$ from $\sigma$ to $\sigma+\delta$ and other way round. Then the synchronization error $E$ varies according to this change (Fig.~\ref{fig:app_results_binary}(b)). Due to change in the $E$, the message is restored (Fig.~\ref{fig:app_results_binary}(c)).

There are more communication applications using the synchronization mechanism e.g. using hyperchaotic systems~\cite{dedieu1993,kocarev1996c,peng1996, carroll1996} or volume-preserving maps~\cite{carroll1998}. The common idea of all these given approaches is the CS phenomena, negative conditional Lyapunov exponent between the systems are needed to exhibit of the synchronization~\cite{pecora1990,pecora1991}. 

\subsection{Secure Communication Based on Phase Synchronization}

Security is an important issue for communication approaches. As might be expected, some methods were improved and reported to break the CS based communication schemes \cite{perez1995,short1994,dedieu1997,parlitz1996c}. Then more secure communication scheme demonstrated by means the PS \cite{chen2003}.

The scheme based on the PS possesses three chaotic R\"ossler systems ($\bm{x}_{1,2,3}$). The transmitter of the scheme consists of two weakly coupled identical systems $\bm{x}_1$ and $\bm{x}_2$ over their $x$-components Eq.~(\ref{eq:app_comm_ps_ill}) and the receiver $\bm x_3$ has slightly different dynamics. In this case, we couple the systems with using their phases Eq.~(\ref{eq:app_comm_ps_ill}) as presented in Ref.~\cite{chen2001}. The phase definition for R\"ossler system is given by Eq.~(\ref{eq:phase_roessler}). The mean of  two systems' phases $\phi_1$ and $\phi_2$ in transmitter can be used as a spontaneous phase signal $\phi_m$ to couple the third system as in Eq.~(\ref{eq:app_comm_ps_ill}). As distinct from the CS based schemes, we have three systems and the reason behind these to improve the security. The return maps of the phase $\phi_m$ is way more complex than $\phi_1$ (or $\phi_2$), this makes to break dynamics not trivial \cite{chen2003}. 

\begin{figure}[h]
\centerline{\includegraphics[width=0.5\linewidth]{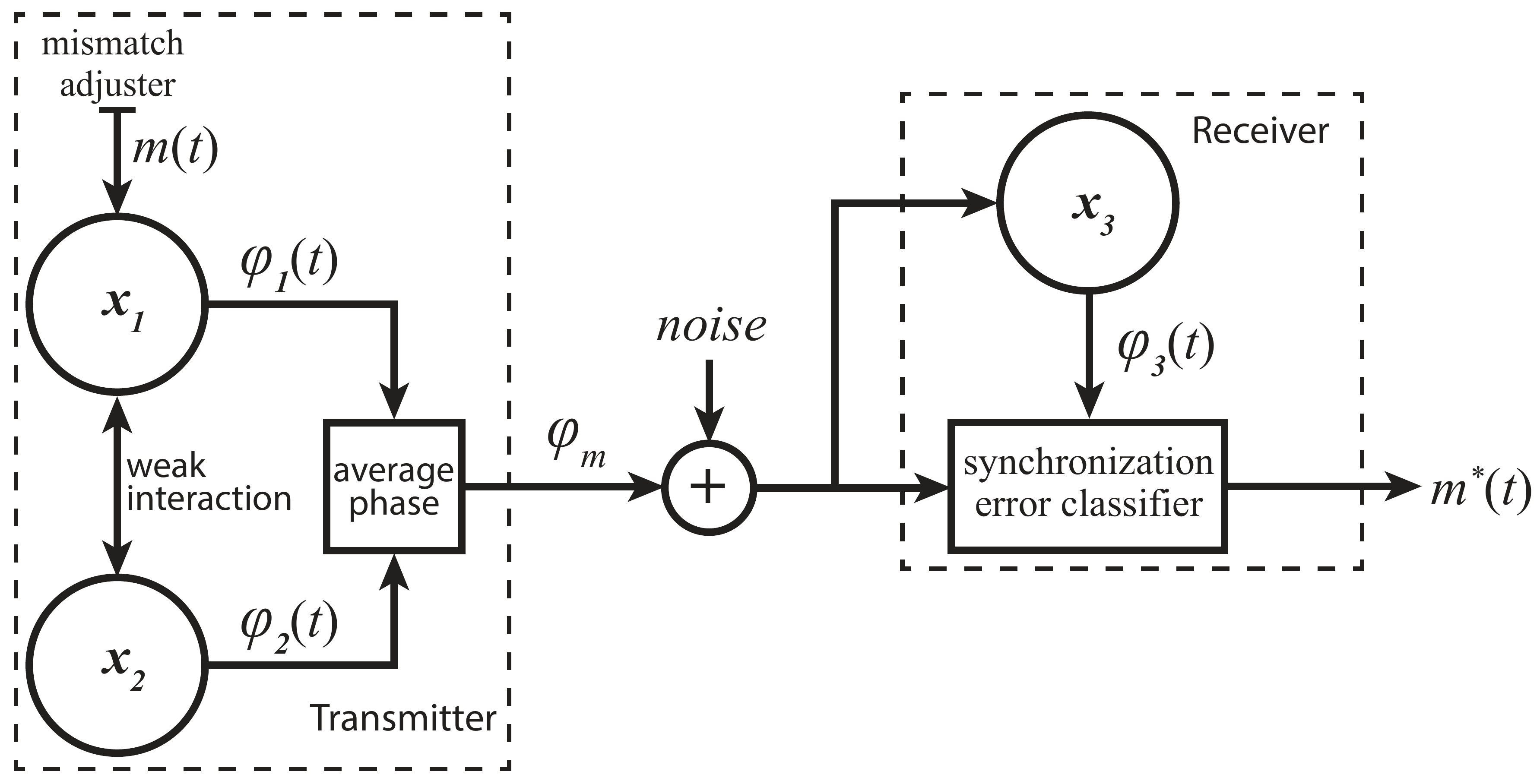}}
\caption{A secure communication scheme by phase synchronization: hiding a digital message on a chaotic signal. Changing a parameter of transmitter causes ...}
\label{fig:app_ps_scheme}
\end{figure} 

We illustrate this application by 
\begin{equation}
\begin{array}{rcl}
\dot{x}_{1,2} &=& -(\omega+\Delta \omega) y_{1,2} - z_{1,2} +\epsilon(x_{2,1}-x_{1,2})  \\
\dot{y}_{1,2} &=& x_{1,2} + ay_{1,2}   \\
\dot{z}_{1,2} &=& b + z_{1,2}(x_{1,2}-c) \\
\dot{x}_{3} &=& -y_{3} - z_{3} + \alpha(r_3 \cos{ \phi_m} - x_3)  \\
\dot{y}_{3} &=& x_{3} + ay_{3}   \\
\dot{z}_{3} &=& b + z_{3}(x_{3}-c) 
\end{array}
\label{eq:app_comm_ps_ill}
\end{equation}
where constant $\omega =1$ and standard parameters of the R\"ossler system $a=0.15$, $b=0.2$ and $c=10.0$.  Coupling constants $\epsilon=5\times10^{-3}$ is between $\bm{x}_1$ and $\bm{x}_2$, and $\alpha$ is between the transmitter and the receiver. $r_3$ is the amplitude of the response system given by Eq.~(\ref{eq:amplitude_roessler}). $\Delta\omega$ is the adjustable mismatch parameter, for this illustration we select
$$
\Delta\omega = \left\{
\begin{array}{ll}
0.01  &\mbox {if bit digit} = 1 \\
-0.01 &\mbox {if bit digit} = 0.
\end{array}
\right.
$$


Similar to digital communication by the CS (see Section~\ref{sec:app_cs}), the modulation of parameters in the transmitter would hide a binary message $m(t)$ on $\phi_m$.  The PS will exhibit between $\phi_m$ and $\phi_3$. Due to the changes on the adjustable control parameters $\Delta \omega$, the phase difference between  $\phi_m$ and $\phi_3$ varies. In other words, the phases are locked on different phase shifts. The message can be retrieved from different phase locking values (Fig.~\ref{fig:app_ps_scheme}).
 
Because of the weak coupling, the CS never occurs Fig.~\ref{fig:app_ps_res}(a). Every 10 time unit we switch $\Delta\omega$ parameter to create a digital message $m(t)$ (010101...) Fig.~\ref{fig:app_ps_res}(b). The hidden message on chaotic signal can be restored from the receiver using the phase difference between $\phi_m$ and $\phi_3$. If the message digit is 0, then the phase difference oscillates most time below 0, otherwise above 0 Fig.~\ref{fig:app_ps_res}(c). Therefore it is easy to restore associated message $m^*$ Fig.~\ref{fig:app_ps_res}(d).
\begin{figure}[h]
\centerline{\includegraphics[width=0.4\linewidth]{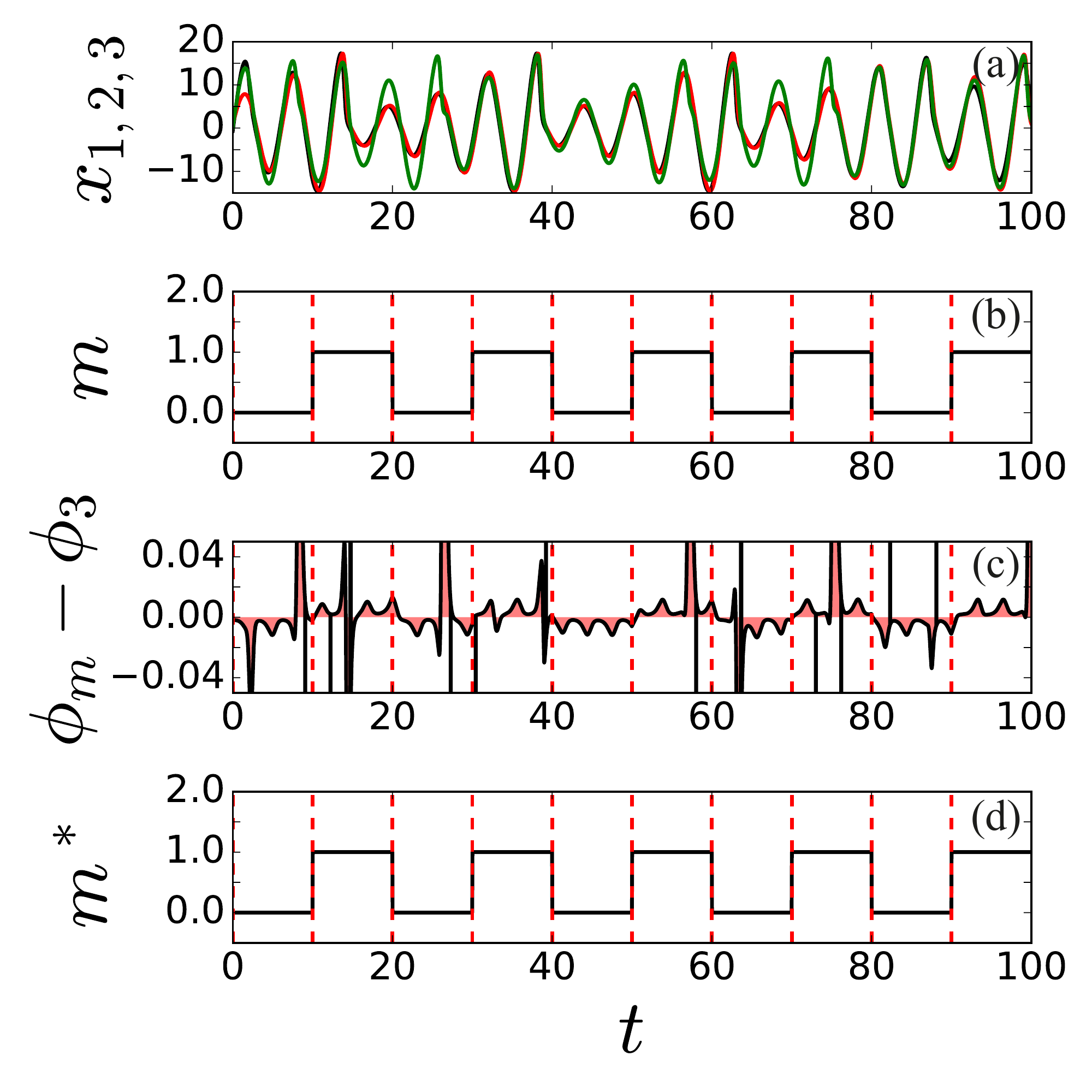}}
\caption{An illustration for the secure communication by phase synchronization: hiding a digital message on a chaotic signal. Changing a parameter of transmitter causes ...}
\label{fig:app_ps_res}
\end{figure} 


In real world examples, it is almost not possible to create identical systems, and the noise is always an issue to deal. The phase locking can be still preservable under effect of a certain level of noise.


\subsection{Parameter Estimation and Prediction}

Now we have data and we want to learn about the system that generated the data. Thus we will be able to predict the future behaviours and critical transitions. The determining equations of the system are known however the parameters are not. The goal is to find these unknown parameters with using synchronization phenomena. So, we blend the data with equations. 
The data is then used to drive the equations. If they and coupled in a proper way (Sec.~\ref{sec:master_slave}), the equations can synchronize with data. 
The key point is the following: if the parameters of slave system are identical with the master whose produced driving signal, then the CS exhibits (synchronization error $E = 0$) otherwise no CS (synchronization error $E > 0$).  Therefore, it is possible to estimate the parameters by a strategy to minimize the synchronization error $E \to 0$ such as POWELL technique \cite{powell1964}.   


We assumed that we have a limited data and we want to predict the future of the system. After the parameters are estimated, the  state of the synchronized slave matches the data. Because the solution of the equations are then the same as the data, we can use the equations to predict future dynamics.

\begin{figure}[h]
\centerline{\includegraphics[width=0.5\linewidth]{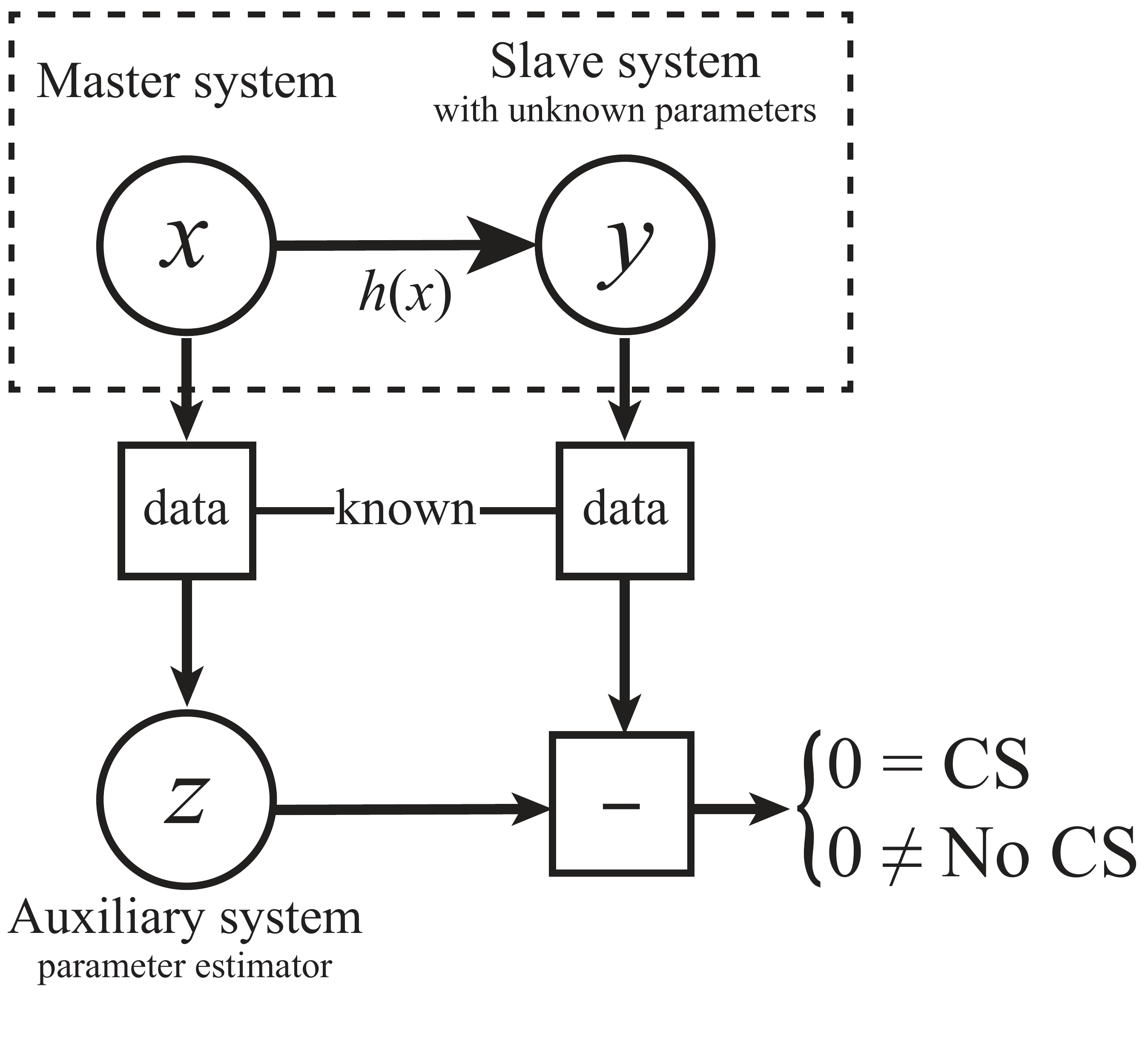}}
\caption{Only two data sets are known from a master-slave system, into the dashed rectangle, without any info about the parameters. An auxiliary system is driven by the data from the master and measure the amplitude difference between the auxiliary and the data from the slave system. If the difference is 0, then there is a CS that means the parameters of the slave and the auxiliary are identical.}
\label{fig:generalized_parest}
\end{figure}

The second approach is to estimate a slave system's parameters of a master-slave system. In this case, the dynamics of master and slave is distinct. We assume that we have two given datasets: one of them from master system and the other one is from the slave. The governing dynamics of the master-slave system is given
\begin{eqnarray}
\dot{\bm x}&=& \bm f(\bm{x}) \nonumber \\
\dot{\bm y}&=& \bm g(\bm{y}) + \bm{h}(\bm{x,y}) \nonumber
\end{eqnarray}
where $\bm x$ and $\bm y$ are the states of master and slave systems respectively. We aim to estimate the parameters of $\bm g$. Here we cannot use $\bm y$ data to drive another $\bm g$ system directly as in previous approach since $\bm y$ is driven by $\bm x$. If we know that master-slave system is in the GS and the coupling function $h$ is known, then we can apply the auxiliary system approach which is the master system drives an auxiliary (copy of the slave) $\bm z$ Fig.~\ref{fig:generalized_parest}. We expect that the CS exhibits between the slave $\bm y$ and auxiliary $\bm z$ systems if the parameters are identical.
Using the GS idea, the problem turned into the CS problem. From now on, minimizing the synchronization error $E \to 0$ technique can be used to estimate the unknown parameters. Similar to the previous approach, the future of the system can be predicted as well.

{\bf Example:} Consider a Lorenz system with classical parameters is driven by a R\"ossler system. We have only two data sets, $x_1$-component of the R\"ossler (Eq.~(\ref{eq:roessler_system})) and $y_1$-component of the Lorenz (Eq.~(\ref{eq:lorenz_system})). Then we drive an auxiliary system $\bm z$ by $x_1$ as
\begin{eqnarray}
\dot{z}_1 &=& \sigma_e ( z_2 -z_1 ) + \alpha(x_1-z_1) \nonumber \\
\dot{z}_2 &=& z_1 ( \rho_e  - z_3) - z_2 \nonumber \\
\dot{z}_3 &=&-\beta_e z_3 + z_1z_2. \nonumber
\label{eq:app_response}
\end{eqnarray}
The goal is to find the parameters of slave system. The spontaneous synchronization error is 
\begin{equation}
E(t) = \|z_1-y_1\|.
\label{eq:sim_err}
\end{equation}
Adjusting the parameters of $\bm{z} (\sigma_e,\rho_e,\beta_e)$ we minimize the simultaneous error $E(t)$ by Powell's algorithm \cite{powell1964}. This method allows us to estimate the parameters of the slave system (Fig.~\ref{fig:app_ps_res}).
\begin{figure}[h]
\centerline{\includegraphics[width=0.5\linewidth]{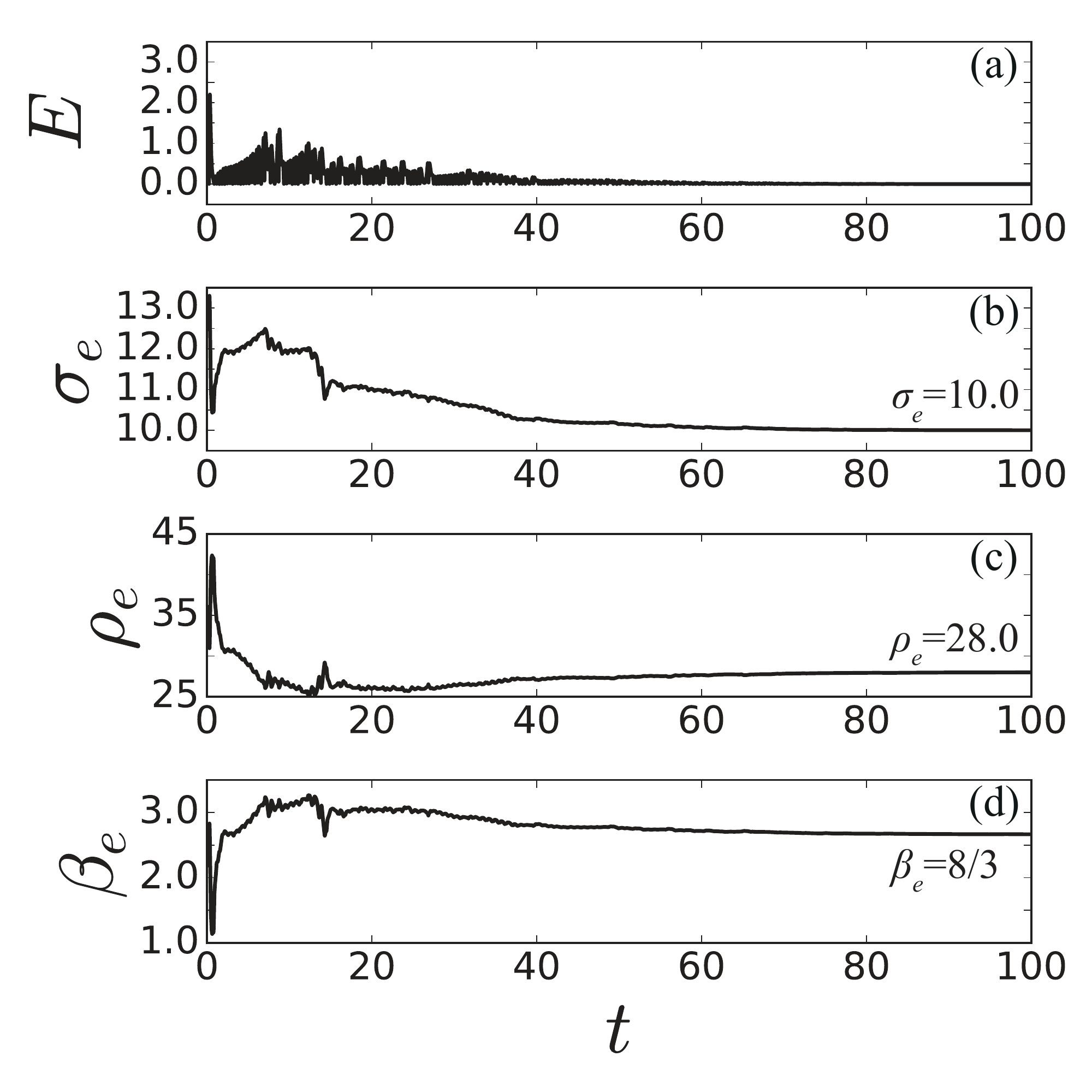}}
\caption{Illustration of parameter estimation estimation. Standard parameters of Lorenz system.}
\label{fig:app_ps_res}
\end{figure} 

\subsection{Chaos Anticipation}
\label{sec:anticipating}
Chaos is unpredictable but  synchronization can help predicting the state of a chaotic system ahead of time. Anticipating synchronization (AS) is a good approach for the future prediction since the slave system synchronizes with the upcoming states of the master system at time $t+\tau$ where $\tau$ is a time delay. The occurrence of AS depends on the coupling constant $\alpha$. Therefore it is not dependent on isolated dynamics or time delay $\tau$ and regarding to type of desired application higher dimensional chaotic systems can be used for an arbitrary time delay. This anticipation of chaos can be used or is used in applications such as semiconductor lasers with optical feedback, secure communications \cite{masoller2001}.

\begin{figure}[h]
\centerline{\includegraphics[width=0.5\linewidth]{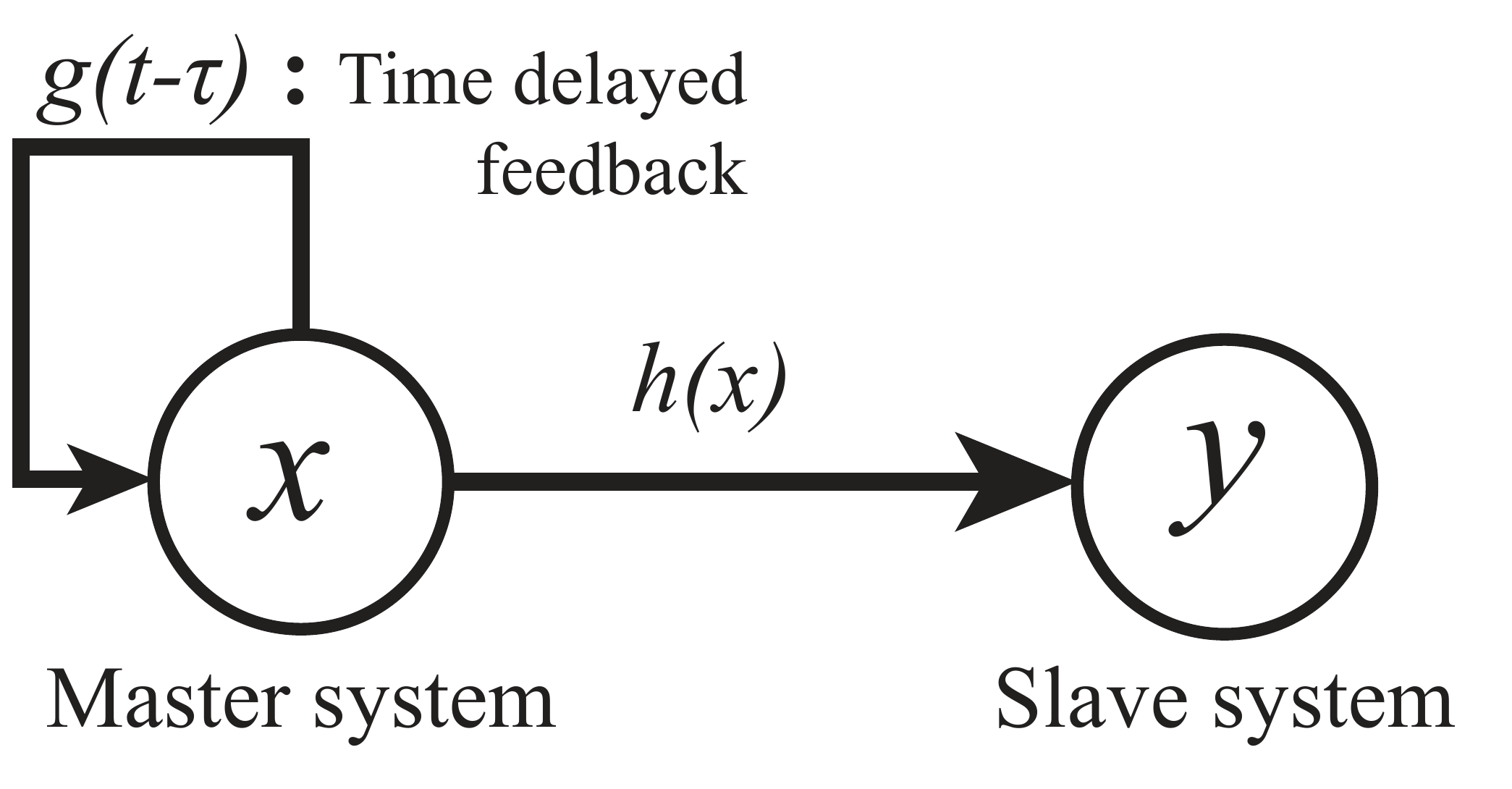}}
\caption{Scheme of anticipating synchronization with memory in the driver systems}
\label{fig:anticipating_x_feedback}
\end{figure}

Consider two chaotic systems in a master-slave interaction and the master has a certain delay $\tau$ feedback Fig.~\ref{fig:anticipating_x_feedback}.
Because of the internal delay feedbacks, it may well happen that the master $\bm{x}$ and slave $\bm{y}$ synchronize but with some time delay
$$
\bm{x}(t) = \bm{y}(t-\tau)
$$
When this happens we have
$$
\bm{y}(t) = x(t + \tau).
$$
Hence, although the system $x$ is fully chaotic, we can precisely predict its future state from the system $y$. In other words the slave system anticipates the master system. This kind of synchronization is called anticipated synchronization (AS). 

{\bf Example:} We consider two coupled Ikeda equations,
\begin{equation}
\begin{array}{rcl}
\dot{x} &=& -\alpha x - \mu\sin x_\tau  \\ 
\dot{y} &=& -\alpha y - \mu\sin x 
\end{array}
\label{eq:anticipating_eq}
\end{equation}
We use the notation $\bm{y}_{\tau} = \bm{y}(t-\tau)$.
The scheme of the system is given in Fig.~\ref{fig:anticipating_x_feedback}. The synchronization error for delayed system is given by,
\begin{equation}
z=x-y_{\tau} \nonumber
\end{equation}
and to show that synchronization is attractive we analyze the first variational equation
\begin{eqnarray}
\dot{z} &=& \dot{x} - \dot{y}_{\tau} \nonumber \\
&=& -\alpha x - \mu \sin x_{\tau} - (-\alpha y_{\tau} - \mu \sin x_{\tau})  \nonumber \\
&=& -\alpha z.
\end{eqnarray}
The solution is  $z(t) = z_0 e^{-\alpha t}$ and 
 for  $\alpha>0$ the synchronization is globally  exponentially stable. 

\begin{figure}[h]
    \centering
    \begin{subfigure}[t]{0.4\linewidth}
        \centering
        \includegraphics[width=\linewidth]{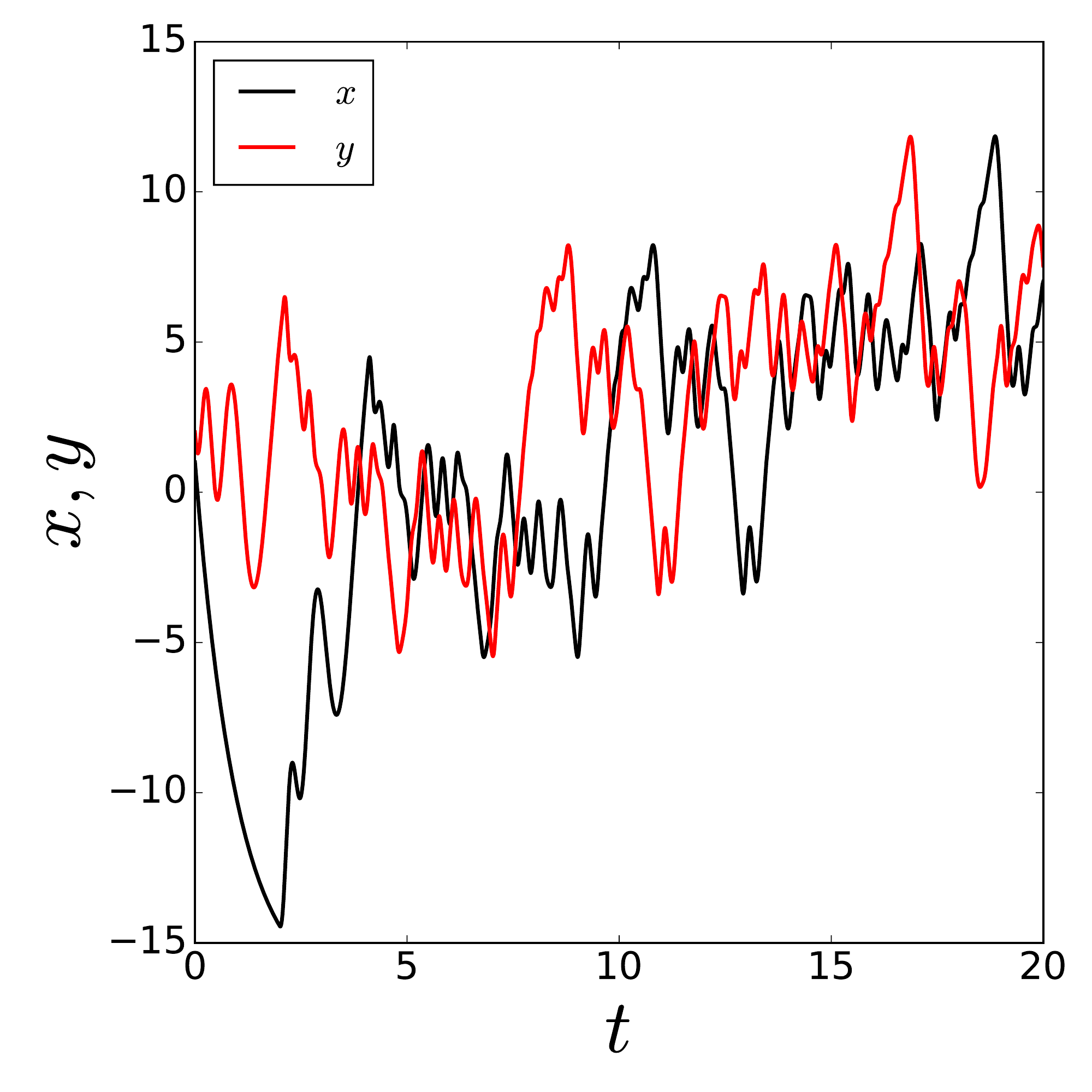}
        \caption{Time series of coupled Ikeda equations.}
    \end{subfigure}%
    ~
    \begin{subfigure}[t]{0.4\textwidth}
        \centering
        \includegraphics[width=\linewidth]{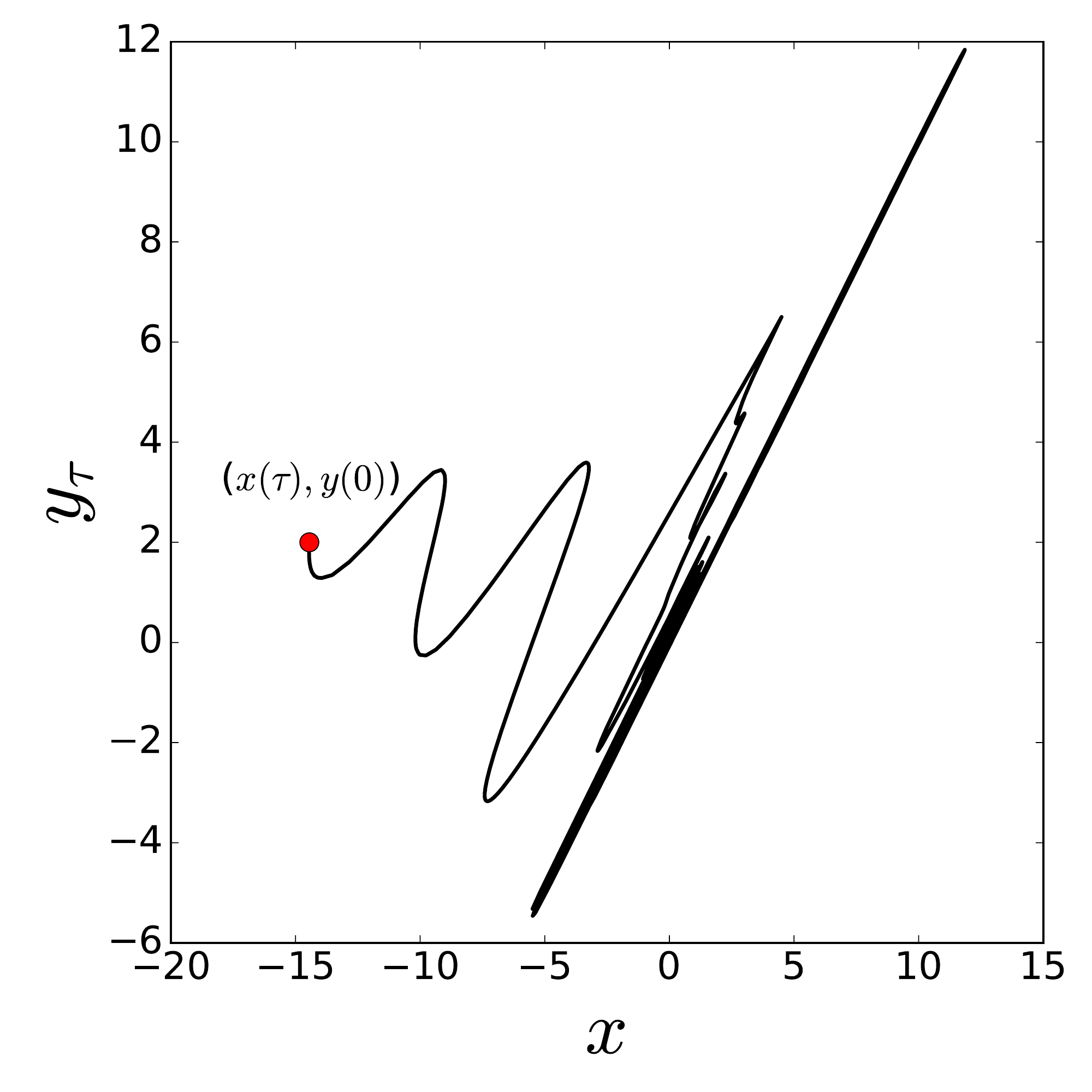}
        \caption{Phase space and synchronization manifold for $x$ and $y_{\tau}$.}
    \end{subfigure}
    \caption{Anticipating chaotic synchronization. In the beginning, the systems are not in harmony. After a transient time both systems are getting into a time-delay $\tau$ synchronization. For this illustration $\tau=2$. (a) time series of the systems (b) phase space and synchronization manifold of the system. The red circle is the initial condition for the trajectory of $(x,y_\tau)$.}
\label{fig:anticipating_sycn}
\end{figure}

To illustrate AS, we simulate Eq.~(\ref{eq:anticipating_eq}) with a fourth order Runge-Kutta integrator for the delay-differential equations for the parameters $\alpha=1$, $\mu=20$ and $\tau=2$. The simulation starts from a random initial condition. After enough transient time $t$, the master $x$ and the slave $y$ systems synchronize with a time delay ($\tau$) Fig.~\ref{fig:anticipating_sycn}(a). The transient time can be observed from the phase space of $x$ and $y_\tau$. The initial condition is given by a red circle in Fig.~\ref{fig:anticipating_sycn}(b), the trajectory converges to the synchronization manifold ($\bm{x}=\bm{y}_{\tau}$).

\begin{figure}[h]
\centerline{\includegraphics[width=0.5\linewidth]{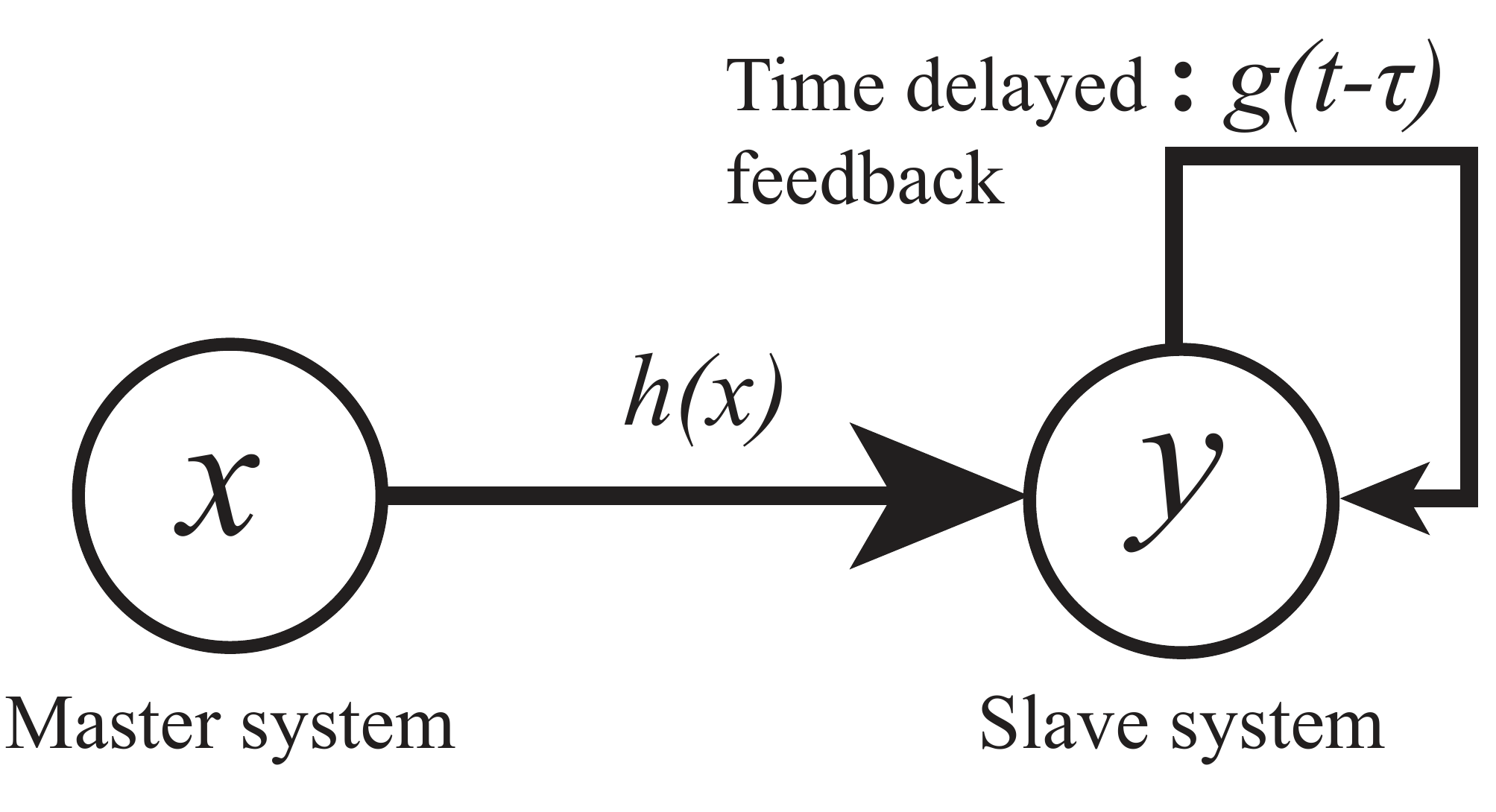}}
\caption{Scheme of anticipating synchronization without memory in the driver systems}
\label{fig:anticipating_y_feedback}
\end{figure}

\begin{figure}[h]
    \captionsetup[subfigure]{justification=centering}
    \centering
    \begin{subfigure}[t]{0.4\linewidth}
        \centering
        \includegraphics[width=\linewidth]{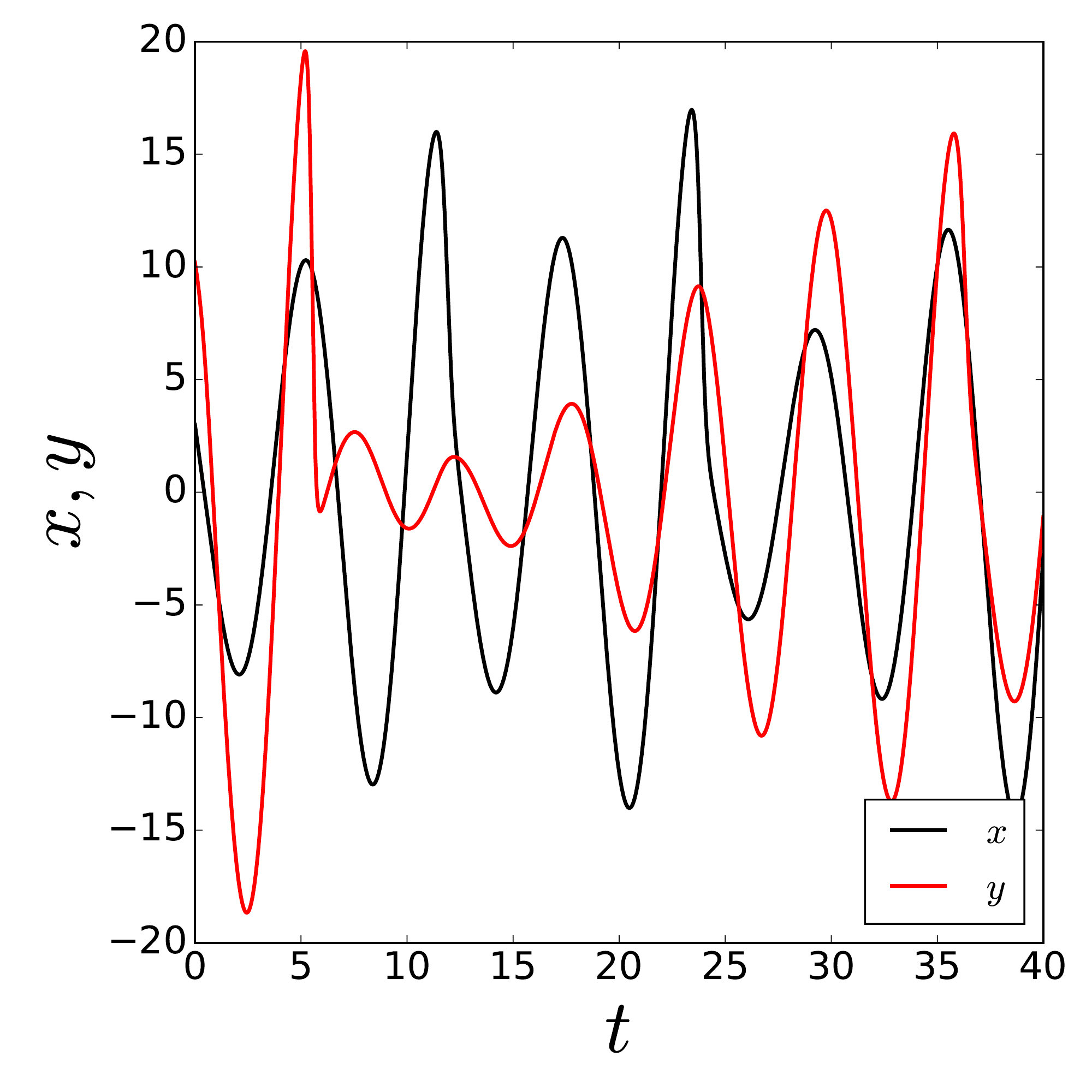}
        \caption{Time series, no synchronization $\alpha < \alpha_c$}
    \end{subfigure}%
    ~
    \begin{subfigure}[t]{0.4\textwidth}
        \centering
        \includegraphics[width=\linewidth]{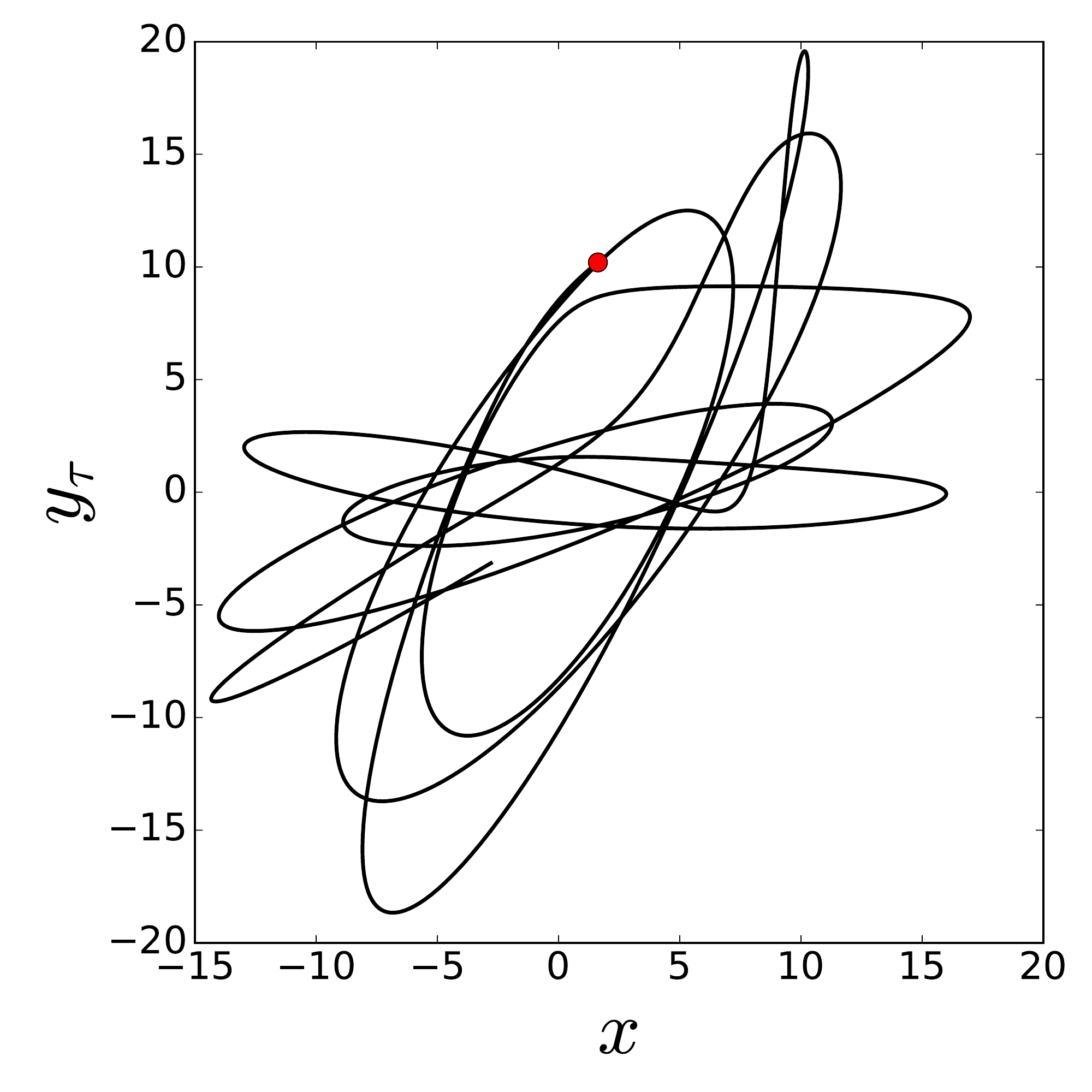}
        \caption{Phase space for $x$ and $y_{\tau}$.}
    \end{subfigure}
     \begin{subfigure}[t]{0.4\linewidth}
        \centering
        \includegraphics[width=\linewidth]{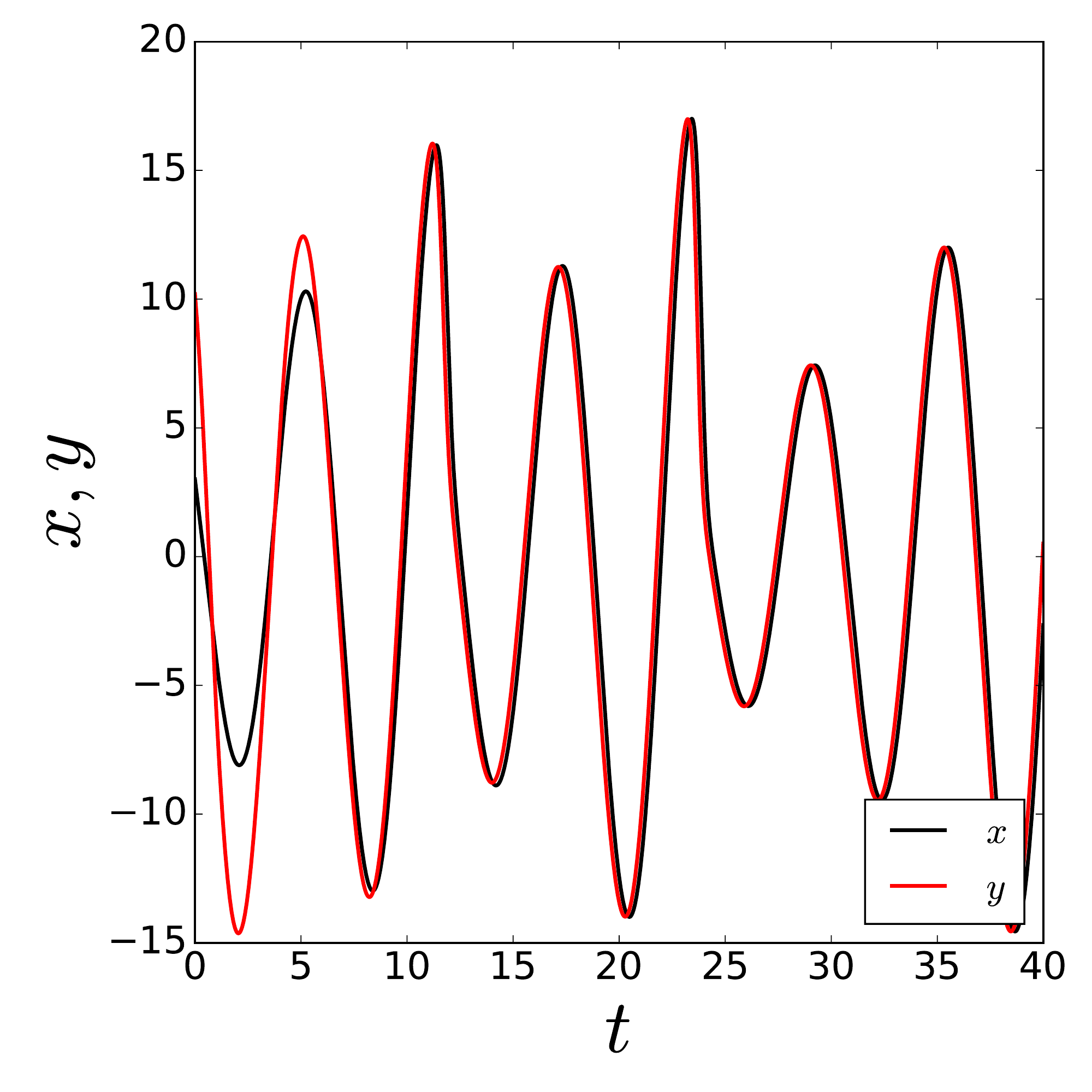}
        \caption{Time series, synchronization $\alpha > \alpha_c$}
    \end{subfigure}%
    ~
    \begin{subfigure}[t]{0.4\textwidth}
        \centering
        \includegraphics[width=\linewidth]{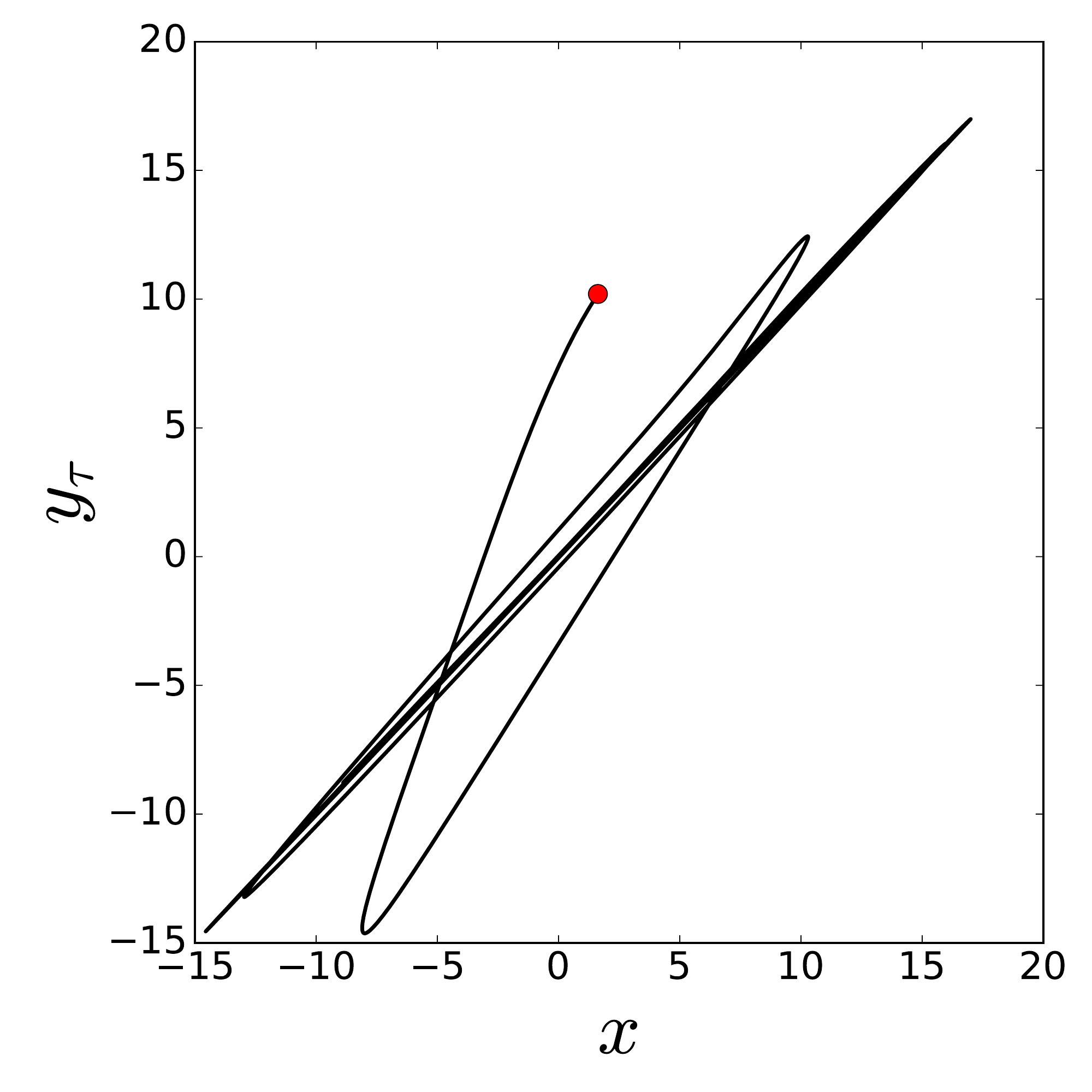}
        \caption{Phase space and synchronization \\ manifold for $x$ and $y_{\tau}$.}
    \end{subfigure}
    \caption{Anticipating synchronization does not obtain for ($\alpha < \alpha_c$), otherwise ($\alpha > \alpha_c$) the AS exhibits. For this illustration the delay is selected as $\tau=0.2$. (a) time series of the systems where $\alpha < \alpha_c$ and (b) phase space. (c) time series of the systems where $\alpha > \alpha_c$ and (d) phase space and synchronization manifold of the system. The red circle is the initial condition for the trajectory of $(x,y_\tau)$.}
\label{fig:anticipating_sims}
\end{figure}

{\bf Example: } The AS can occur for higher dimensional chaotic system. For such cases the critical coupling constant can be positive ($\alpha_c > 0$). The AS can be obtained without delayed state in the master system, that is, without memory in the master system. The scheme of this model is given in Fig.~\ref{fig:anticipating_y_feedback}. 
We can demonstrate this case with R\"ossler system, the governing equations are given by
\begin{eqnarray}
\begin{array}{rcl}
\dot{x}_1 &=& -x_2-x_3  \\
\dot{x}_2 &=& x_1+ax_2  \\
\dot{x}_3 &=& b+x_3(x_1-c)   \\
\dot{y}_1 &=& -y_2-y_3+\alpha(x_1-y_{1,\tau})  \\
\dot{y}_2 &=& y_1+ay_2  \\
\dot{y}_3 &=& b+y_3(y_1-c)  
\end{array}
\label{eq:anticipating_nomemory}
\end{eqnarray}
where the parameters are $a=0.15$, $b=0.2$ and $c=10$. We simulate Eq.~(\ref{eq:anticipating_nomemory}) for no AS ($\alpha < \alpha_c$) Fig.~\ref{fig:anticipating_sims} and AS ($\alpha > \alpha_c$)  Fig.~\ref{fig:anticipating_sims} cases. In this memoryless AS approach, the synchronization is also dependent on time delay $\tau$. 



\section{Synchronization in complex networks}
\label{sec:networks}

Synchronization is commonly found in networks of  natural and mankind-made systems such as  neural dynamics \cite{singer1999,belykh2005,gregoriou2009}, cardiovascular systems \cite{lotric2000, stefanovska2001, shiogai2010}, power grids \cite{motter2013}, superconducting Joseph junctions \cite{wiesenfeld1996}.  The theory we presented in the previous chapters can be generalized to understand certain aspects of synchronization in these complex systems.


We will 
focus on diffusively interacting identical oscillators, so the dynamics of the coupled system reads as
\begin{eqnarray}
\frac{ d \bm{x}_i}{dt} &=& \bm{f}(\bm{x}_i) + \alpha \sum_{j=1}^N A_{ij}  [\bm H(\bm{x}_j) - \bm H(\bm{x}_i)], 
\label{eq:motion_adj} 
\end{eqnarray}
where $\bm{f}:\mathbb{R}^n \rightarrow \mathbb{R}^n$ describes the dynamics of the isolated system, $\alpha$ is the overall coupling strength, $N$ is the number of oscillators, $\bm{H}:\mathbb{R}^n \rightarrow \mathbb{R}^n$ is the coupling function. Finally, $A_{ij}$ dictates who is interacting with whom. 
$A_{ij}=1$ if $i$ receives a connection from $j$ and $0$ otherwise. The matrix $A$ (the dimension is $N\times N$) provides the network linking structure and it is called adjacency matrix. This matrix will play an clear role in the analysis.

The network dynamics of diffusively coupled system in Eq.~(\ref{eq:motion_adj}) models many physical systems. A few specific examples are:


%
%


\begin{description}
\item[Electronic Circuits with Resistive interaction.] Electric circuits, e.g., Chua, Roessler-like, Lorenz-like, can be coupled over their resistors then Eq.\label{eq:motion_adj} models this system \cite{pecora1990}. Another case, only one electric circuit can be driven by an external signal as master-slave system (for details, see Sec.~\ref{sec:master_slave})\cite{baptista2003}.
%

\item[Neuron Networks with Electrical Coupling.] In brain network, $\bm f$ can be the isolated neuron dynamics modelled by  differential equations with chaotic or periodic behaviour and having different time-scales, that is, the isolated can have burst and single regime \cite{hodgkin1952} and $H$ the electrical synapses $H(\bm{x}_1 - \bm{x}_2) = (x_1- x_2, 0,0)$. 

\item[Stable Biological System.] In biological systems when the isolated dynamics has a stable periodic motion then typically one can rephrase the network dynamics in terms of our model. For instance, the heart consists of millions of pacemaker cells. Each cell when isolated has its own rhythm, but when put together these cells interact and behave in unison to deliver the strong electrical pulse that make our heart beat \cite{strogatz2003}. The dynamics of the pacemaker cells are modelled by phase oscillators $\phi_i$ with distinct frequencies $\omega_i$ and the coupling function is a simple sine function $H(\phi_1 - \phi_2) = sin(\phi_1-\phi_2)$\cite{schafer1998,shiogai2010}.

\item[Laser Arrays.]   Lasers can be arranged arrays or complex networks. In this case, one is interested in the electrical field dynamics. Such electrical field is govern by equations with interaction akin to diffusion  Ref.~\cite{murphy2010}. So, the approach presented here can be extrapolated to such lasers under slight changes.
\end{description}

In fact, when we considering periodic oscillators \footnote{Or Roessler type oscillator where the phases are well defined and the coupling between chaotic amplitudes and phases are small} the 
above model is a normal form for the networked system. That is,  the isolated system has a periodic exponentially attracting orbit,  we couple the system, and in the weak coupling regime, the amplitudes will change only slightly but the phases can change by large amounts. So the dynamics can be described only in terms of the phases. The phase description will again fit in our Eq.~(\ref{eq:motion_adj}). 

Our synchronization results given in the previous sections are exponentially stable. In other words, if once the trajectories are into the synchronization subset, the solution is robust and persistent to the perturbations. For $N$ coupled nonidentical systems ($\bm f_1 \ne \bm f_2 \ne \dots \ne \bm f_N$), complete synchronization is not possible. However, because of exponentially stable solutions, highly coherent state around the synchronization subset can be still observed \cite{sun2009, pereira2013}.



\subsection{Interactions in terms of Laplacian}

Because of the diffusive nature of the interaction, it is possible to represent the coupling in terms of the Laplacian matrix $\bm{L}$. Indeed,
\begin{eqnarray}
\sum_{j=1}^N A_{ij}  [\bm{H}(\bm{x}_j) - \bm{H}(\bm{x}_i)] &=&  \sum_{j=1}^N A_{ij} \bm{H}(\bm{x}_j) - \bm{H}(\bm{x}_i) \sum_{j=1}^N A_{ij} \nonumber \\
&=&  \sum_{j=1}^N A_{ij} \bm{H}(\bm{x}_j) - k_i \bm{H}(\bm{x}_i) \nonumber \\
&=&  \sum_{j=1}^N ( A_{ij}  - \delta_{ij} k_i ) \bm{H}(\bm{x}_j) \nonumber 
\end{eqnarray}
where $k_i = \sum_{j=1}^N A_{ij}$ is the degree of the $i$th node, $\delta_{ij}$ is the Kronecker delta, and recalling that $L_{ij} = \delta_{ij} k_i - A_{ij}$ we obtain
\begin{eqnarray}
\frac{ d \bm{x}_i}{dt} &=&\bm{f}(\bm{x}_i) - \alpha \sum_{j=1}^N L_{ij} \bm{H}(\bm{x}_j). 
\label{eq:motion_lap}
\end{eqnarray}
Our results will depend on this representation and on the spectral properties of $\bm{L}$.

Notice that 
$$
\bm{x}_1(t)=\bm{x}_2(t)= \cdots = \bm{x}_N(t) = \bm{s}(t), 
$$ 
is an invariant state for all coupling strength $\alpha$, because the laplacian is zero row sum.  When $\alpha=0$ the oscillators are decoupled, and  Eq. (\ref{eq:motion_lap}) describes $N$ copies of the same oscillator with distinct initial conditions.  Since, the  chaotic behavior leads to a divergence of nearby trajectories, without coupling, any small perturbation on the globally synchronized motion will grow exponentially fast, and lead to distinct behavior between the node dynamics.
We wish to address the local stability of the globally synchronized state. That is, if all trajectories start close  together would become synchronized 
$$
\lim_{t \rightarrow \infty}\|\bm{x}_i(t) - \bm{x}_j(t) \| = 0
$$
The goal of the remaining exposition is to answer  this question. 
%
%
Before, we continue with the analysis, we will briefly review some examples and constructions of graphs and discuss the relevant aspects
necessary to tackle for problem.

\subsection{Relation to other types of Synchronization}

We will focus on the transition to complete synchronization, which is mainly related to Sec. \ref{sec:complete_sync}. This is no severe restriction as in certain scenarios other types of synchronization can often be formulated in terms of our model.   \\

\noindent
{\it Extension to Phase Synchronization.} As we discussed in the introduction of Sec.~\ref{sec:networks}, if the dynamics of $\bm f$ is periodic then we can introduce a phase variable which will follow our main system of equations Eq.~(\ref{eq:motion_adj}) as the phase reduction tells us that generically the interaction between phases are diffusive.  Moreover, because our results will yield robust transition to synchronization, if the oscillators are nearly identical the phase synchronization will persist. \\

\noindent
{\it Extension to Generalized Synchronization.} Roughy speaking a form of generalized synchronization in networks is the so-called pinning control, where one tries to control the behaviour of synchronized trajectories by driving the network with external nodes. 
One extends the network to include the driver node.  Therefore, the theory necessary to tackle this problem is  the same as presented here. The main question is how to connect the driver nodes in such a way that control is effective.


%
%

\subsection{Complex Networks}
\label{sec:adj_lap_mat}
%
%
%
%
%
%
%

\begin{figure}[h]
\centerline{\includegraphics[width=0.6\linewidth]{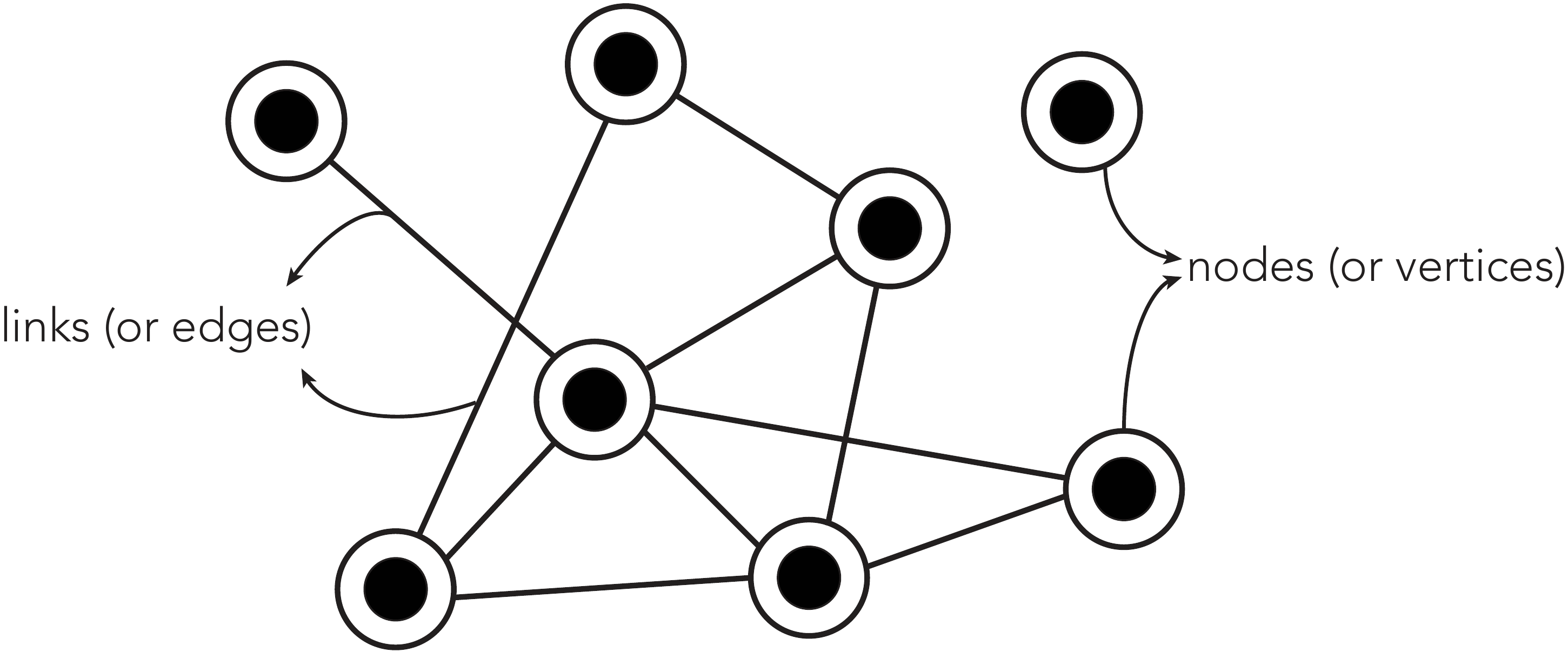}}
\caption{A network visualisation with eight nodes and ten links.}
\label{fig:network}
\end{figure}

A \textit{network}, also called graph $G$ in mathematical literature, is a set of $N$ elements, called \textit{nodes} (or vertices), connected by a set of $M$ \textit{links} (or edges) Fig.~\ref{fig:network}. Networks represent interacting  elements and are all around from biological systems, e.g. swarm of fireflies, food webs or brain networks, to mankind made systems, e.g. the world wide web, power grids, transportation networks or social networks.

\begin{figure}[h]
\centerline{\includegraphics[width=1.0\linewidth]{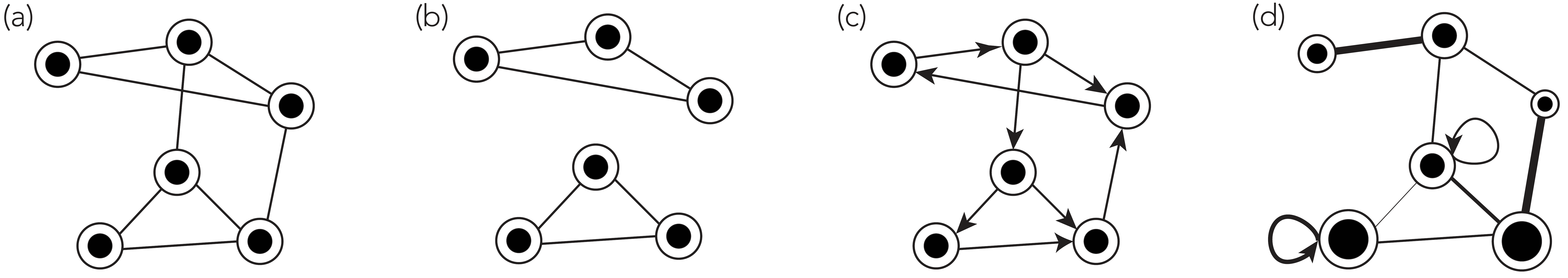}}
\caption{Visualization of network types: (a) an unweighted simple network, (b) a disconnected network, (c) a directed network, (d) a weighted network with self-loops.}
\label{fig:network_types}
\end{figure}

\begin{figure}[h]
\centerline{\includegraphics[width=1.0\linewidth]{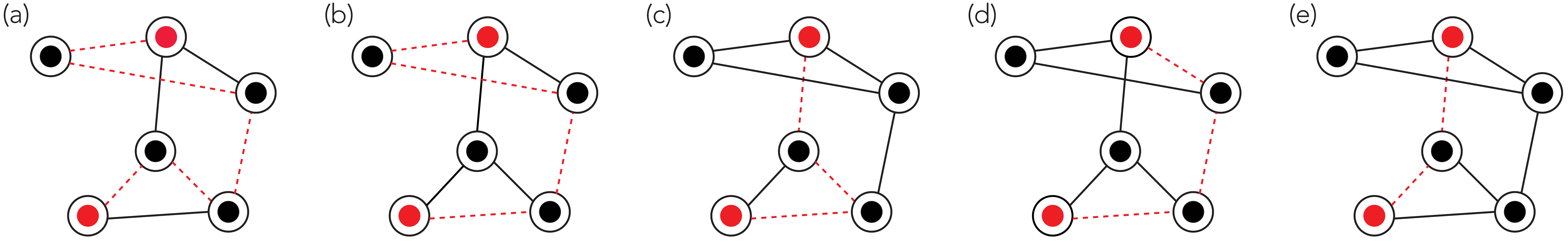}}
\caption{Visualization of paths (red dashed) between two selected nodes (red) in a network: path length of (a) is five, (b) is four, (c) and (d) are three and (e) is two. Therefore the shortest path length for these two red nodes is two.}
\label{fig:paths}
\end{figure}

\begin{description}
\item[{\it A network is called  simple}] if the nodes do not have self-loops (i.e., nodes have connections to themselfs). 
An illustration of a simple network is found in Fig.~\ref{fig:network_types}(a). A nonsimple network, therefore, containing connections is depicted in Fig.~\ref{fig:network_types}(d). We need a bit of technology from graph theory to make sense of our networks. A few basic notions are as follows

\item[{\it A network is undirected}] if  there is no distinction between the two nodes associated with each link Fig.~\ref{fig:network_types}(a). 

\item[\it{ A path in a network}] is a route (sequence of edges) between connected nodes without repeating i.e. a path can visit a node only once. The length of a path is the number of links in the path. See further details in Ref.~\cite{bollobas1998,newman2010}. In Fig.~\ref{fig:paths} we illustrate the paths of two selected (red) nodes in a network. Between two red nodes there are five different paths and each path is given in subplots of Fig.~\ref{fig:paths}. The length of paths are five for Fig.~\ref{fig:paths}(a), four for Fig.~\ref{fig:paths}(b), three for Fig.~\ref{fig:paths}(c) and (d), and two for Fig.~\ref{fig:paths}(e). Therefore the shortest path length, also called \textit{geodesic path}, between these two red nodes is illustrated in Fig.~\ref{fig:paths}(e).

\item[{\it The network diameter $d$}] is the longest length of the shortest path between all possible pairs of nodes. In order to compute the diameter of a graph, first we find the shortest path between each pair of nodes. The longest length of all these geodesic paths is the diameter of the graph. If there is an isolated node (a node without any connections) or disconnected network components, then the diameter of the network is infinite. A network of finite diameter is called {\bf connected} (Fig.~\ref{fig:network_types}(a)), otherwise {\bf disconnected} (Fig.~\ref{fig:network_types}(b)).

\item[{\it A network is directed}] if the links transmit the information towards only associated direction Fig.~\ref{fig:network_types}(c). If the graph is directed then there are two connected nodes say, $u$ and $v$, such that $u$ reachable from $v$,  but $v$ is not reachable from $u$.  See Fig.~\ref{fig:network_types}(c) for an illustration. 

\item[{\it A network is weighted}] if links have different importance from each other or the links may be carry different amount of information. Such graphs are called \textit{weighted} networks Fig.~\ref{fig:network_types}(d). Moreover, a network may have self-loops, that is, a node can affect itself as well Fig.~\ref{fig:network_types}(d).

\end{description}

A graph can be described by its \textit{adjacency matrix} $\bm{A}$ with $N\times N$ elements $A_{ij}$. The adjacency matrix $\bm{A}$ encodes the topological information, and is defined as $A_{ij}=1$ if $i$ and $j$ are connected, otherwise $0$. Therefore, the adjacency matrix of an undirected network is symmetric, $A_{ij} = A_{ji}$. The {\it degree} $k_i$ of the  $i$th node is the number of edges belongs to the node, defined as
$$
k_i = \sum_j A_{ij}.
$$ 
The \textit{Laplacian matrix} $\bm{L}$ is another way to represent the network, defined as
\begin{equation}
L_{ij} = \left\{
\begin{array}{ll}
k_i  &\mbox {if } i=j \\
-1 &\mbox{if } i \mbox{ and } j  \mbox{ are connected} \\
0 &\mbox{otherwise }.
\end{array}
\right.
\label{eq:laplacian_matrix} 
\end{equation}
There is a direct relationship between the Laplacian $\bm{L}$ and the adjacency matrix $\bm{A}$. In a compact form  it reads 
$$
L_{ij} = \delta_{ij}k_i-A_{ij}
$$
where $\delta_{ij}$ is the Kronecker delta, which is 1 if $i=j$ and 0 otherwise. We demonstrate some example network sketches with their adjacency $\bm{A}$ and Laplacian $\bm{L}$ matrices in Fig.~\ref{fig:matrices}.

\begin{table}[h!]
	\centering
	\begin{tabular}{ccc}
	\begin{minipage}{0.2\textwidth}
		\includegraphics[width=1.\linewidth]{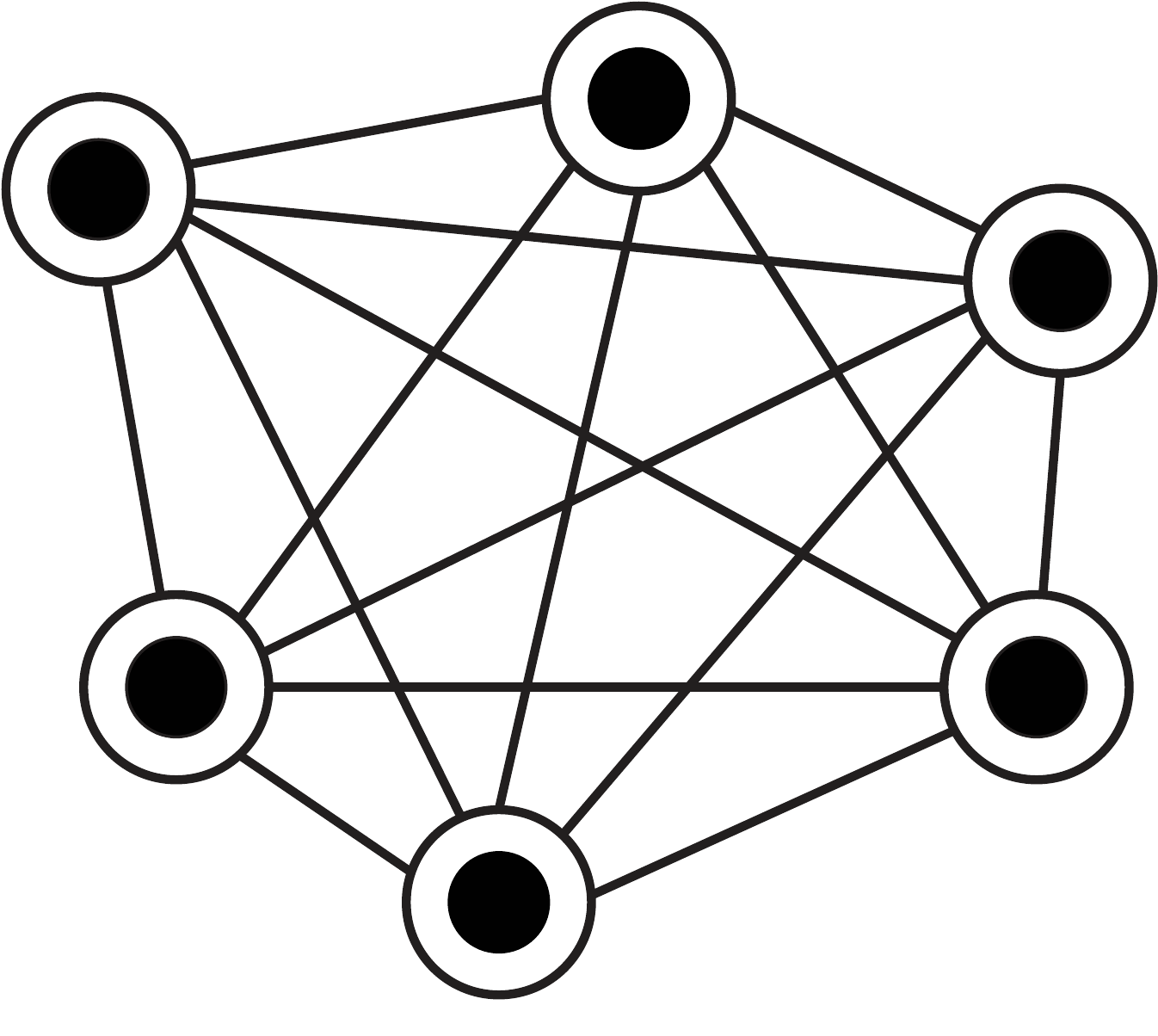}
		\captionof*{figure}{complete graph}
	\end{minipage}
	&
	 $\bm{A} = \left(\begin{array}{cccccc}0 & 1 & 1 & 1 & 1 & 1 \\1 & 0 & 1 & 1 & 1 & 1 \\1 & 1 & 0 & 1 & 1 & 1 \\1 & 1 & 1 & 0 & 1 & 1 \\1 & 1 & 1 & 1 & 0 & 1 \\1 & 1 & 1 & 1 & 1 & 0\end{array}\right) $
	 & 
	 $\bm{L} = \left(\begin{array}{cccccc}5 & -1 & -1 & -1 & -1 & -1 \\-1 & 5 & -1 & -1 & -1 & -1 \\-1 & -1 & 5 & -1 & -1 & -1 \\-1 & -1 & -1 & 5 & -1 & -1 \\-1 & -1 & -1 & -1 & 5 & -1 \\-1 & -1 & -1 & -1 & -1 & 5\end{array}\right)$ \\ \\ \\ 
	 \begin{minipage}{0.2\textwidth}
	\includegraphics[width=1.\linewidth]{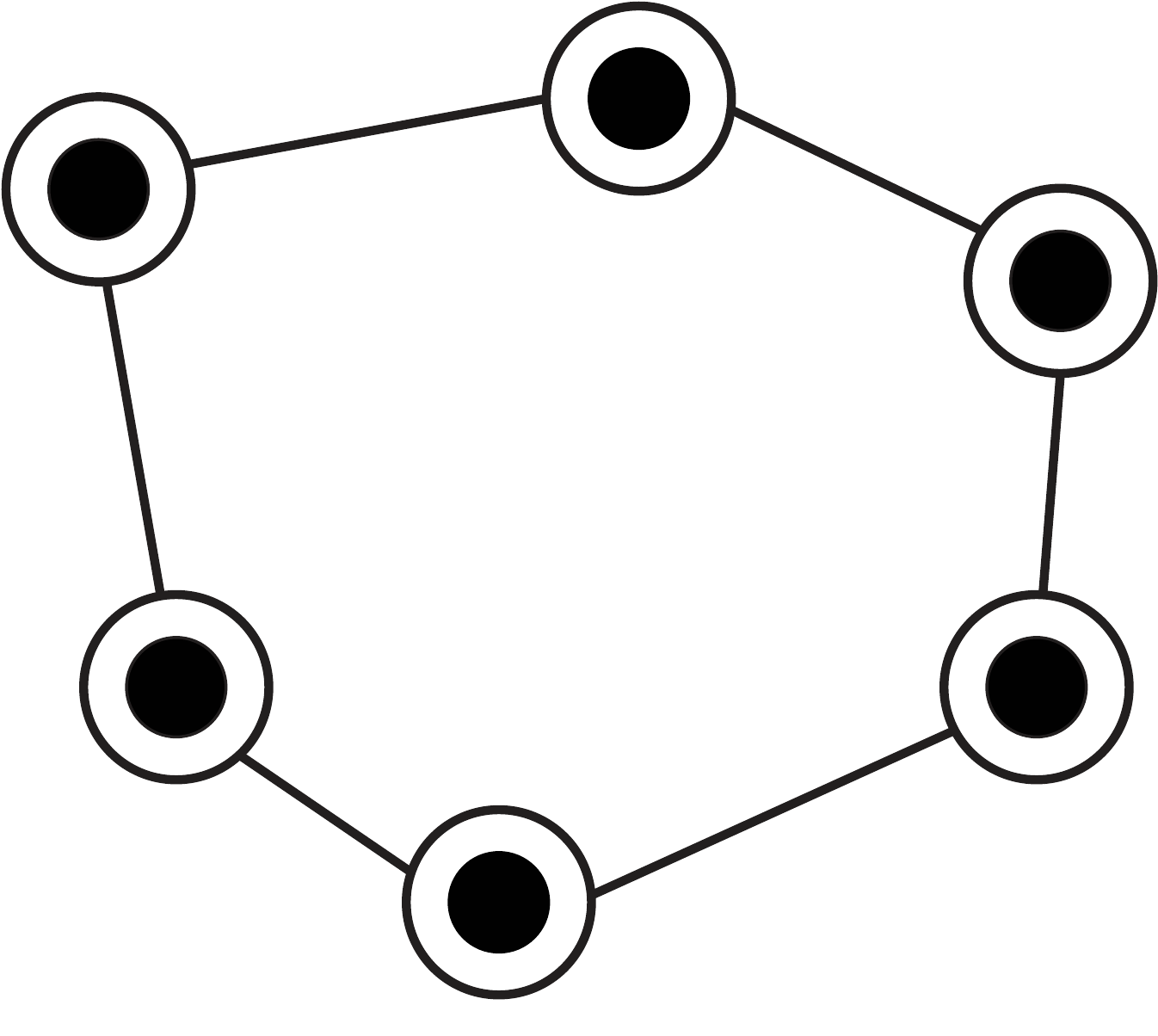}
	\captionof*{figure}{ring (cycle) graph}
	\end{minipage}
	&
	 $\bm{A} = \left(\begin{array}{cccccc}0 & 1 & 0 & 0 & 0 & 1 \\1 & 0 & 1 & 0 & 0 & 0 \\0 & 1 & 0 & 1 & 0 & 0 \\0 & 0 & 1 & 0 & 1 & 0 \\0 & 0 & 0 & 1 & 0 & 1 \\1 & 0 & 0 & 0 & 1 & 0\end{array}\right) $
	 & 
	 $\bm{L} = \left(\begin{array}{cccccc}2 & -1 & 0 & 0 & 0 & -1 \\-1 & 2 & -1 & 0 & 0 & 0 \\0 & -1 & 2 & -1 & 0 & 0 \\0 & 0 & -1 & 2 & -1 & 0 \\0 & 0 & 0 & -1 & 2 & -1 \\-1 & 0 & 0 & 0 & -1 & 2\end{array}\right)$ \\ \\ \\ 
	 \begin{minipage}{0.2\textwidth}
	\includegraphics[width=1.\linewidth]{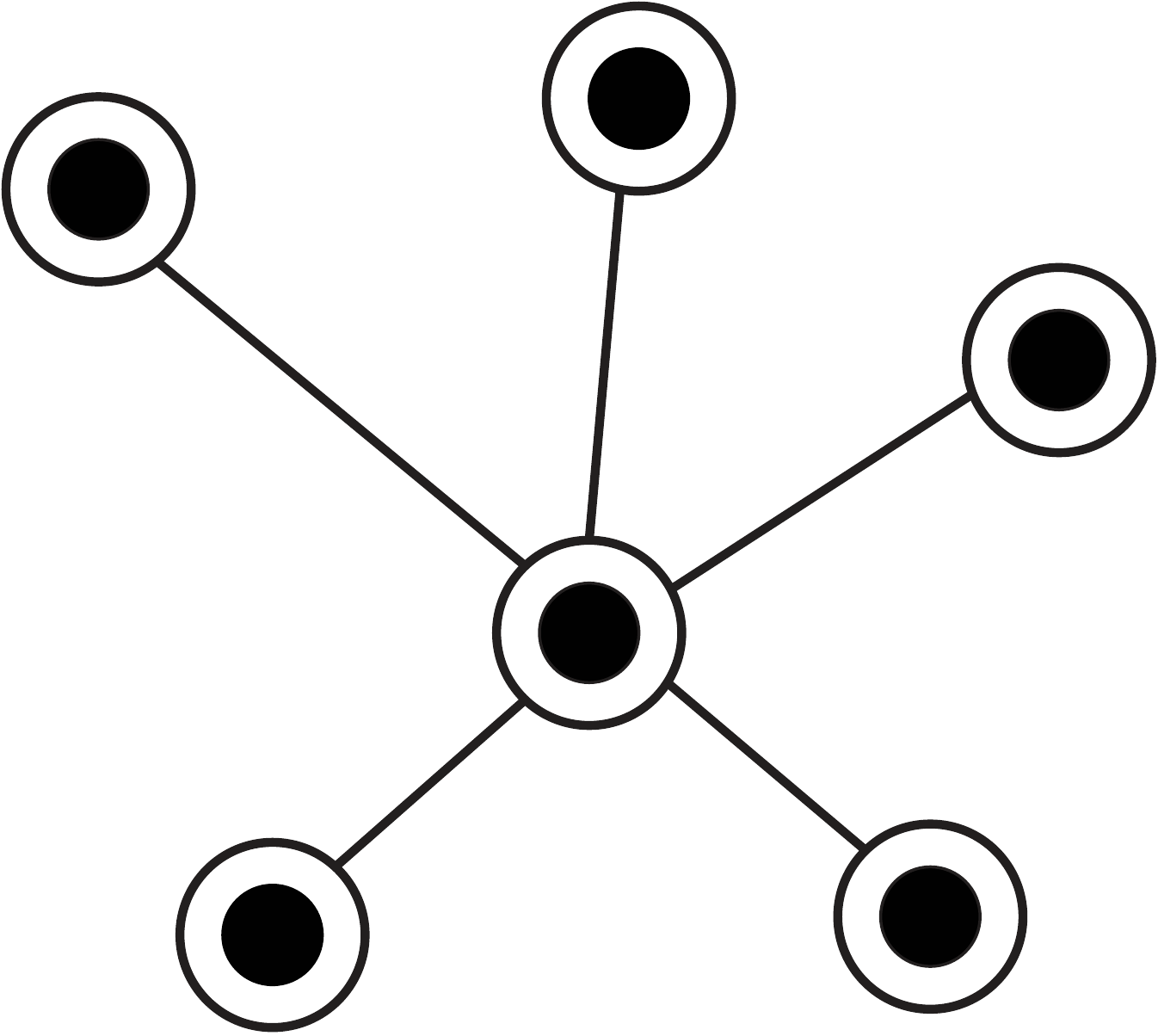}
	\captionof*{figure}{star graph}
	\end{minipage}
	&
	 $\bm{A} = \left(\begin{array}{cccccc}0 & 1 & 1 & 1 & 1 & 1 \\1 & 0 & 0 & 0 & 0 & 0 \\1 & 0 & 0 & 0 & 0 & 0 \\1 & 0 & 0 & 0 & 0 & 0 \\1 & 0 & 0 & 0 & 0 & 0 \\1 & 0 & 0 & 0 & 0 & 0\end{array}\right) $
	 & 
	 $\bm{L} = \left(\begin{array}{cccccc}5 & -1 & -1 & -1 & -1 & -1 \\-1 & 1 & 0 & 0 & 0 & 0 \\-1 & 0 & 1 & 0 & 0 & 0 \\-1 & 0 & 0 & 1 & 0 & 0 \\-1 & 0 & 0 & 0 & 1 & 0 \\-1 & 0 & 0 & 0 & 0 & 1\end{array}\right)$ \\
\end{tabular}
\captionof{figure}{Various network examples with six nodes. Their adjacency $\bm{A}$ and Laplacian  $\bm{L}$ matrices.}
\label{fig:matrices}
\end{table}

The networks we encounter in real applications have a wilder connection structure. Typical examples are cortical networks,  the Internet, power grids and metabolic networks \cite{newman2010, newman2003}. These networks don't have a regular  structure of connections such as the ones presented in Fig. \ref{fig:matrices}.  We say that the network is {\it complex} if it does not possess a regular connectivity structure. 

One of the goals is the understand the relation between the topological organization of the network and its relation functioning such as its collective motion. 

\begin{description}

\item[2$k$ Regular Graph] is a standard graph model where each node has $2k$ links then the total number of links is $M=kN$ where $N$ is total number of nodes Fig.~\ref{fig:small_world}~(a). This model is rather important one since the connections of spatiotemporal graphs, in general, connected to the nearest neighbours. $2k$ regular graph is an alternative representation of such models. It is important to mention that the graph model is fixed with given $k$ and $N$ therefore all properties of the graph is known analytically. 

\item[Erd\"os-R\'enyi] network is generated by setting an edge between each pair of nodes  with equal probability $p$, independently of the other edges Fig.~\ref{fig:complex_networks}~(a). If $ p \gg \ln N / N$, then a the  network is almost surely connected, that is, as $N$ tends to infinity, the probability  that a graph on $n$ vertices is connected tends to $1$. The degree is pretty  homogeneous, almost surely  every node has the same expected degree \cite{chung2006}.

\item[Small World] network is a random graph model which possesses the small-world properties; i.e the average path length is short and clustering is large. The network is generated from a $2k$ regular graph, each link of the graph is rewired with a probability $p$. Therefore if $p=0$ then the there is no rewiring and the graph is $2k$ regular. For $p=1$ then each link is rewired i.e the graph is approaching to Erd\"os-R\'enyi network with $p=\frac{kN}{2 \binom {N} {2}}$. The small-world properties come true between $0<p<1$ Fig.~\ref{fig:small_world}. In many real world networks, the properties of small-world topology can be obtained. 

\begin{figure}[h]
    \centering
    \begin{subfigure}[t]{0.3\linewidth}
        \centering
        \includegraphics[width=\linewidth]{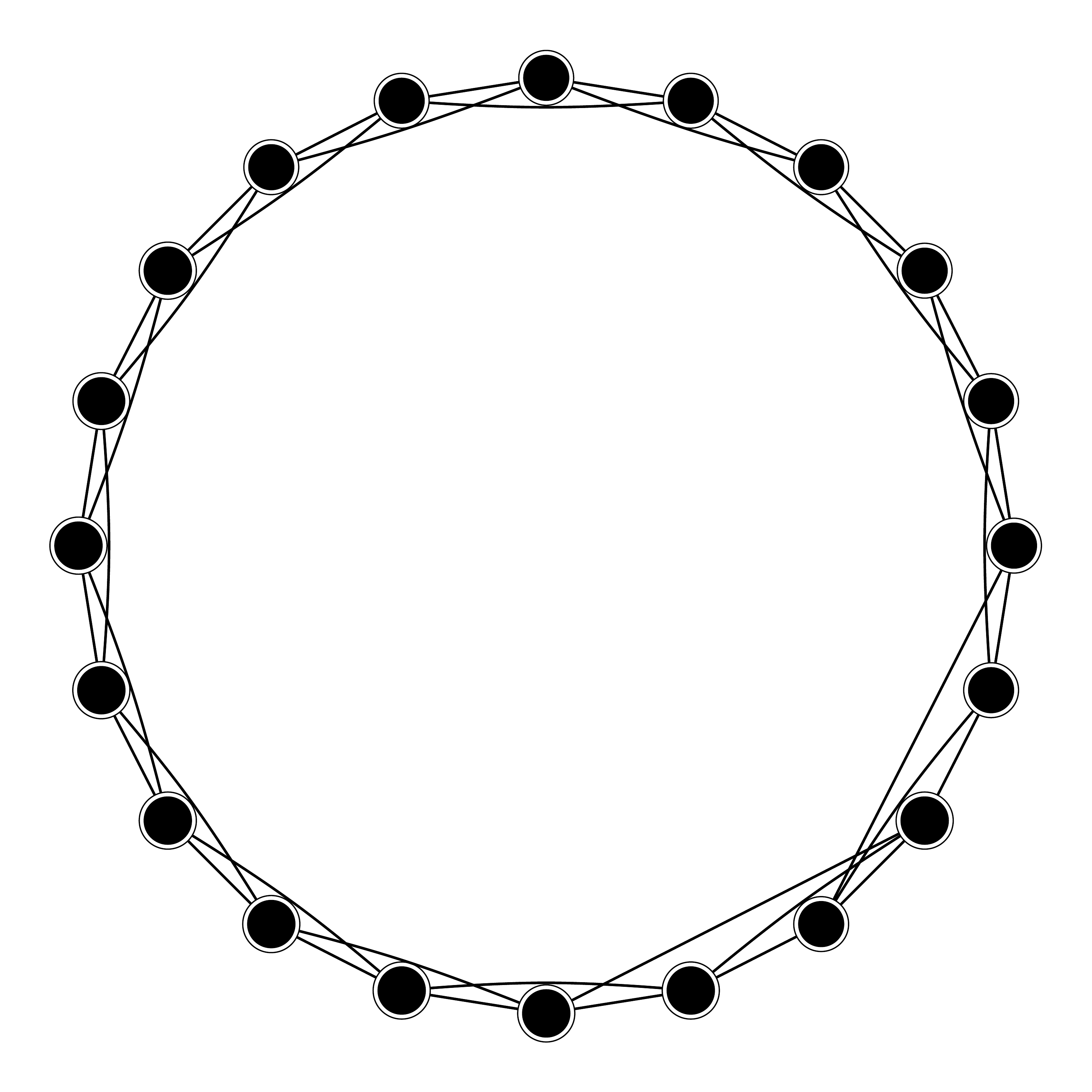}
        \caption{2k-regular graph}
    \end{subfigure}%
    ~ 
    \begin{subfigure}[t]{0.3\textwidth}
        \centering
        \includegraphics[width=\linewidth]{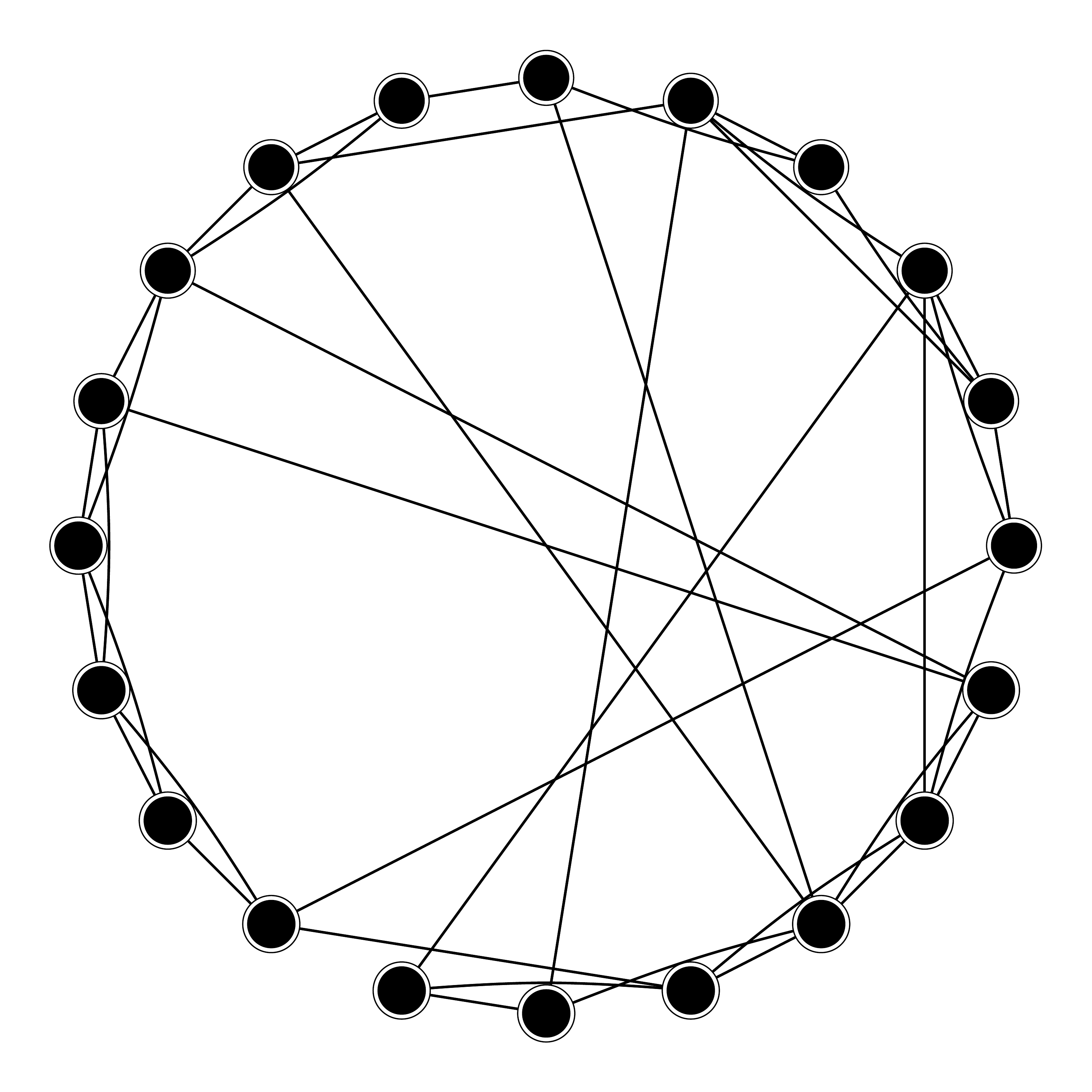}
        \caption{Small-World}
    \end{subfigure}
    ~
       \begin{subfigure}[t]{0.3\textwidth}
        \centering
        \includegraphics[width=\linewidth]{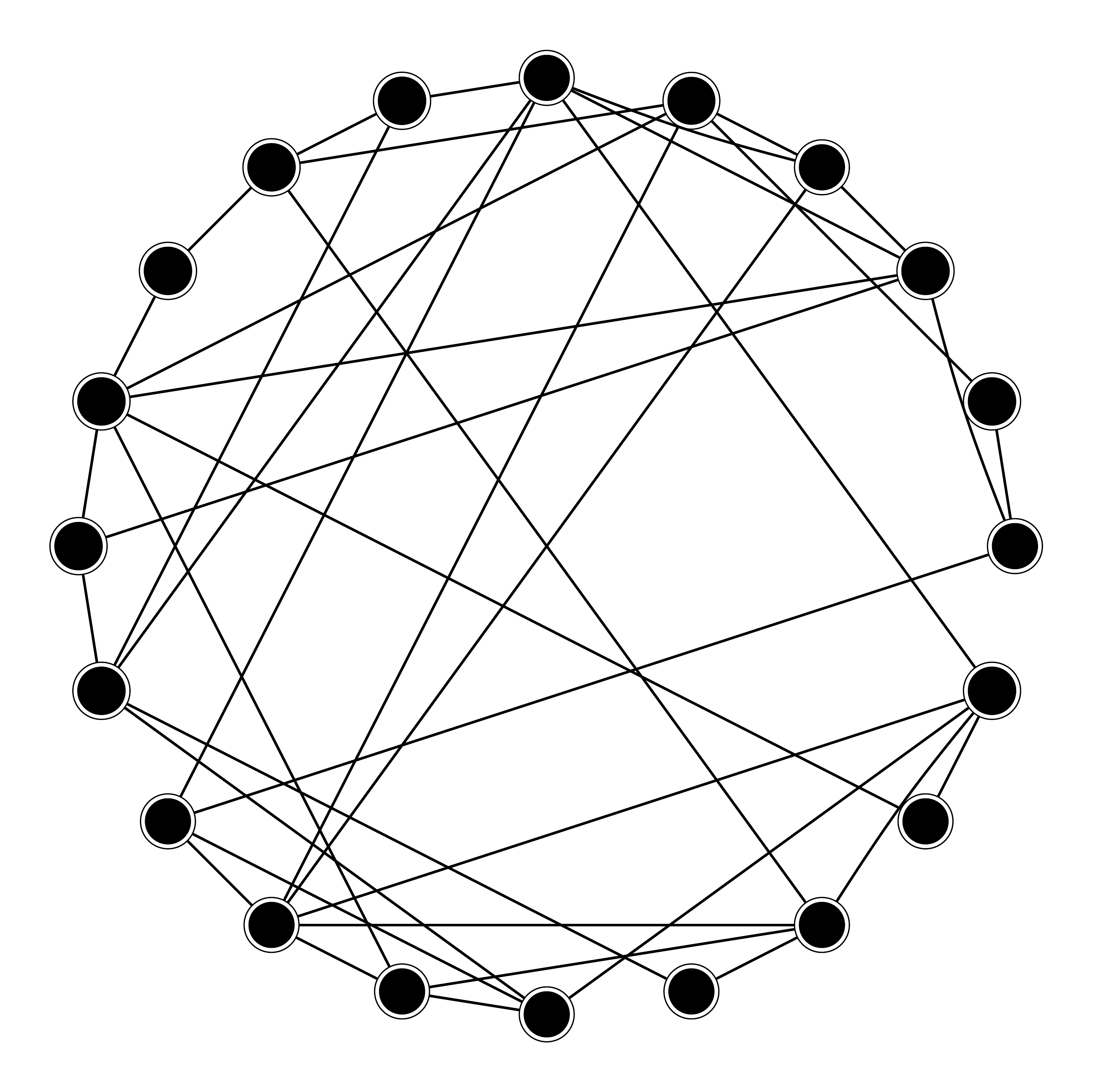}
        \caption{Random (Erd\"os-Renyi)}
    \end{subfigure}
    \caption{Watts - Strogatz random network approach}
\label{fig:small_world}
\end{figure}

\item[The Barabasi-Albert] network possesses a great deal of heterogeneity in the node's  degree, while most nodes have only a few connections, some nodes, termed hubs,  have many connections Fig.~\ref{fig:complex_networks}~(a). These networks do not arise by chance alone.  The network is generated by means of the cumulative advantage principle -- the rich gets  richer. According to this process, a node with many links will have a higher probability  to establish new connections than a regular node.  The number of nodes of degree $k$ is  proportional to $k^{-\beta}$. These networks  are called  scale-free networks \cite{newman2010,newman2003}. Many graphs arising in various real world networks display similar structure as the  Barabasi-Albert network \cite{barabasi1999,albert2000,albert2002}. 

\begin{figure}[h!]
    \centering
        \begin{subfigure}[t]{0.3\textwidth}
        \centering
        \includegraphics[width=\linewidth]{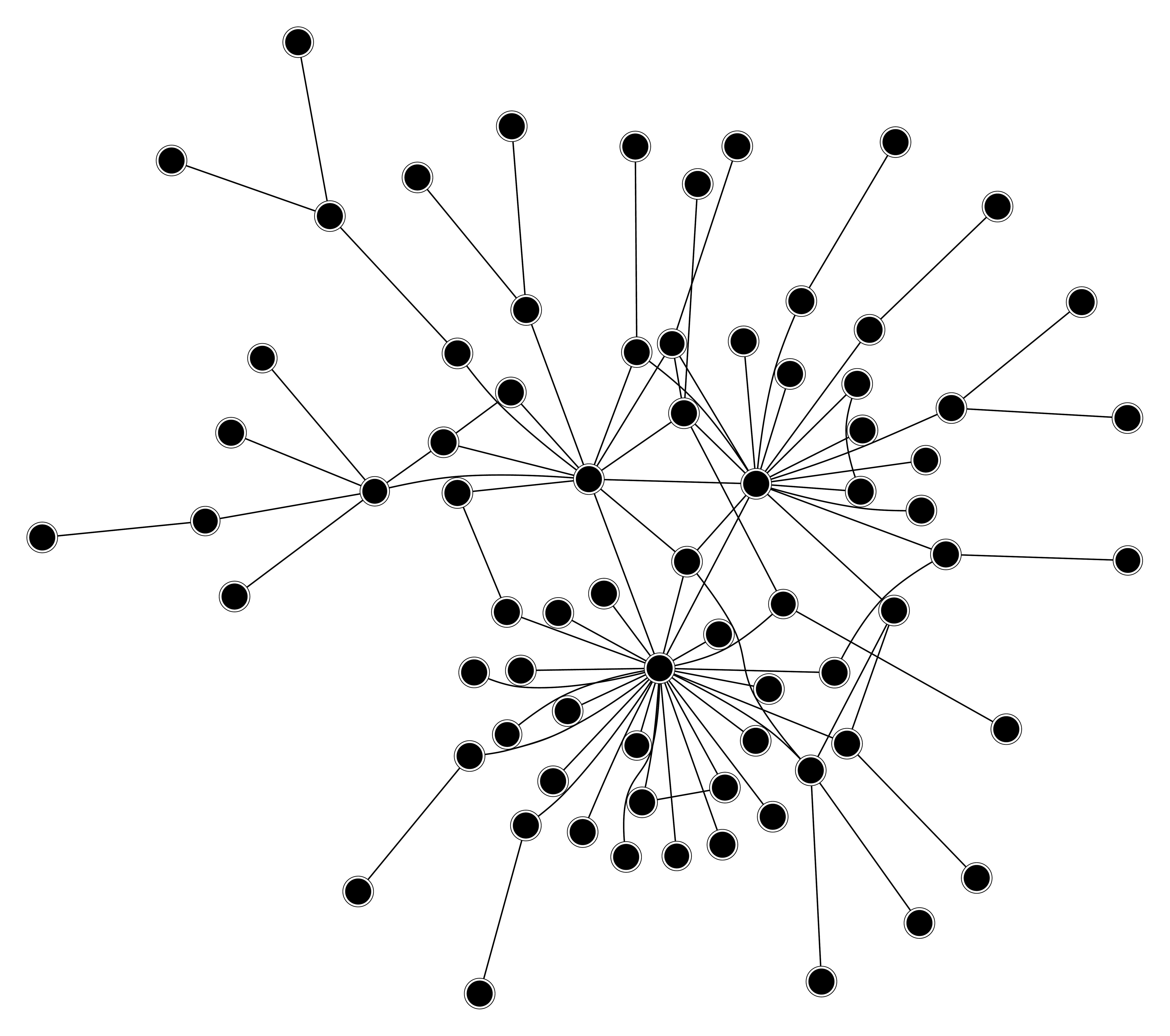}
        \caption{Barabasi-Albert network}
    \end{subfigure}
    ~
    \begin{subfigure}[t]{0.3\linewidth}
        \centering
        \includegraphics[width=\linewidth]{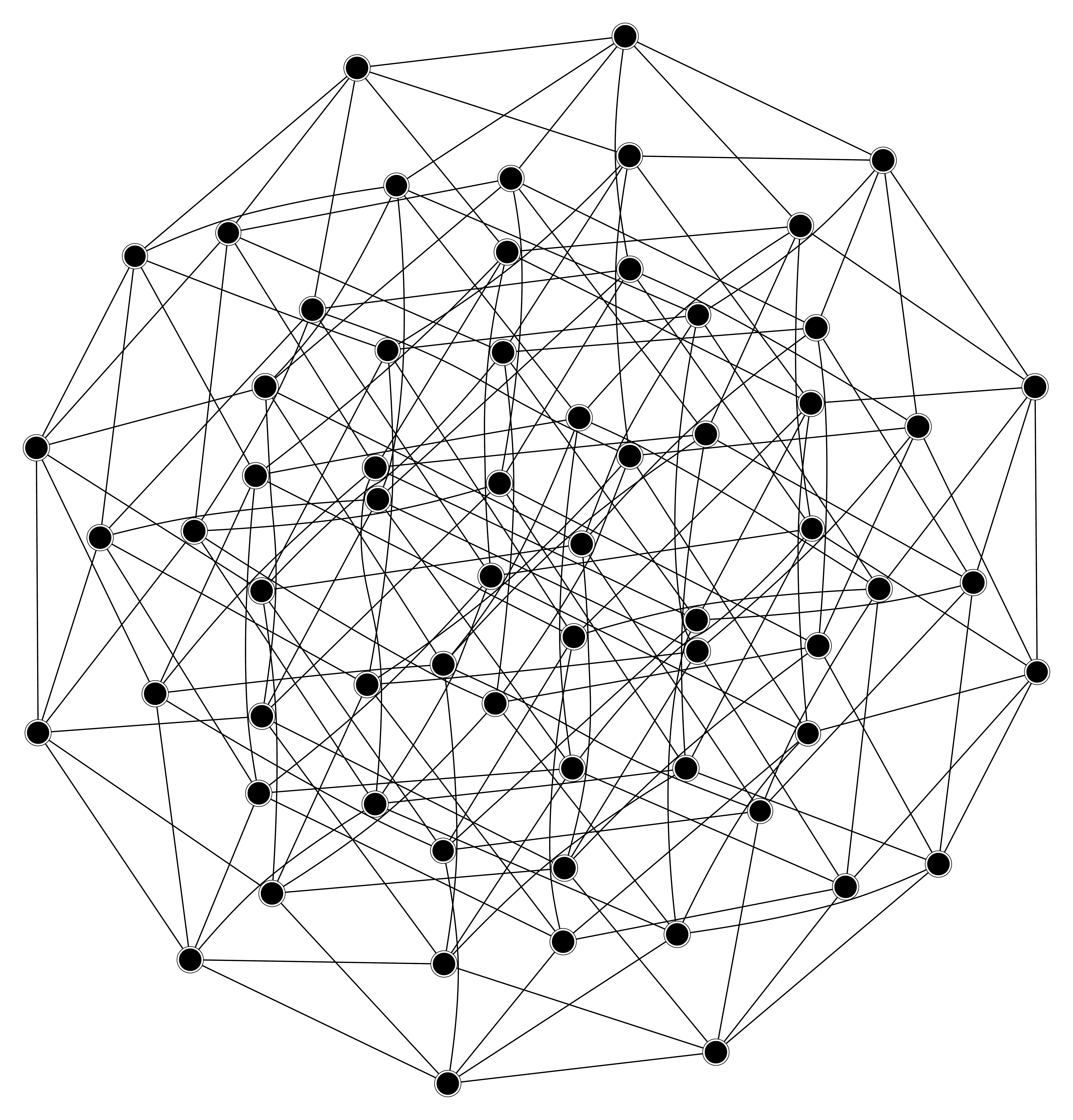}
        \caption{Hypercube network}
        \label{fig:hypercube}
    \end{subfigure}%
    \caption{Some examples of complex networks.}
\label{fig:complex_networks}
\end{figure}

\item[Hypercube graph] $Q_m$ is a $m$-dimensional hypercube formed regular graph (Fig.~\ref{fig:complex_networks}~(b)). It is a regular graph since each node has $m$ neighbours and total number of nodes is $2^m$ and edges is $2^{m-1}$.


\end{description}

Random networks serve as a proxy to many applications as well as a surrogate. There are many nice ways to construct random network 

\begin{description}
\item[Configuration Model] is a random network model created by the degree distribution $P(k)$. If the degree distribution of a graph is known, then the nodes with associated number of links are known however the connection structure between nodes is unknown. The nodes can be drawn with their stubs (half links) Fig.~\ref{fig:configuration_model}~(a), then randomly these stubs can be linked and two stubs create a proper link Fig.~\ref{fig:configuration_model}~(b). This process is a random matching, obviously different network structures can arise from this random process.  

\begin{figure}[h!]
    \centering
    \begin{subfigure}[t]{0.3\linewidth}
        \centering
        \includegraphics[width=\linewidth]{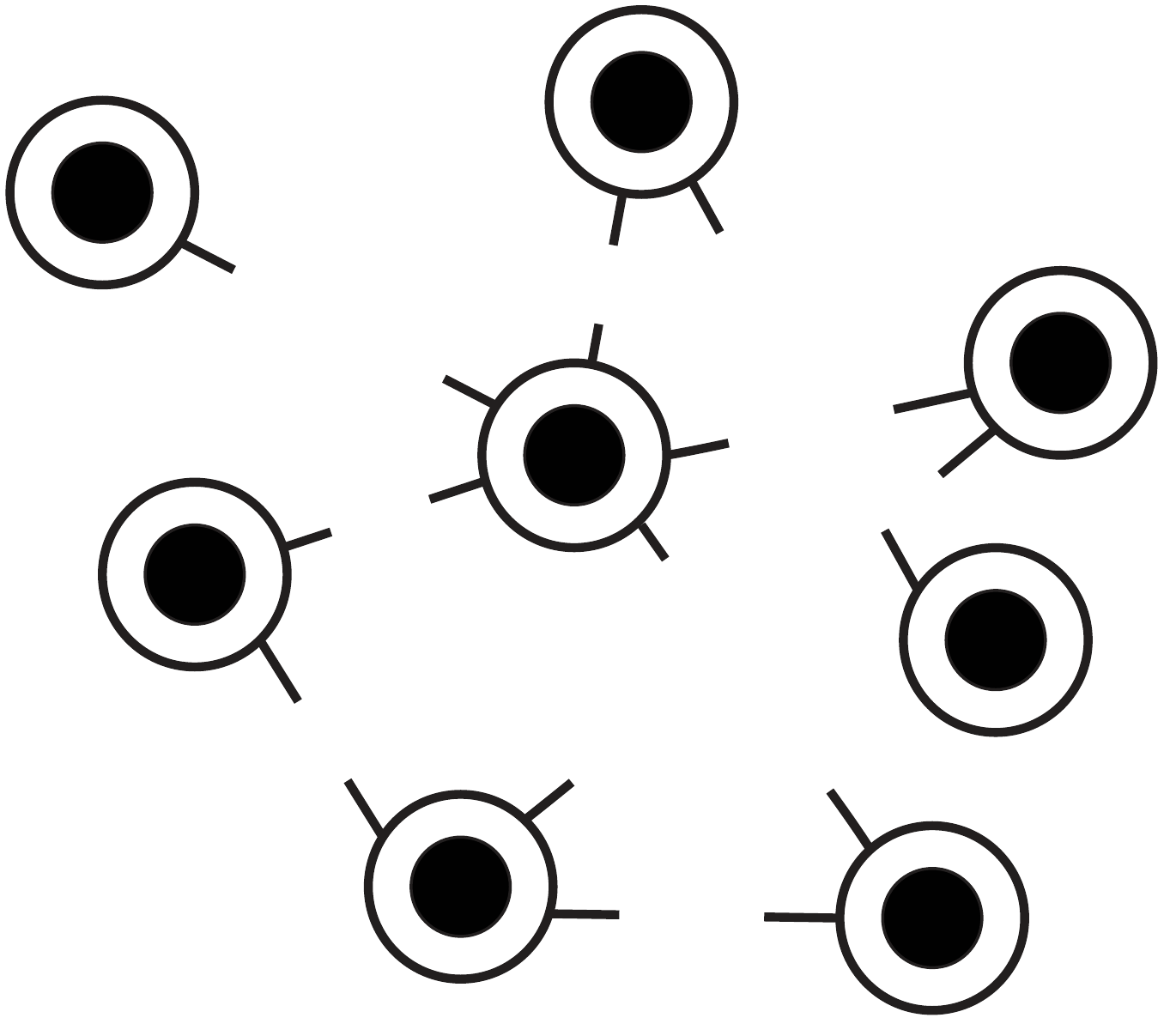}
        \caption{stubs with associated nodes}
    \end{subfigure}%
    ~ ~ ~~~~~~~~
    \begin{subfigure}[t]{0.3\textwidth}
        \centering
        \includegraphics[width=\linewidth]{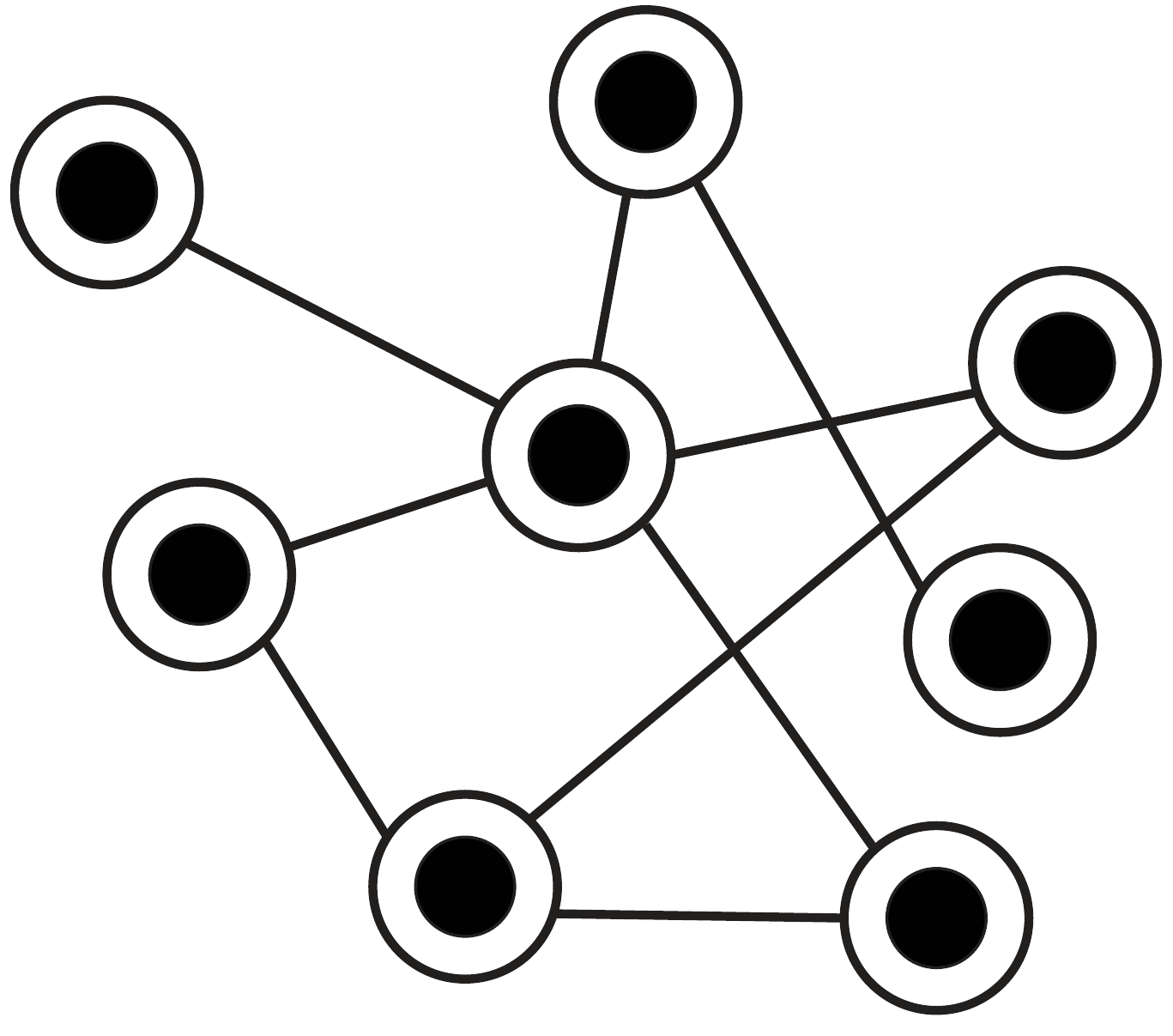}
        \caption{randomly matched stubs}
    \end{subfigure}
    \caption{Configuration model.}
\label{fig:configuration_model}
\end{figure}

\item[Expected degrees.] Fix a network of  $N$  nodes and 
consider a vector 
$$
{\bm w} = (w_1 , w_2 ,  \cdots, w_N ),
$$
In this model,  each link $A_{ij}$ between nodes 
$i$ and $j$ is an independent Bernoulli variable taking value $1$ with success probability 
$$
p_{ij} = w_i w_j \rho,
$$ 
and 0 with probability $1 - p_{ij}$, where  
$$\rho = \frac{1}{\sum_{i=1}^N w_i}.$$ To ensure that $p_{ij} \le 1$
it assumed that $\bm{w}=\bm{w}(N)$ is chosen so that 
\begin{equation}
\label{DeltaRho}
\Delta^2  \rho \le 1.
\end{equation}
The {\em degree} $k_i = \sum_j A_{ij}$ of the $i$th is a 
random variable. An interesting property of this model is that under this construction $w_i$
is the expected value of $k_i$, that is, 
$$
\mathbb{E}_{\bm{w}}(k_i) = \sum_j \mathbb{E}_{\bm{w}}(A_{ij}) =  w_i  
$$ 
\end{description}
So, the different to the configuration model is that we do not fix the node degree, but rather obtain the 
degree probabilistically. This model also have many desirable concentration properties in the large $N$ limit.

\subsection{Spectral Properties of the Laplacian}
\label{sec:lap_spec}

The eigenvalues and eigenvectors of $\bm{A}$ and $\bm{L}$ tell us a lot about  the network structure. The eigenvalues of $\bm{L}$ for instance are related to how well  connected is the graph and how fast a random walk on the graph could spread. In particular, the smallest nonzero eigenvalue of $\bm{L}$ will determine the synchronization properties  of the network. Since the graph is undirected the matrix $\bm{L}$ is  symmetric its eigenvalues are real,  and $\bm{L}$ has a complete set of orthonormal  eigenvectors \ref{LA:ThmDiag}. The next result characterizes important properties of the Laplacian

\begin{Teorema}
Let $G$ be an undirected network and  $\bm{L}$ its associated Laplacian. Then: 
\begin{itemize}
\item[a)] \bm{L} has only real eigenvalues, 
\item[b)] $0$  is an eigenvalue and a corresponding 
eigenvector is $\bm{1} =  (1, 1, \cdots , 1)^*$, where $^*$ stands for the transpose. 
\item[c)] \bm{L} is positive semidefinite, its eigenvalues  enumerated 
in increasing order and repeated according to their multiplicity satisfy
$$
0= \lambda_1 \le \lambda_2 \le \cdots \le \lambda_N
$$ 
\item[d)] The multiplicity of $0$ as an eigenvalue of $\bm{L}$ equals the 
number of connect components of G. 
\end{itemize}
\end{Teorema}

Therefore, $\lambda_2$ is bounded away from zero  whenever the network is connected. The smallest non-zero eigenvalue is known as  algebraic connectivity, and it is often called the Fiedler value.  The spectrum of the Laplacian is  also related to some other topological invariants.  One of the most  interesting connections is its relation to the diameter, size and degrees. 
\begin{Teorema} 
Let $G$ be a simple network of size $N$ and $L$ its associated Laplacian. Then: 
\begin{enumerate}
\item  \cite{mohar1991a} $ \lambda_2 \ge \displaystyle \frac{4}{Nd} $
\item[]
\item \cite{fiedler1973} $ \lambda_2 \le  \displaystyle \frac{N}{N-1} k_1 $
\end{enumerate}
\label{boundl}
\end{Teorema}

We will not present the proof of the Theorem here, however, they can be found in references we provide in the theorem. We suggest the reader to see further bounds on the spectrum  of the Laplacian in Ref. \cite{mohar1991b}. Also Ref. \cite{mohar1997} presents many applications of the  Laplacian eigenvalues to diverse problems. One of the main goals in spectral graph theory is  the obtain better bounds by having  access to further information on the graphs.

For a fixed network size, the magnitude of $\lambda_2$ reflects how well connected is graph.
\begin{sidewaystable}
\caption{Network of $N$ nodes with $m$ mean degree. Examples of nonrandom networks are depicted in Fig. \ref{fig:matrices} and random networks in Fig. \ref{fig:complex_networks}. For random networks, the mean degree is given by $m =\sum_k kP(k)$ where $P(k)$ is the degree distribution of the network.  The randomness of the networks arises via probability $p$. Consider the degree distribution for the SF network $P(k) = k^{-\gamma}$. $C$ is constant.}
\begin{center}
\label{tab:networks}
\begin{tabular}{ccccccc}
\hline
Network 	&	$\lambda_2$ & $\lambda_N$  &	 $k_{\max} $ & $k_{\min}$ &	$d$  & Reference \\
\hline
\hline
complete	&	 $ N $ &	N &$N-1$ & 	$N-1$	&	$1$&  \\
~\\
ring	&	 $ \displaystyle 2  - 2\cos \left( \frac{2 \pi }{N}\right)$ &  $4$ &$2$	& $2$	&		
$\begin{array}{c}
(N+1)/2 \mbox{ if  $N$ is odd} \\
N/2  \, \, \mbox{ \, \, \, \, if  $N$ is even} \\
\end{array}
$ 
&  \cite{arenas2008}\\
\\
star & $\displaystyle 1$&   $N$ &  $N-1$  & $1$  & 	2  &\\
\\
hypercube (Q$_m$) & $2$& $\frac{2\ln(N-2)}{m}\left(\frac{m}{N}\right)$ & $m$& $m$& $m$&\\
\\
2$k$-regular & $\displaystyle 2k+1- \frac{\sin\left[(2k+1)\frac{\pi}{N}\right]}{\sin\left(\frac{\pi}{N}\right)}$& $\displaystyle 2k+1- \frac{\sin\left[(2k+1)\frac{\pi(N-1)}{N}\right]}{\sin\left(\frac{\pi(N-1)}{N}\right)}$  &  $2k$ &  $2k$ & $\begin{array}{c}
\lceil(N+1)/2k\rceil \mbox{ if  $N$ is odd} \\
\lceil N/2k\rceil  \, \, \mbox{ \, \, \, \, if  $N$ is even} \\
\end{array}
$ 
&  \cite{arenas2008}\\ 
\\
SW network & $\sim \displaystyle 2k+1- \frac{\sin\left[(2k+1)\frac{\pi}{N}\right]}{\sin\left(\frac{\pi}{N}\right)}+ Cp$& $\sim \displaystyle 2k+1- \frac{\sin\left[(2k+1)\frac{\pi(N-1)}{N}\right]}{\sin\left(\frac{\pi(N-1)}{N}\right)}+Cp$ & $\sim 2k$&$\sim 2k$ & $O(\log N)$& \cite{newman2010} \\ \\
SF network& $\sim m_0\left(1-\frac{1}{(\gamma-1)}\right)$& $\sim m_0 N^{\frac{1}{\gamma-1}}$ & $\sim m_0$   & $\sim m_0N^{\frac{1}{\gamma-1}}$  & $O(\frac{\log N}{\log \log k})$  &\cite{wu2003}\\ \\
ER network &$\sim Np$& $\sim Np$  & $pN(1+O(1))$&$(1-p)(1+O(1))$ & $\frac{\log N}{2\log Np}$&\cite{riordan2000}\\ \\
\hline
\end{tabular}
\end{center}
\end{sidewaystable}


\section{Stability of Synchronized Solutions}

We will state two results on network synchronization. The first assumes that the coupling function $\bm{H}$ is linear and the network is undirected. This assumption facilitates the discussion of the main ideas. Then, we will discuss the case where the coupling is nonlinear.  Basically, the results are the same with an additional complication as the latter involves the theory of Lyapunov exponents.

\begin{Teorema} \label{MainThm}Consider the diffusively coupled network model
$$
\bm{x}_i = \bm{f}(\bm{x}_i) - \alpha \sum_{j=1}^N L_{ij} \bm{H}(\bm{x}_j),
$$ 
on an undirected and connected network. Assume that the $\bm{H}$ is a positive definite linear operator. The reason of the assumption will appear in Step 5 of the theorem.
Then, there is $\Gamma = \Gamma(\bm{f},\bm{H})$ such that for any 
$$
\alpha > \frac{\Gamma}{\lambda_2},
$$
the global synchronization is uniformly asymptotically stable. Moreover, the transient to the globally 
synchronized behavior is given the spectral gap $\lambda_2$, that is, for any $i$ and $j$  
$$
\| \bm{x}_i(t) - \bm{x}_j(t) \| \le C e^{- (\alpha \lambda_2  - \Gamma) t},
$$
where $C$ is a constant.
\end{Teorema}

The above result relates the  threshold coupling  for synchronization in contributions coming  solely 
from dynamics $\Gamma$, and network structure $\lambda_2$. 

\begin{Definicao}[Synchronization Threshold] We call 
$$
\alpha_c(\bm{f},\bm{H},G) = \frac{\Gamma(\bm{f},\bm{H})}{\lambda_2(G)} 
$$
the critical synchronization coupling. 
\end{Definicao}

Therefore, for a fixed node dynamics $\bm{f}$ and coupling $\bm{H}$ we can analyze how distinct network facilitates or inhibits global synchronization. To continue our discussion we need the following

\begin{Definicao}[Better Synchronizable]
We say that the network $G_1$ is more synchronizable than $G_2$ if for fixed $\bm{f}$ and $\bm{H}$ we have
$$
\alpha_c(G_1) < \alpha_c (G_2)
$$ 
Likewise, we say that $G_1$ has better synchronizability than $G_2$.
\end{Definicao}

\subsection{Which networks synchronize best}
\label{sec:synchronizability}
In this setting, the coupling function $\bm{H}$ is positive definite and the network is undirected, the  synchronizability depends only on the spectral gap. Using the previous study on the properties of various networks presented in Sec.~\ref{sec:lap_spec}. We consider networks of $N$ nodes then

\begin{itemize}
\item {\it A complete graph} is the most synchronizable. In fact, $\alpha_c \approx 1/N$. So, the larger the graph the less coupling strength is necessary to synchronize the network.

\item {\it A path} or {\it ring} are poorly synchronizable. For these networks,  $\alpha_c \approx N^2$.

\item {\it 2$k$-regular graphs} share the same properties for also poorly synchronizable when $k \ll N$.

\item {\it Erd\"os-Renyi graphs} the synchronization properties depend only on the mean degree $\alpha_c \approx 1/<k>$

\item {\it Small world networks} are better than regular but worse than random graphs. In the limit of large graphs $\alpha_c \approx 1/s$ where $s$ is the fraction of added random links.  In general, adding links to a network favours synchronization.
\end{itemize}

~\\

\subsection{Proof of the Stable Synchronization}

Now we present the proof of Theorem \ref{MainThm}. We omit some details that are not relevant for the understanding of the proof. A full discussion of the proof can be found in \cite{pereira2014}. We will show that whenever the nodes start close 
together they tend to the same future dynamics, that is, $\| \bm{x}_i (t)  - \bm{x}_j (t) \| \rightarrow 0$, 
for any $i$ and $j$. For pedagogical purposes we split the proof into six main steps.

{\bf Step 1: Kronecker Form}.  We have $n$ coupled equations each has dimension $m$. Because of the nice structure of the interaction we can use the Kronecker product to write them as a single block.  Given two matrices $\bm{A} \in \mathbb{R}^{m \times n}$ and $\bm{B} \in \mathbb{R}^{r \times s}$, the Kronecker Product of the matrices $A$ and $B$ and defined as the matrix
$$
\bm{A} \otimes \bm{B} = 
\left(
\begin{array}{ccc}
A_{11} \bm{B} & \cdots &  A_{1n} \bm{B} \\
\vdots & \ddots & \vdots \\ 
A_{m1} \bm{B}  & \cdots & A_{mn} \bm{B} 
\end{array}
\right),
$$
\noindent
we introduce the following notation
$$
\bm{X} = \mbox{col}( \bm{x}_1 , \cdots , \bm{x}_N ),
$$
where col denotes the vector formed by stacking the columns vectors $\bm{x}_i$ into a single  vector. Similarly 
$$
\bm{F}(\bm{X}) = \mbox{col}( \bm{f}(\bm{x}_1) , \cdots, \bm{f}( \bm{x}_N) ).
$$
Then Eq. (\ref{eq:motion_lap}) can be arranged into a compact form
\begin{equation}
\label{comp}
\frac{ d \bm{X}}{dt} = \bm{F}(\bm{X}) - \alpha (\bm{L} \otimes \bm{H}) \bm{X},
\end{equation}
where  $\otimes $ is the Kronecker product. The easiest way to check that this is correct is to compute the $ith$ block of dimension $m$ and compare with the equation for the ith node.

\bm{ Step 2: Tranversal Laplacian Eigenmodes.}  The Kronecker product has many nice properties 
such as 
\begin{equation}\label{KronProd}
( \bm{A} \otimes \bm{B})(\bm{C} \otimes \bm{D})=\bm{A}\bm{C} \otimes \bm{B}\bm{D}.
\end{equation}
And this holds whenever the matrix multiplication make sense. A nice consequence of the multiplication result in  Kronecker form is that if 
$$\bm{A} \bm{v}_i = \lambda_i \bm{v}_i \,\,\, \mbox{ and } \,\,\, \bm{B} \bm{u}_j = \mu_j \bm{u}_j  \,\,\, \mbox{   then   } \,\,\,
\bm{A}\otimes \bm{B} (\bm{v}_i \otimes \bm{u}_j) =  \lambda_i \mu_j \bm{v}_i \otimes \bm{u}_j
$$

Since we are assuming that $\bm{L}$ is undirected and $\bm{H}$ is positive definite the eigenvectors $\{\bm{v}_i\}_{i=1}^N$ of $\bm{L}$ form a basis of $\mathbb{R}^N$. Likewise, the eigenvectors of $\bm{H}$ form a basis of $\mathbb{R}^m$.  This implies that the eigenvectors of $\bm{L}\otimes \bm{H}$ form a basis of $\mathbb{R}^{Nm}$. Using this fact we can represent $\bm{X}$ as
\begin{eqnarray}
\bm{X} &=&  \sum_{i=1}^N \bm{v}_i \otimes \bm{y}_i \nonumber
\end{eqnarray}
where $\bm y_i$ is the coordinates of $\bm X$ in the Kronecker basis. For sake of simplicity we call $\bm{y}_1 = \bm{s}$, and remember that  
$\bm{v}_1 = \bm{1}$ is an eigenvector.  Hence
$$
\bm{X} = \bm{1} \otimes \bm{s} + \bm{U},
$$
where 
$$
\bm{U} = \sum_{i=2}^N \bm{v}_i \otimes \bm{y}_i.
$$
In this way we split the contribution in the direction of the global synchronization 
and \bm{U}, which accounts for the contribution of the transversal. Note that if 
$\bm{U}$ converges to zero then the system completely synchronize, that is 
$\bm{X}$ converges to $\bm{1} \otimes \bm{s}$ which clearly implies that
$$
\bm{x}_1 = \cdots = \bm{x}_N = \bm{s}
$$

The goal then is to obtain conditions so that $\| \bm{U} \| \rightarrow 0$.
\\

\bm{ Step 3: Variational equations for the Transversal Modes.} The equation of motion in terms of the 
Laplacian modes decomposition reads
\begin{eqnarray}
\bm{1} \otimes \frac{d \bm{s}}{dt } + \frac{d \bm{U}}{dt }&=& \bm{F} (\bm{1} \otimes \bm{s} + \bm{U})  - \alpha(\bm{L} \otimes \bm{I})\left(  \bm{1} \otimes \bm{s} + \bm{U}\right), \nonumber
\end{eqnarray} 
We assume that $\bm{U}$ is small and perform a Taylor expansion about the synchronization manifold. 
$$
\bm{F} (\bm{1} \otimes \bm{s} + \bm{U})  = \bm{F} (\bm{1} \otimes \bm{s})  + D \bm{F} (\bm{1} \otimes \bm{s}) \bm{U}  + \bm{R}(\bm{U}), 
$$
where $\bm{R}(\bm{U})$ is the Taylor remainder $\| \bm{R} (\bm{U}) \| = O(\| \bm{U} \|^2) $.
Using the Kronecker product properties Eq.~\ref{KronProd} and  the fact that $\bm{L} \bm{1} = \bm{0}$, together with 
$$
\bm{1} \otimes \frac{d \bm{s}}{dt } = \bm{F} (\bm{1} \otimes \bm{s}) = \bm{1} \otimes \bm{f}(\bm{s})
$$
and we have
\begin{eqnarray}
\frac{d \bm{U}}{dt }= [ D \bm{F} (\bm{1} \otimes \bm{s})  - \alpha (\bm{L} \otimes \bm{I}) ]   \bm{U} + \bm{R}(\bm{U})
\end{eqnarray} 
and likewise
$$
 D \bm{F} (\bm{1} \otimes \bm{s}) \bm{U}  = [  \bm{I}_N \otimes D \bm{f} (\bm{s}) ]  \bm{U}, 
$$
therefore, the first variational equation for the transversal modes reads
\begin{eqnarray}
\frac{d \bm{U}}{dt } = [\bm{I}_N \otimes D \bm{f} (\bm{s}) - \alpha \bm{L} \otimes \bm{I}_m] \bm{U}. 
\label{eq:var_eq}
\end{eqnarray} 

\bm{ Step 4: Decoupling of  Transversal  Modes.} Instead of analyzing the full set of equations, 
we can do much better by projecting the equation  into subspace $W_i =$ span$\{\bm{v}_i \otimes \bm{I}\}$. Let $P_i : \mathbb{R}^N \otimes \mathbb{R}^m \rightarrow W_i$ be a projection operator given by $P_i = \bm{v}_i \bm{v}_i^* \otimes \bm{I}_m$, it follows that  $P_i$ is an orthogonal projection since $\bm{v}_i$'s are orthonormal. Using Eq. (\ref{eq:var_eq}) and the identity Eq. \ref{KronProd} we obtain
\begin{eqnarray}
P_i \frac{d\bm{U}}{dt} &=&  [ \bm{v}_ i\bm{v}_i^* \otimes D \bm{f} (\bm{s}) - \alpha (\bm{v}_ i\bm{v}_i^* \bm{L}) \otimes \bm{I}_m] \bm{U},  \\
&=& [ \bm{v}_ i\bm{v}_i^* \otimes D \bm{f} (\bm{s}) - \alpha \lambda_i (\bm{v}_ i \bm{v}_i^*) \otimes \bm{I}_m] \bm{U}
\end{eqnarray}
where in the last passage we used that the network is undirected implying $\bm{v}_i* \bm{L} = \lambda_i \bm{v}_i^*$. Using  and the fact that  $\bm{v}_j^* \bm{v}_i = \delta_{ij}$, where is $\delta_{ij}$ the Kronecker delta,  we have that 
$ \bm{v}_ i \bm{v}_i^* \otimes \bm{I}_m \bm{U} = \sum_{j=2}^N \bm{v}_ i \delta_{ij} \otimes \bm{y}_i$. Moreover, since $P_i$ does not depend on time $P_i \dot{\bm{U}} = \dot{(P_i \bm{U})}$
$$
\sum_{j=2}^N \bm{v}_ i \delta_{ij} \otimes \frac{d\bm{y}_i}{dt}  =  \sum_{j=2}^N \bm{v}_i \delta_{ij} \otimes [ D \bm{f} (\bm{s}) - \alpha \lambda_i  \bm{I}_m] \bm{y}_i
$$
the nonzero coefficients give the dynamics in $W_i$. Hence,
$$
 \frac{d\bm{y}_i}{dt}  =   [ D \bm{f} (\bm{s}) - \alpha \lambda_i  \bm{I}_m] \bm{y}_i
$$ 

All blocks have the same form which are different only by $\lambda_i$, the $i$th eigenvalue of  $L$. 
We can write all the blocks in a parametric form 
\begin{equation}
\label{para}
\frac{d {\bm{u}}}{dt} = \bm{K}(t)  \bm{u},
\end{equation}
where 
$$
\bm{K}(t) = D\bm{f}(\bm{s}(t)) - \kappa \bm{I}_m
$$
with $\kappa \in \mathbb{R}$. Hence if $\kappa = \alpha \lambda_i$ we have the equation for the $i$th block. 
This is just the same type of equation we encounter before in the example of the two coupled oscillators,  see Eq. (\ref{var2}).

{\bf Step 5. Stability}. Because $\bm{H}$ is positive definite we can first solve the homogeneous equation 
$\dot{\bm{u}} = - \kappa \bm{H} \bm{u}$. This equation has an globally attracting trivial solution. Then, we incorporate $D\bm f$ in terms of the variation of constants formula. So, first notice that 
$$
\bm{u}(t) = e^{- \kappa \bm{H} t } \bm{u}_0 \,\,\,\, \Rightarrow \,\,\,\, \|  \bm{u}(t) \| \le K \| \bm{u}_0 \|e^{- \kappa \lambda_{\min}(\bm{H})}
$$
where $ \lambda_{\min}(\bm{H})$ is the smallest eigenvalue of $\bm{H}$. So, by the variation of constants formulate
$$
\bm{u}(t)  = e^{- \kappa \bm{H} t } \bm{u}_0 + \int_0^t e^{- \kappa \bm{H}( t  - \tau)} D\bm{f}(\bm{s}(\tau)) d\tau
$$
taking norms 
$$
\| \bm{u}(t) \| =  K \| \bm{u}_0 \| + \int_0^t \|e^{- \kappa \lambda_{\min}(\bm{H}) ( t  - \tau)} \| D\bm{f}(\bm{s}(\tau)) \|d\tau
$$
where $K$ is a constant and defining $M_f = \sup_t \| D\bm{f} (\bm{s}(t)) | $
 by Gronwal inequality 
$$
\| \bm{u}(t) \| =  K_1  \| \bm{u}_0 \|  \|e^{(- \kappa \lambda_{\min}(\bm{H})+ M_f) t}.
$$
The trivial solution will be exponentially stable when 
$$
- \kappa \lambda_{\min}(\bm{H})+ M_f < 0 \,\,\,\, \Rightarrow \,\,\,\, \kappa > \Gamma = \frac{M_f}{ \lambda_{\min}(\bm{H})}
$$

Recall that taking $\kappa = \alpha \lambda_i > \Gamma$ we are stabilizing the equation for the $i$th block.  
But, once we stabilize the second block all blocks will be stable (because $\lambda_2$ is the smallest nonzero eigenvalue) 
$$
\alpha \lambda_N \ge \cdots \ge \alpha \lambda_3 \ge \alpha \lambda_2 \ge \Gamma
$$   
Hence, the stability condition so that all blocks have exponentially stable trivial solution is    
   $$
\alpha  > \frac{\Gamma}{\lambda_2}
$$


\bm{ Step 6: Norm Estimates and Nonlinearities.} Using the bounds for the blocks it is easy to obtain a 
bound for the norm of the evolution operator. Indeed, note that 
\begin{eqnarray}
\|  \bm{U}  \|_2 &=& \left\| \sum_{i=2}^N \bm{v}_i \otimes \bm{y}_i \right\|_2 \le \sum_{i=2}^N \| \bm{v}_i \| \| \bm{y}_i \|_2 \nonumber 
\end{eqnarray}
 Therefore, 
\begin{eqnarray}
\|  \bm{U}  \|_2 & \le& \sum_{i=2}^N \| \bm{v}_i \| K_i  e^{-\eta_i (t-s)} \| \bm{ y}_i(s)\| \nonumber
\end{eqnarray}

Now using that $e^{-\eta_i (t-s)} \le e^{ - (\alpha \lambda_i - \Gamma)(t-s)}$, so

$$
\|  \bm{U} (t) \|_2 \le  K_2 e^{ - \eta (t-s)}
$$
with $\eta  = \alpha \lambda_2 - \Gamma$ for any $t\ge s$. 

Because the trivial solution is exponentially stable (uniformly in $\bm{s}(t)$) 
by the principle of linearization, we conclude that the nonlinearities coming Taylor 
remainder does not affect the stability of the trivial solution, which correspond to the 
global synchronization. 

The claim about the transient is straightforward, because all norms are equivalent in finite dimensions we can take
$$
\| \bm{X}(t) - \bm{1} \otimes \bm{s}(t) \|_{\infty} \le K_3 e^{ - \eta (t-s)} \| \bm{U}(s) \|_{\infty}
$$
implying that $\max_{i}\| \bm{x}_i(t) - \bm{s}(t) \|_2 \le K_3 e^{-\eta(t-s)} \| \bm{U}(s) \|_{\infty}$  and in virtue of the triangular triangular inequality
$$
\|  \bm{x}_i(t) -  \bm{x}_j(t) \|_{\infty} \le \|  \bm{x}_i(t) -  \bm{s}(t) \|_{\infty} + \|  \bm{x}_i(t) -  \bm{s}(t) \|_{\infty}
$$
and using the previous bound, we concluding the proof.  $\Box$

\section{General Diffusive Coupling and Master stability function}
\label{sec:msf}

Until now we have considered linear coupling functions which are positive definite. 
This assumption can be relaxed and thereby we are generalize our previous results. 
The statement will then become rather technical and will be beyond the scope of our review. 
So, here we will discuss the main ideas but will not give much details on the technical issues. 
Consider the function
$g: \mathbb{R}^m \times \mathbb{R}^m \rightarrow \mathbb{R}^m$. We say that $g$ is diffusive if  
$$
g(x,x) = 0 \mbox{      and         } g(x,y) = - g(y,x)
$$

Hence, we can extend the model to a general diffusive coupling

\begin{equation}
\dot{\bm{x}}_i =\bm{f}(\bm{x}_i) + \alpha \sum_{j=1}^N A_{ij}  g(\bm{x}_j,\bm{x}_j). 
\label{eq:motion_lap}
\end{equation}

We perform the analysis close to synchronization $x_i = s + \xi_i$ so
$$
g(\bm{x}_j,\bm{x}_j) = g(s+ \xi_j, s + \xi_i) = g(s,s) + D_1 g(s, s) \xi_j + D_2 g(s, s) \xi_i 
$$
but because the coupling is diffusive
$$
D_2 g(s,s) = - D_1 g(s,s)
$$
we obtain
$$
g(\bm{x}_j,\bm{x}_i) =  G(s)(  \xi_j - \xi_i)   + R(\xi_i,\xi_j)
$$
where $G(s)  = D_1 g(s, s) $ and  $R(\xi_i,\xi_j)$ contains quadratic terms. So, the first variational equation about the synchronization manifold

\begin{eqnarray}
\dot{\xi}_i &=& D\bm{f}(\bm{s}(t)) \xi_i + \alpha \sum_{j=1}^N A_{ij} G(s) (\xi_j - \xi_i) \\
&=&D\bm{f}(\bm{s}(t)) \xi_i - \alpha  G(s) \sum_{j=1}^N L_{ij} \xi_j. 
\label{Dif}
\end{eqnarray}

Now we can perform the same steps as before. In fact, Steps $1-4$ remain unchanged. The change difference is Step 5, which concerns the stability of the modes. Performing all the steps 1 to 4 we obtain the parametric equation for the modes
$$
\dot{u} = [Df(s(t)) - \kappa G(s(t))] u
$$ 
And we can no longer apply the {\it trick} of using the coupling function $\bm{H}$ to solve the equation. Here, $G(s(t))$ depends on time 
and this generality tackling the stability of the trivial solution is challenging.

The main idea is to fix $\kappa$ compute the maximum Lyapunov exponent  $\Lambda(\kappa)$ as 
$$
\| u(t) \| \le C e^{\Lambda(\kappa) t}
$$
 see Appendix \ref{sec:lyapunov_exp}.  The
map 
\begin{equation}\label{msf}
\kappa \mapsto \Lambda(\kappa)
\end{equation}
is called Master Stability Function. Notice that if  $\Lambda (\kappa) <0$ when $\kappa \in (\alpha_c^1 , \alpha_c^2)$ then 
$\|u\| \rightarrow 0$. 

The stability condition then become
$$
\alpha_c^1 \le \alpha \lambda_2 \le \cdots \le \alpha \lambda_N \le  \alpha_c^2 
$$
Or 
\begin{equation}
\label{eq:sta_cond}
\frac{\alpha_c^2}{\alpha_c^1} \ge \frac{\lambda_N}{\lambda_2} 
\end{equation}

This is a well studied condition. Much energy has been devoted to study the master stability function Eq. \ref{msf}, see e.g., \cite{huang2009,schultz2016}.

\subsection{Examples of Master Stability Functions} 

Now let us consider coupled R\"ossler systems which are coupled through their $x$--coordinates:
\begin{align}
\dot{x}_i &= - y_i - z_i  + \alpha\sum\limits_{j=1}^N A_{ij} (x_j - x_i)\\
\nonumber\dot{y}_i &= x_i + a y_i \\
\nonumber\dot{z}_i &= b + z_i (x_i - c) \;.
\end{align}

In order to compute $\Lambda_{\max}(\kappa)$, we find that $D\bm{f}$ and $D\bm{H}$ are given by

\begin{align}\label{Ros}
D\bm{f}(\bm{s}) =
\begin{pmatrix}
0 & -1 & -1 \\ 1 & a & 0 \\ z^\ast & 0 & x^\ast-c
\end{pmatrix} \;\; \;\;\; \; \mbox{  and  }
D\bm{H} = \bm{H} = 
\begin{pmatrix}
1 & 0 & 0 \\ 0 & 0 & 0 \\ 0 & 0 & 0
\end{pmatrix} \;,
\end{align}
\noindent
$x$ and $z$ are the components of $\bm{s}$. The constants are $a=0.2$, $b=0.2$ and $c=5.7$. 

\begin{figure}[h]
\centerline{\includegraphics[width=0.4\linewidth]{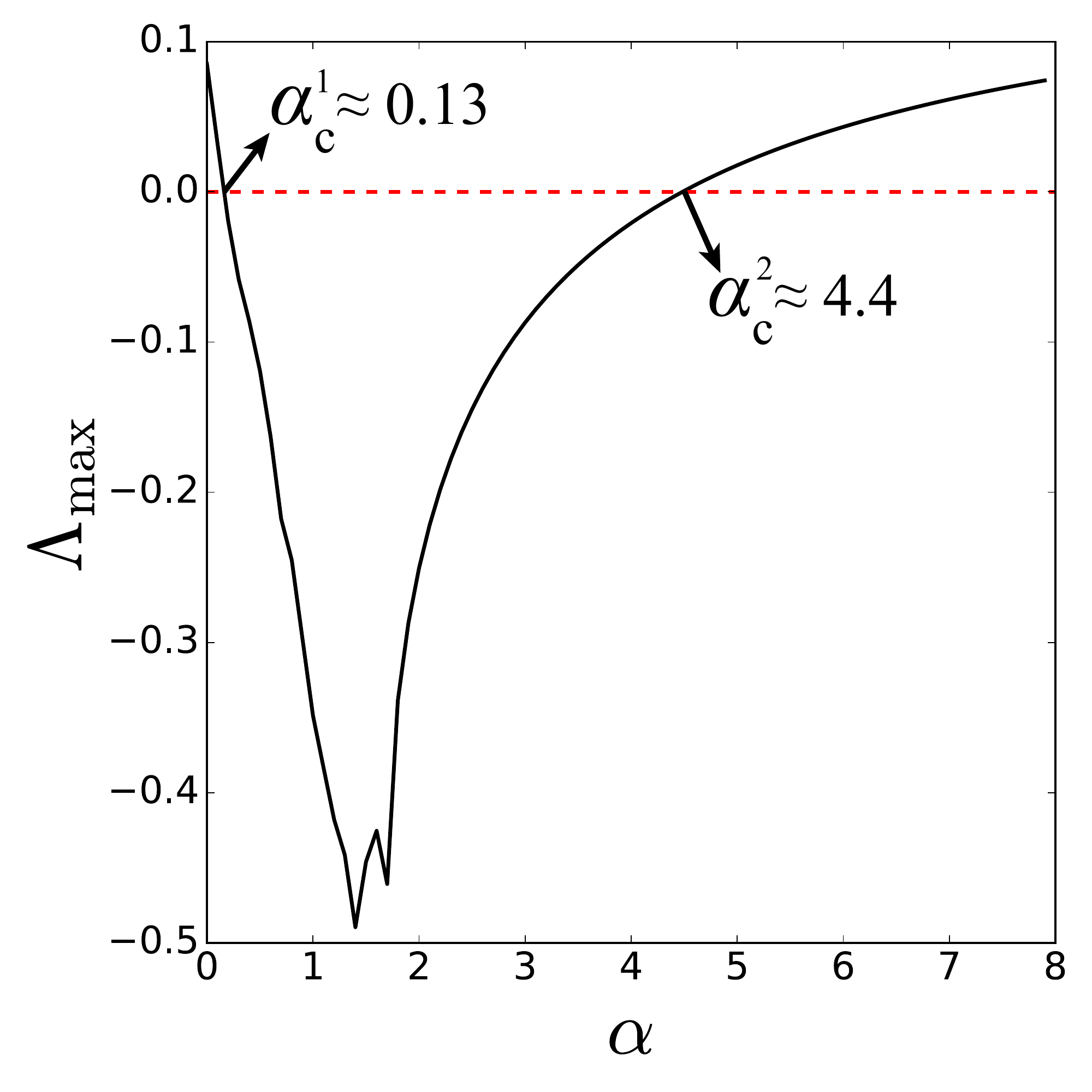}}
\caption{Master stability function for x-coupled R\"ossler attractors on a network structure.}
\label{fig:msf_roessler}
\end{figure}

To compute $\Lambda(\kappa)$, we first simulate the isolated dynamics $\dot{s} = f(s)$ and obtain the trajectory $s(t)$, 
then we feed this trajectory to 
$\dot{u} = [Df(s(t)) - \kappa H] u$ and then for each $\kappa$ estimate the  maximal Lyapunov exponent $\Lambda(\kappa)$. 
The result is depicted in Fig.~\ref{fig:msf_roessler}.  

Stability region where $\Lambda$ is negative is bounded between $\alpha_c^1\approx0.13$ and $\alpha_c^2\approx 4.4$. 
So, if the network is a complete graph then $\lambda_2 = \cdots = \lambda_N = N$, so the network synchronization when 
$$
\frac{\alpha_c^2}{N} > \alpha > \frac{\alpha_c^1}{N}  
$$

\subsection{Synchronization conditions and Synchronization Loss}
In Sec~.\ref{sec:synchronizability}, we discussed the synchronizability of network types when there is only one critical threshold in other words $\alpha_c^2 \rightarrow \infty$. This case is true when $\bm H$ is a positive matrix. In the general case, there are two finite critical couplings Eq.~\ref{eq:sta_cond} and the master stability function has a finite stability region between these critical points (as discussed in the above section).  Now we discuss this synchronization scenario for various network types. To quantify the synchronizability, we will use the network properties given in Table.~\ref{tab:networks}.

\begin{description}
\item[2$k$-regular graphs -- Diameter driven synchronization loss.]  When $k << N$, the mean geodesic distance (shortest path length) between nodes are increasing very fast as $N$ increasing. Hence, while the diameter of the network is increasing $d = 2N/k$, speed of information exchange between the nodes is decreasing drastically. In this case, the complete synchronization is not stable and it is visible from the Laplacian spectrum of the graph. For $N\to\infty$ and $k << N$, the extremal eigenvalues given in Table~\ref{tab:networks} can be extended to Taylor expansion and major role playing part can be rewritten as
\[
 \lambda_2 = \frac{4\pi^2(k+\frac{1}{2})^3}{3N^2}  \mbox{  and   }  \lambda_N = 2k\frac{N}{N-1}
\]
The ratio between the largest $\lambda_N$ and second smallest eigenvalue $\lambda_2$ of the Laplacian is not growing in the same scale,
$$
\frac{\lambda_N}{ \lambda_2} \approx N^2/k^2
$$
and the condition in Eq. (\ref{eq:sta_cond}) is never satisfied.

\item[ER graphs -- Optimal Synchronization.] In this case, the extremal eigenvalues $\lambda_2$ and $\lambda_n$ of the Laplacian matrix increase in the same scale. The diameter of the network increases very slowly $d\approx \log N$ when the network size $N$ increases Table~\ref{tab:networks}. Therefore the synchronization is stable for any scale of size.

\item[BA networks -- Heterogeneity driven synchronization loss.]  When the network is too heterogeneous the complete synchronization is unstable, 
this is because the extremal eigenvalues $\lambda_2$ and $\lambda_N$ grow in different scales and the condition in Eq. (\ref{eq:sta_cond}) is never met. For instance, consider a BA network. Then, the eigenvalues satisfy  
\[
\lambda_2 \approx m \mbox{  and   }  \lambda_N \approx m N^{1/2}
\]
where $m$ is the mean degree. Hence, the eigenration becomes
\[
\frac{\lambda_N }{\lambda_2} \approx N^{1/2}
\]
this should be compared to the stability interval given by the master stability function. Lets consider the example in the above section with the R\"ossler Eq.~(\ref{Ros}). The master stability function gives (as seen in Figure \ref{fig:msf_roessler}) an stability interval $\alpha_c^2 / \alpha_c^1 \approx 34$. The stability conditions Eq. (\ref{eq:sta_cond}) reads as 
$$
N^{1/2} < 34
$$
So, when the BA network is large enough it is not possible to synchronize the system. In particular the critical system size to be able to synchronization a network of R\"ossler as in Eq.~(\ref{Ros}) is, therefore,   
$$
N \approx 10^3.
$$
\end{description}

\subsubsection{Extensions}
 \label{sec:extensions}
There are a few extensions of the model. Here we discuss a few directions.   

\begin{description} 
\item[Directed Networks.] The major problem considering directed networks is that they may not 
be diagonalizable. So, the decoupling of transversal modes by projection is a nontrvial steps. 
There are a few ways to overcome this. The first, using Jordan decomposition of the Laplacian.
The other possibility is to perturb the Laplacian to make the eigenvalues simples. This must be done in such 
a way that the perturbation does not spoil the stability. In both cases, the stability condition remain unchanged. Only 
the transients may be longer. 

When the network is nondiagonalizable, small perturbations in the network may lead to large perturbations in the eigenvalues (the 
eigenvalues in this case are not differentiable functions of the perturbations) \cite{nishikawa2010}. Moreover, structural improvements in the network 
may lead to desynchronization \cite{pade2015}.

\item[Nonidentical Nodes] Here we consider $f(x) \mapsto f(x) + r_i(x,t)$, where $r_i$ is either a perturbation 
of the vector field or a signal playing the role of noise. When $r_i$ is very small synchronization will persist \cite{belykh2003}. 
For general networks, Bollt and co-workers \cite{sun2009} extended the master stability function approach when $r_i$ is a perturbation of the vector field. 
Pereira and co-workers \cite{pereira2013} study the effect of general perturbations $\| r_i \| \le \delta$ and the role that the network structure plays in suppressing the fluctuations. In the case where $\bm{H}$ is positive definite and the network is undirected, they showed that the synchronization error 
$$
E = \frac{1}{n(n-1)} \sum_{ij} \|x_i - x_j \|
$$ 
behaves as 
$$
E \le K \frac{\delta}{\alpha \lambda_2 - \alpha_c}
$$

For example, if the oscillators where uncoupled and $r_i$ independent noise then the Central Limit theorem would yield
$
E= O(N^{-1/2}).
$
For complete networks, the interaction and synchronization yields
$
E  = O( N^{-1})
$
which is a large improvement over the naive application of the Central Limit theorem. In certain sense, 
this shows that interacting maybe better then isolation.

\item[Nonidentical Coupling Functions] In many applications the coupling function are not identical and has fluctuating components \cite{stankovski2015}. Consider an undirected networks of identical oscillators and coupling function 
$$
H_{ij}(x_i - x_j,t) = H(x_i - x_j) + P_{ij}(x_i - x_j,t)
$$
where $\|P_{ij}(x_i - x_j,t) \| \le \eta$. In this case, the network structure will play a major on the size of perturbation $\eta$. 
If the network is random and the degree distribution homogeneous, then even for large perturbations $\delta$ synchronization 
is stable. If the network is heterogeneous degrees such as Barabasi-Albert then typically $\delta_c = O(N^{-\beta})$ is the critical perturbation size. If $\delta > \delta_c$ synchronization becomes unstable, solely because of the interaction between  network structure and perturbations in the coupling function. 

\item[Cluster Synchronization] According to similarities in coupled systems, such as symmetries in network topology or identical dynamics in a diverse population or equally time-delayed nodes in differently distributed feedbacks, the partial or cluster synchronization can emerge. In order to enlighten the reason of these cluster synchronization cases, many techniques are developed and experimental observations are analyzed\cite{zhou2006b,belykh2008,sorrentino2007,dahms2012,fu2013,williams2013}. 
\begin{figure}[h]
\centerline{\includegraphics[width=0.4\linewidth]{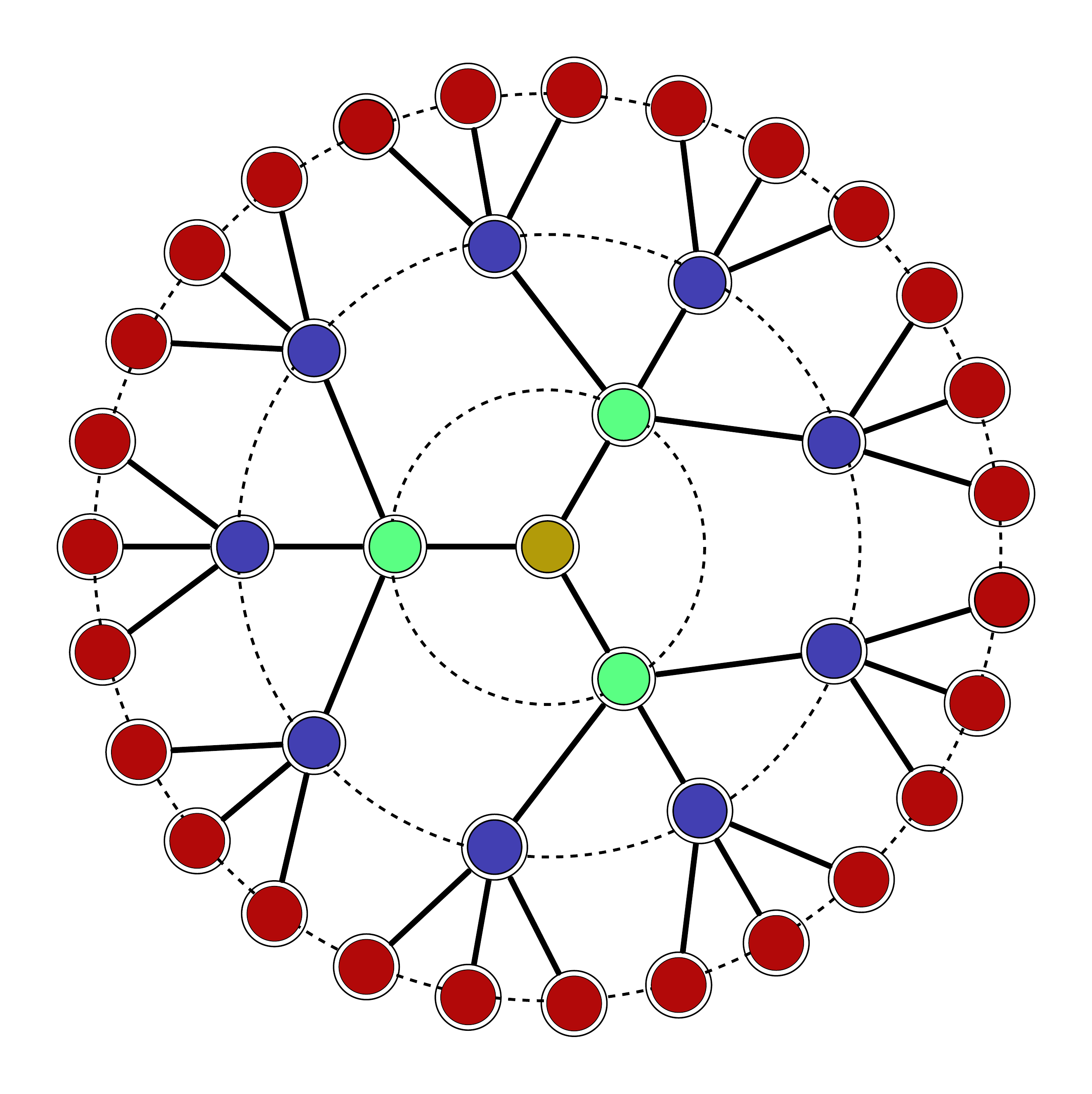}}
\caption{Bethe lattice graph: an example of symmetries in network structure. Nodes in the same level of the Bethe graph are exactly symmetric to each other. Therefore same color nodes constitute a cluster.}
\label{fig:cluster_sync}
\end{figure}

The symmetries are easy to detect for some network geometries for instance Bethe lattice is a regular graph which grows from a root (parent) node by $m$-nodes for $\ell$-levels, an example of Bethe lattice given in Fig.~\ref{fig:cluster_sync} for $m=3$ and $\ell=3$. The nodes in the same level of the Bethe graph, they all symmetric to each other. In Fig.~\ref{fig:cluster_sync}, the levels of the graph are given in the same color and the cluster synchronization occurs for each level. 

Recently, Pecora {\it{et al.}} put forward that all the symmetries in network structures are not visible directly. They developed a computational group theory based method to reveal these hidden symmetries and predicted possible synchronization patterns \cite{pecora2014,sorrentino2016}.

\item[Time Delay Coupling] Simultaneous coupling is not always possible for real world applications in other words some time delays can occur in the interaction process. Therefore it is important to investigate synchronizability and stability of coupled time-delayed systems. The necessary conditions for time delayed synchronization is analytically shown by Pyragas \cite{pyragas1998}. 
The finding of time-delay synchronization is used as an application for the anticipating synchronization (see Section \ref{sec:anticipating}).
 \end{description}

\section*{Acknowledgements}
The authors gratefully acknowledge the support of the European Research Council  (ERC AdG grant no 339523 RGDD), CNPq (grant no 400360/2014-4), EU Horizon2020 Innovative Training Network CRITICS (grant no 643073), a FAPESP-Imperial College SPRINT grant and Lobachevsky State University of Nizhni Novgorod grant (grant no 14-41-00044 RSF).

\newpage

\appendix
\section{List of frequently used notions and abbreviations}
\subsection*{Table of notions}
 \begin{tabular}{ll}
$| \cdot |$ & absolute value\\
$\| \cdot \|$ & norm \\
$\delta_{ij}$ & Kroneker delta\\
$G$ & graph \\
$\bm{L}$ & Laplacian matrix \\
$\bm{A}$ & adjacency matrix \\
$\bm{I}$ & identity matrix \\
$\bm{H}$ & coupling function \\
$\bm{1}$ & vector whose every components is 1 \\
$\alpha$ & coupling strength \\
$\alpha_c$ & critical coupling strength for synchronization \\
$\bm{f}$ & isolated dynamics (vector field) \\
$\Lambda$ & Maximum Lyapunov exponent \\
$\lambda_2$ & spectral gap: the second minimum eigenvalue of Laplacian matrix\\
$D\bm{f}$ & Jacobian matrix of $\bm{f}$\\
$\phi$ & phase \\
$t$ & time \\
$a,b$ and $c$ & parameters of R\"ossler system \\
$\sigma,\rho$ and $\beta$ & parameters of Lorenz system \\
$m$ and $n$ & dimensions of vector fields \\
$i$ and $j$ & natural numbers \\
$N$ & system size of networks \\
$M$ & total number of links \\
$k_i$ & degree of $i$-th node\\
$d$ & diameter of a network\\
 \end{tabular}
 
\subsection*{Table of abbreviations} 
 \begin{tabular}{ll}
CS & complete synchronization\\
PS & phase synchronization \\
GS & generalized synchronization \\
AS & anticipating synchronization \\
ER & Erd\"os - Renyi network \\
SF & Scale-free network \\
SW & Small world network \\
BA & Barabasi - Albert network
 \end{tabular}
 
\section{Lyapunov exponent}
\label{sec:lyapunov_exp}
Sensitive dependence on initial conditions is one of the main characteristics of chaotic systems. The main idea is that nearby orbits diverge at an exponential rate. This rate is called Lyapunov exponents. In this Appendix, we provide the basic notions on the theory of Lyapunov exponents. 

If we have a nonlinear equation we can study the properties of a given solution $\bm{s}$ by linearizing the dynamics around the orbit, as we have done in Sec \ref{sec:complete_sync}. This procedure leads to a linear nonautonomous equation 

\[
v^{\prime}=A\left(  t\right)  v
\]
where $A$ is continuous and bounded matrix function. 
The goal is to study the behaviour of solutions. Typically solving the equation explicitly is impossible. 
So the theory of Lyapunov exponents plays a major role.

Let  $v:\mathbb{R}\rightarrow\mathbb{R}^{n}$ be a solution $v^{\prime}=A\left(  t\right)  v$ and $T(t,s)$ is the fundamental matrix. The Lyapunov exponent of the solution is defined as 
\[
\lambda\left(  v\right)  =\overline{\lim_{t\rightarrow\infty}}\frac{1}{t}%
\ln\left\Vert T(  t,s)  v(s)  \right\Vert
\] 

We also define 
$\lambda\left(  0\right)  =-\infty$. The largest Lyapunov exponents is our main 
object of study and is given by 
\[
\Lambda=\overline{\lim_{t\rightarrow\infty}}\frac{1}{t}\ln\left\Vert
\Pi\left(  t,s\right)  \right\Vert
\]
it determines the behaviour of solutions asymptotically because for any solution of the equation we have
\[
\left\Vert v\left(  t\right)  \right\Vert <C_{\varepsilon}e^{\left(
\Lambda+\varepsilon\right)  t}%
\]

If $\Lambda<0$, the trivial solution $v\left(  t\right)  =0$ is asymptotically stable. 
Lyapunov exponents generalizes stability criteria for autonomous (given by eigenvalues) and periodic equations (given by Floquet exponents).

\begin{lemma}
Let $A\in Mat\left(  n\right)$ and $v$ be an eigenvector of $Av = \beta v$. Then 
$\lambda\left(  v\right) = \beta$.
\end{lemma}

If all $\lambda\left(  v\right)  <0$, we have $\max_{v}\left\{
\lambda\left(  v\right)  \right\}  =\Lambda<0$, and the trivial solution is asymptotically stable. 
The Lyapunov exponent also generalizes the Floquet exponents. 

\begin{lemma}
Let $A\left(  t\right)$ be a periodic matrix by the Floquet representation we have $T(  t,s)  =P(  t,s)
e^{(  t-s)  Q(  s)  }$. Let $v$ be an eigenvector of $Q\left(  s\right)  $, then  $\lambda\left(  v\right)  $ is an 
eigenvalue of $Q(s)$.
\end{lemma}

Hence, Lyapunov exponents are the eigenvalues of the monodromy matrix $Q$. 
Although, for the synchronization analysis we care about the maximal Lyapunov exponents, it is important to know that there are at most $n$ distinct Lyapunov exponents because the  
set $X=\left\{  v\left(  t\right)  |\lambda\left(  v\right)
\leq\alpha\right\}  $ is a vector space.

\section{Lyapunov Function}

One of the main techniques to tackle stability of nonlinear system is the Lyapunov function method. The method by Lyapunov allows us to obtain the stability without finding the trajectories by studying properties  of the Lyapunov function. We consider a dynamical system is modelled by a differential equation
\begin{equation}
\label{eq:system_lyap_func}
\dot{\bm x}=\frac{d\bm x}{dt}=\bm f(\bm x)
\end{equation}

We will study notions relative to connected nonempty subsets $\Omega$ of 
$\mathbb{R}^m$. A function $V: \mathbb{R}^m \rightarrow \mathbb{R}$ is said to be positive 
definite with respect 
to the set $B$ if $V(\bm{x})>0$ for all $\bm{x} \in \mathbb{R}^q\backslash \Omega$. 
It is radially unbounded  if 
$$
\lim_{\|\bm{x}\| \rightarrow \infty }V (\bm{x}) = \infty.
$$ 
Note that this condition guarantees that all level sets of $V$ are bounded. This fact plays 
a central role in the analysis.  We also define $V^{\prime} : \mathbb{R}^m \rightarrow \mathbb{R}$ as
$$
V^{ \prime}( \bm{x} ) = \nabla V (\bm{x})  \cdot \bm{f}(\bm{x}).
$$
where $\cdot$ denotes the Euclidean inner product. 
This definition agrees with the time derivative along the trajectories. That is, if $\bm{x}(t)$
is a solution of Eq. (\ref{eq:system_lyap_func}), then by the chain rule we have
$$
\frac{d V(\bm{x}(t))}{dt} = V^{\prime}(\bm{x}(t)).
$$
Notice that applying the chain rule we obtain
$$
V^{\prime}(\bm{x}(t)) = \nabla V(\bm{x}(t)) \cdot \bm{f}(\bm{x}(t))
$$
This has a nice geometric interpretation. 
Since $ \nabla V(\bm{x}(t))$ is perpendicular to the level set of $V$ if $V^{\prime}(\bm{x}(t))<0$ it means 
that the vector field is point inwards the level set and trajectories will enter the level set and never leave it. Repeating the argument we obtain stability as the following statement shows

\begin{Teorema}[Lyapunov]
Let $V: \mathbb{R}^m \rightarrow \mathbb{R}$ be radially unbounded and positive definite with 
respect to the set $\Omega \subset D$. Assume that 
$$
V^{\prime}(\bm{x}) < 0 \mbox{  for all  }  \bm{x} \in \mathbb{R}^m \backslash \Omega
$$  
Then all trajectories of Eq. (\ref{eq:system_lyap_func}) eventually enter the set $\Omega$, 
in other words, the system is dissipative. 
\label{lyap}
\end{Teorema}

There are also converse Lyapunov theorems \cite{lyapunov1992}. Typically if the system is dissipative (and have nice properties)
then there exists a Lyapunov function. Although the above theorem is very useful, since we don't need knowledge of the trajectories, the 
drawback is the function $V$ itself. There is no recipe to obtain a function $V$ fulfilling all these properties. 
One could always try to guess the function, or go for a general form such as choosing a quadratic function $V$.
We assume that the Lyapunov function is given. 

\section{Chaos in Lorenz system}
\label{sec:lorenz_system}


In order to understand the behaviour of a continuous system, we can use the concept of a Poincar\'{e} section -- a transversal surface to the flow.  This method was developed by Henri Poincar\'{e} in 1890s. The crossing points are a set of discrete numbers and this number sequence is called Poincar\'e map. We can study the structure of the crossings of the trajectory to the surface. This reduces the dimension of the system by $1$.  The structure of crossing points between the plane and the trajectory determines the behaviour of the system. For example, if the trajectory cross the section always at same $k$-coordinate points and repeat these points in the same order then the system is periodic so-called period-$k$.

\begin{figure}[h!]
    \centering
    \begin{subfigure}[t]{0.4\linewidth}
        \centering
        \includegraphics[width=\linewidth]{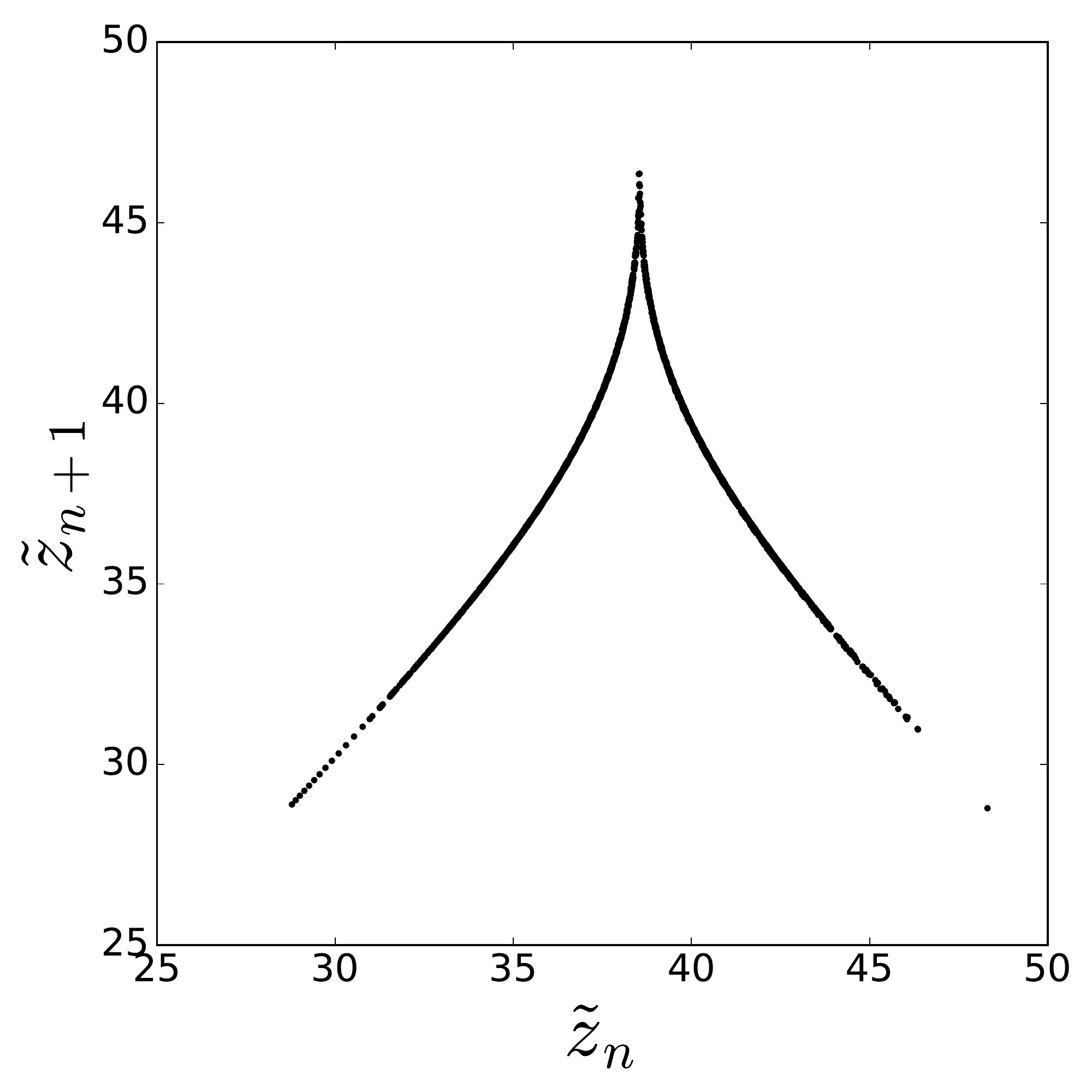}
        \caption{Lorenz system}
    \end{subfigure}%
    ~ ~ ~~
    \begin{subfigure}[t]{0.4\textwidth}
        \centering
        \includegraphics[width=\linewidth]{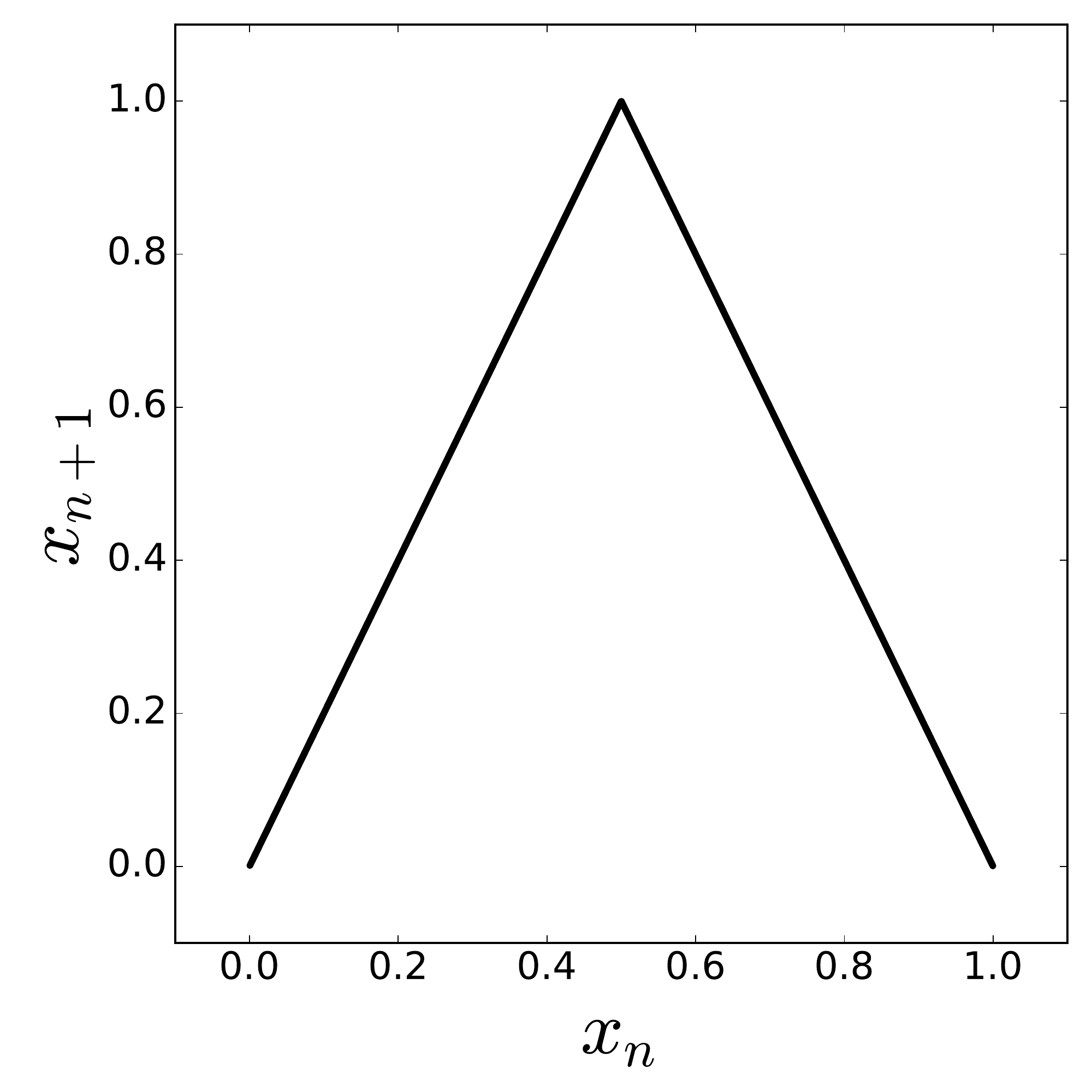}
        \caption{Tent map}
    \end{subfigure}
    \caption{Similarity between Poincar\'e map of Lorenz system and the tent map.}
\label{fig:poincare_map}
\end{figure}

The maxima of $z$-component of the Lorenz system, which is Poincar\'e section of velocities, graphically show the chaotic regime clearly. The governing equation of the Poincar\'e map ($\{z_n\}$) can be plotted as   ${z_n}$ vs ${z}_{n+1}$ (Fig.~\ref{fig:poincare_map}~(a)) which resembles the tent map function (Fig.~\ref{fig:poincare_map}~(b)).  The tent map is given by
\begin{equation}
f(x_n) = x_{n+1} = \left\{
\begin{array}{ll}
 2x_n  & 0 \le x_n \le 1/2 \\
2-2x_n & 1/2 < x_n \le 1. \nonumber
\end{array}
\right.
\label{eq:tent_map} 
\end{equation}
Lyapunov exponent of the tent map
\begin{equation}
\Lambda = \lim_{t\rightarrow\infty}\frac{1}{t}\ln\|Df(x) \|
\end{equation}
where $Df$ is the Jacobian of $f$ and $ \|Df(x)\| = 2 \mbox{ for all } x\ne1/2$ since the function is not differentiable at $x=1/2$. Therefore the Lyapunov exponent is $\Lambda = \ln2$ and according to the positive Lyapunov exponent, the behaviour of the system is chaotic.

The definition of chaos given by Devaney is the the following let $X$ be a metric space and a continuous map $f:X \to X$ is chaotic if
\begin{enumerate}
\item $f$ is transitive (indecomposability); that is any non-empty intervals $U,V \subset X$ there exist a natural number $k$ such that $f^k(U) \cap V$. The transitivity condition ensures the irreducibility of $X$, means that $X$ is an open invariant set and cannot split into two open invariant sets. 

\item the periodic trajectories of $f$ are dense in X (an element of regularity); that is every point in the attractor of $f$ will return back to neighbourhood of the initial point. In other words, any trajectory  initiating from an interval $I \subset X$, will be back in $I$ in sufficiently long but finite time. Therefore the subset $I$ contains infinitely many periodic points. 

\item $f$ has sensitive dependence on initial conditions (unpredictability); that is if there is an infinitesimal distance $\delta_0$ between any two point $x,y \in X$ and there exists a nonnegative number $k$ such that after $n$ iterations the distance between $f^n(x)$ and $f^n(y)$ is larger than $\delta_n > \delta_0$. These nearly started orbits diverge from each other at a rate $\Lambda$ (see Appendix \ref{sec:lyapunov_exp}). 

More details can be found in the reference \cite{hirsch2004}.
\end{enumerate}

%
%
%
%

\section{Mathematical Structure of Generalized Synchronization}

For completeness, we include this brief discussion of the mathematical structure of GS  and it may be skipped without harm to the remaining sections. 
Again lets consider the $\psi : \mathbb{R}^m \rightarrow \mathbb{R}^n$ and the manifold 
$$
M = \{ (x,y) \in \mathbb{R}^n \times \mathbb{R}^m : y = \psi (x) \} \subset \mathbb{R}^{n+m}.
$$
GS corresponds to the case where $M$ is normally attracting. 
Lets review the notion of normally attracting invariant manifold (NAIM). $M$ is normally attracting if it is invariant under the flow $\Phi$ (of the full system) and the dynamics in the directions normal to $M$ is contracting stronger
than in direction tangential to $M$. That is, 
\begin{enumerate}
\item (Invariance)  $\Phi^t(M) = M$ for all $t$,
\item (Normal Contraction) There is a continuous splitting of the tangent space 
$$
\forall x\in M : \,\,\,\, T_{x} \mathbb{R}^{n+m} = T_x M \oplus E_x^s
$$
which is left invariant by $D\Phi^{t}$. Moreover, there is $\eta>0$ such that
$$
\forall t \ge 0, x\in M, v\in E_x^s : \,\,\, \| D \Phi^t(x) v  \| \le C e^{-\eta} \| v \|
$$
\item (Contraction in the normal directions dominate the tangential) There exist $r>1$, $\lambda>0$ and $C>0$ such that 
$$
\forall t\ge 0, x\in M : \,\,\, \| D\Phi^t(x) \left|_{E_x^s} \right. \| \le Ce^{-\lambda t} \| D\Phi^{-t}(x) \left|_{T_x M} \right. \| ^{-r}
$$
\end{enumerate}

If $M$ is a NAIM for the system $F$. Then there exist locally invariant stable manifolds $W_{loc}^s(M)$ 
such that  $W_{loc}^s(M)$ is tangent to $TM\oplus E_s$ at $M$ and $W_s (M) \in C^r$. 
Moreover, $W_{loc}^s(M)$ consists of all points near $M$ whose forward orbit converges to $M$ at rate $e^{-\eta t}$.  For each $y \in W_s(M)$ shadows a point  $x\in M$ such that $y\in W^s_{loc}(x)$ and 
\begin{equation}\label{15}
\| \Phi^t(y) - \Phi^t(x) \| \le Ce^{-\eta t} \| y - x\|
\end{equation}

Since the orbits of points $x_0, x_1 \in M$ cannot approach each other that fast, we can characterize points  $y \in W_{loc}^s (x)$ precisely as those that satisfy Eq. (\ref{15}). Lets consider consider the straithening of the manifold. That is, we introduce new coordinates
$$
u = y - \psi(x)
$$
in this coordinates the manifold corresponds to the $x$ axis and $u$ are the normal directions to $M$. 
Lets take two points $u_1,u_0 \in W_{loc}^s(x)$ then 
$$
u_i = y_i - \psi(x) \Rightarrow u_1 - u_0 = y_1 - y_0
$$ 
Hence, if 
$$
\| y_1 - y_0 \| \le K e^{-\eta t}
$$
and $\eta$ is larger than the smallest Lyapunov exponents of the driver in modulus the manifold $M$ will be  normally attracting, according to condition (3). In fact, $\psi$ will be differentiable.  Another important fact to the mention is that NAIM persist under small perturbations. For us this means that 
once we obtain GS small perturbations such as increasing the coupling strength will be destroy GS \cite{eldering2013,fenichel1972}. If the condition is not satisfied then $\psi$ won't be a NAIM. However, it may still happen that when $r=0$ and $M$ is attracting. In this case, $\psi$ is only continuous. 
This is called strong and weak generalized synchronization \cite{hunt1997}.

{\bf Back to our Diffusively driven oscillator}. In Sec. \ref{XX}  
we showed the contraction rate between two nearby trajectories is 
$$\eta = \alpha \lambda_{min} - \| Dg \| $$

On the other hand, the smallest Lyapunov exponent of the driver is at most 
$- \| Df \|$, hence the condition for normal attraction is 
$$
\eta > \| Df\| \Rightarrow \alpha > \frac{\| Df \| + \| Dg \|}{\lambda_{min}}
$$
This gives the bound for $M$ to be NAIM.

\bibliographystyle{apa}
\bibliography{sync_rev}

\end{document}